\documentclass{article}
\usepackage{amsmath,amssymb,amsfonts,amsthm}
\usepackage[all]{xy}
\usepackage{color}

\usepackage{rotating}

\begin{document}

\newtheorem{theorem}{Theorem}[subsection]
\newtheorem{proposition}[theorem]{Proposition}

\newtheorem{corollary}[theorem]{Corollary}
\newtheorem{observation}[theorem]{Observation}

\newtheorem{theoremapp}{Theorem A.\!\!}
\newtheorem{propositionapp}{Proposition A.\!\!}

\newtheorem{corollaryapp}{Corollary A.\!\!}
\newtheorem{observationapp}{Observation A.\!\!}
\newtheorem{lemmaapp}{Lemma A.\!\!}

\theoremstyle{definition}
\newtheorem{definition}[theorem]{Definition}
\newtheorem{dfn}[theorem]{definition}
\newtheorem{remark}[theorem]{Remark}
\newtheorem{example}[theorem]{Example}

\newtheorem{definitionapp}{Definition A.\!\!}
\newtheorem{dfnapp}{definitionapp}
\newtheorem{remarkapp}{Remark A.\!\!}
\newtheorem{exampleapp}{Example A.\!\!}

\newdir^{ (}{{}*!/-5pt/@^{(}}

\newcommand {\hisham}[1]{{\marginpar{\textcolor{red}{\Huge{$\star$}}}\scriptsize{\bf \textcolor{red}{HISHAM}:}\scriptsize{\ #1 \ }}}

\newcommand {\domenico}[1]{{\marginpar{\textcolor{blue}{\Huge{$\star$}}}\scriptsize{\bf \textcolor{blue}{DOMENICO}:}\scriptsize{\ #1 \ }}}	

\newcommand {\urs}[1]{{\marginpar{\textcolor{green}{\Huge{$\star$}}}\scriptsize{\bf \textcolor{green}{URS}:}\scriptsize{\ #1 \ }}}

\def\proof {{\it Proof.}\hspace{7pt}}

\def\endofproof{\hfill{$\square$}\\}
\def\dfn{Definition}
\def\rmk{Remark}
\def\Cech{{{\v C}ech}}

\def\Loo{{$L_\infty$}}

\title{A higher stacky perspective on Chern-Simons theory}
\author{Domenico Fiorenza, Hisham Sati and Urs Schreiber}

\maketitle	

\begin{abstract}
   The first part of this text is a gentle exposition of some basic constructions and results in 
   the extended prequantum theory of 
    Chern-Simons-type gauge field theories.
   We explain in some detail how the action functional
   of ordinary 3d Chern-Simons theory is naturally localized (``extended'', ``multi-tiered'') 
   to a map on the universal moduli stack of principal connections, a map that 
   itself modulates a circle-principal 3-connection on that moduli stack, and how the 
   iterated transgressions of this extended Lagrangian unify the action functional with
   its prequantum bundle and with the WZW-functional. In the second part we provide a brief review and 
   outlook of the higher prequantum field theory of which this is a first example. 
   This includes a higher geometric description of supersymmetric Chern-Simons theory,
   Wilson loops and other defects,
   generalized geometry, higher Spin structures, anomaly cancellation, 
   and various other aspects of quantum field theory.
\end{abstract}

\tableofcontents

\section{Introduction}

One of the fundamental examples of quantum field theory is 3-dimensional Chern-Simons gauge field theory
as introduced in \cite{Witten}. We give a pedagogical exposition of this from a new, natural, perspective 
of \emph{higher geometry} formulated using \emph{higher stacks} in \emph{higher toposes} along the lines 
of \cite{FiorenzaSatiSchreiberIII} and references given there. Then we indicate how this opens the door to 
a more general understanding of \emph{extended prequantum} (topological) field theory, constituting
a pre-quantum analog of the extended quantum field theory as in \cite{LurieTQFT}, in the sense 
of higher geometric quantization \cite{Rogers}. An indication of the general mechanism by
which the extended prequantum theory described here is supposed to induce an extended quantum field theory can be found in \cite{Nuiten}.

\par
The aim of this text is twofold. On the one hand, we will attempt to dissipate the false belief that higher toposes are an esoteric discipline whose secret rites are reserved to initiates. To do this we will present a familiar example from differential topology, namely \emph{Chern-Simons theory}, from the perspective of higher stacks, to show how this is a completely natural and powerful language in differential geometry. Furthermore,  since any language is best appreciated by listening to it rather than by studying its grammar, in this presentation we will omit most of
the rigorous definitions, leaving the reader the task to imagine and reconstruct them from the context. 
Clearly this does not mean that such definitions are not available: we refer the interested reader to \cite{lurie} for the general theory of higher toposes and to \cite{Hab, survey} for general theory and applications 
of \emph{differential cohesive} higher toposes that can express differential geometry, differential cohomology
and prequantum gauge field theory; the reader
interested in the formal mathematical aspects of the theory might enjoy  looking at \cite{SchreiberShulman}.
\par
On the other hand, the purpose of this note is not purely pedagogical: we show how the stacky approach unifies in a natural way all the basic constructions in classical Chern-Simons theory (e.g., the action functional, the Wess-Zumino-Witten bundle gerbe, the symplectic structure on the moduli space of flat $G$-bundles as well as its prequantization),  clarifies the relations of these with differential cohomology, and clearly points towards ``higher Chern-Simons theories" and their higher and extended geometric prequantum theory. 
A brief survey and outlook of this more encompassing theory is given in the last sections. This is based on 
our series of articles including 
\cite{SSSI, SSSII, SSSIII}, \cite{FSS} and
\cite{FiorenzaSatiSchreiberI, FiorenzaSatiSchreiberII, FiorenzaSatiSchreiberIII}.
A set of lecture notes explaining this theory is \cite{TwistedStructuresLecture}.

\par
We assume the reader has a basic knowledge of characteristic classes and of Chern-Simons theory. 
Friendly, complete and detailed introductions to these two topics can be found in  
\cite{milnor-stasheff} and \cite{CS, FreedCS1, FreedCS2, FreedCS3}, respectively.

\par
In this article we focus on the (extended) \emph{geometric quantization}
of Chern-Simons theory. Another important approach is the (extended) 
\emph{perturbative quantization} based on path integrals in the BV-BRST formalism, as discussed notably 
in \cite{AlekseevBarmazMnev}, based
on the general program of extended perturbative BV-quantization laid out in 
\cite{Cattaneo-Mnev-ReshetikhinI, Cattaneo-Mnev-ReshetikhinII}. 
The BV-BRST formalism -- a description of phase spaces/critical loci in higher (``derived'') geometry --  
is also naturally formulated in terms of the higher cohesive geometry of higher stacks that we consider here, but
further discussion of this point goes beyond the scope of this article.
The interested reader can find more discussion in section 1.2.15.2 and 3.10.8 of \cite{survey}.

\section{A toy example: 1-dimensional $U(n)$-Chern-Simons theory}
\label{toy-example}

Before describing the archetypical 3-dimensional Chern-Simons theory with a compact simply connected gauge group 
\footnote{We are using the term ``gauge group" to refer to the structure group of the theory. 
This is not to be confused with the group of gauge transformations.}
from a stacky perspective, here we first look from this point of view at 1-dimensional Chern-Simons theory with gauge group $U(n)$. Although this is a very simplified version, still it will show in an embryonic way all the features of the higher dimensional theory.\footnote{Even 1-dimensional Chern-Simons theory exhibits a rich structure
once we pass to {\it derived} higher gauge groups 
as in 
\cite{GG}.  This goes beyond the present exposition, but see section 
\ref{Sigma-Models} below for an outlook and section 5.7.10 of \cite{survey} for more details.}
Moreover, a slight variant of this 1-dimensional CS theory shows up as a component of 3d Chern-Simons
theory \emph{with Wilson line defects}, this we discuss at the end of the exposition part in 
section \ref{WithWilsonLoops}.

\subsection{The basic definition}
\label{TheBasicDefinition}

Let $A$ be a $\mathfrak{u}_n$-valued differential 1-form on the circle $S^1$. 
Then $\frac{1}{2\pi i}\mathrm{tr}(A)$ is a real-valued 1-form, which we can integrate over $S^1$ to get a real number. This construction can be geometrically interpreted as a map
\[
\{\text{trivialized $U(n)$-bundles with connections on $S^1$}\}\xrightarrow{\frac{1}{2\pi i}\int_{S^1}\mathrm{tr}}\mathbb{R}\;.
\]
Since the Lie group $U(n)$ is connected, the classifying space $BU(n)$ of principal $U(n)$-bundles is simply connected, and so the set of homotopy classes of maps from $S^1$ to $BU(n)$ is trivial. By the characterizing property of the classifying space, this set is the set of isomorphism classes of principal $U(n)$-bundles on $S^1$, and so every principal $U(n)$-bundle over $S^1$ is trivializable. Using a chosen trivialization to pull-back the connection, we see that  an arbitrary $U(n)$-principal bundle with connection $(P,\nabla)$ is (noncanonically) isomorphic to a trivialized bundle with connection, and so our picture enlarges to
\[
\xymatrix{
\{\text{trivialized $U(n)$-bundles with connections on $S^1$}\}\ar@{>>}[d]\ar[rr]^{\phantom{mmmmmmmmmmmmmmm}\frac{1}{2\pi i}\int_{S^1}\mathrm{tr}}&&\mathbb{R}\\
\{\text{$U(n)$-bundles with connections on $S^1$}\}/\text{iso}&&
}
\]
and it is tempting to fill the square by placing a suitable quotient of $\mathbb{R}$ in the right bottom corner. To see that this is indeed possible, we have to check what happens when we choose two different trivializations for the same bundle, i.e., we have to compute the quantity
\[
\tfrac{1}{2\pi i}\int_{S^1}\mathrm{tr}(A')-\mathrm{tr}(A)\;,
\]
where $A$ and $A'$ are two 1-form incarnations of the same connection $\nabla$
under different trivializations of the underlying bundle. 
What one finds is that this quantity is always an integer, thus giving a commutative diagram 
\[
\xymatrix{
\{\text{trivialized $U(n)$-bundles with connections on $S^1$}\}\ar@{>>}[d]\ar[rr]^{\phantom{mmmmmmmmmmmmmmm}\frac{1}{2\pi i}\int_{S^1}\mathrm{tr}}&&\mathbb{R}\ar@{>>}[d]\\
\{\text{$U(n)$-bundles with connections on $S^1$}\}/\text{iso}\ar[rr]^{\phantom{mmmmmmmmmmm}\exp{\int_{S^1}\mathrm{tr}}}&&U(1)\;.
}
\]
The bottom line in this diagram is the \emph{1-dimensional Chern-Simons action for $U(n)$-gauge theory}.
An 
elegant way of proving that $\frac{1}{2\pi i}\int_{S^1}\mathrm{tr}(A)-\mathrm{tr}(A')$ is always an integer is as follows. Once a trivialization has been chosen, one can extend a principal $U(n)$-bundle with connection $(P,\nabla)$ on $S^1$ to a trivialized  principal $U(n)$-bundle with connection over the disk $D^2$. Denoting by the same symbol $\nabla$ the extended connection and by $A$ the 1-form representing it, then by Stokes' theorem we have 
\[
\tfrac{1}{2\pi i}\int_{S^1}\mathrm{tr}(A)=\tfrac{1}{2\pi i}\int_{\partial D^2}\mathrm{tr}(A)=
\tfrac{1}{2\pi i}\int_{D^2}d\mathrm{tr}(A)=\tfrac{1}{2\pi i}\int_{D^2}\mathrm{tr}(F_\nabla)\;,
\]
where $F_\nabla$ is the curvature of $\nabla$. If we choose two distinct trivializations, what we get are two trivialized principal $U(n)$-bundles with connection over $D^2$ together with an isomorphism of their boundary data. Using this isomorphism to glue together the two bundles, we get a (generally nontrivial) $U(n)$-bundles with connection $(\tilde{P},\tilde{\nabla})$ on 
$S^2=D^2\coprod_{S^1}D^2$, the disjoint union of the upper and lower hemisphere glued along the 
equator,  and
\[
\tfrac{1}{2\pi i}\int_{S^1}\mathrm{tr}(A')-\mathrm{tr}(A)=\tfrac{1}{2\pi i}\int_{S^2}\mathrm{tr}(\tilde{\nabla})=\langle c_1(\tilde{P}), [S^2]\rangle\;,
\]
the first Chern number of the bundle $\tilde{P}$. Note how the generator $c_1$ of the second integral cohomology group 
$H^2(BU(n),\mathbb{Z})\cong \mathbb{Z}$ has come into play. Also notice how, by the above considerations, one could have actually \emph{defined} the 1-dimensional Chern-Simons action as
\[
\nabla\mapsto \exp\int_{D^2}\mathrm{tr}(F_{\tilde{\nabla}}),
\]
where $\tilde{\nabla}$ is any extension of $\nabla$ to $D^2$.
\par
Despite its elegance, the argument above has a serious drawback: it relies on the fact that $S^1$ is a boundary. And, although this is something obvious, still it is something nontrivial and indicates that generalizing 1-dimensional Chern-Simons theory to higher dimensional Chern-Simons theory along the above lines will force limiting the construction to those manifolds which are boundaries. For standard 3-dimensional Chern-Simons theory with a compact simply connected gauge group, this will
actually be no limitation, since the oriented cobordism ring is trivial in dimension 3, but one sees that this 
is a much
less trivial statement than saying that $S^1$ is a boundary. However, in any case, that would definitely 
not be true in general for higher dimensions,  as well as for topological structures on manifolds
beyond orientations.

\subsection{A Lie algebra cohomology approach}\label{lie-algebra-approach}
A way of avoiding the cobordism argument used  in the previous section is to focus on the fact that
\[
\tfrac{1}{2\pi i} \mathrm{tr}: \mathfrak{u}_n\to \mathbb{R}
\]
is a Lie algebra morphism, i.e., it is a real-valued 1-cocycle on the Lie algebra 
$\mathfrak{u}_n$ of the group 
$U(n)$. A change of trivialization for a principal $U(n)$-bundle $P\to S^1$ is given by a gauge transformation $g:S^1\to U(n)$. If $A$ is the $\mathfrak{u}_n$-valued 1-form corresponding to the connection $\nabla$ in the first trivialization, the gauge-transformed 1-form $A'$ is given by
\[
A'=g^{-1}Ag+g^{-1}dg\;,
\]
where $g^{-1}dg=g^*\theta_{U(n)}$ is the pullback of the Maurer-Cartan form $\theta_{U(n)}$ of $U(n)$ via $g$. Since $\frac{1}{2\pi i}\mathrm{tr}$ is an invariant polynomial (i.e., it is invariant under the adjoint action of $U(n)$ on $\mathfrak{u}_n$), it follows that
\[
\tfrac{1}{2\pi i}\int_{S^1} \mathrm{tr}(A')- \mathrm{tr}(A)=\tfrac{1}{2\pi i}\int_{S^1} g^*\mathrm{tr}(\theta_{U(n)})\;,
\]
and our task is reduced to showing that the right-hand term is a ``quantized'' quantity, i.e., that it always assumes integer values. Since the Maurer-Cartan form satisfies the Maurer-Cartan equation
\[
d\theta_{U(n)}+\tfrac{1}{2}[\theta_{U(n)},\theta_{U(n)}]=0\;,
\] 
we see that
\[
d\mathrm{tr}(\theta_{U(n)})=-\tfrac{1}{2}\mathrm{tr}\bigl([\theta_{U(n)},\theta_{U(n)}]\bigr)=0\;,
\]
i.e., $\mathrm{tr}(\theta_{U(n)})$ is a closed 1-form on $U(n)$. As an immediate consequence,
\[
\tfrac{1}{2\pi i}\int_{S^1} g^*\mathrm{tr}(\theta_{U(n)})=\langle g^*[\tfrac{1}{2\pi i}\mathrm{tr}(\theta_{U(n)})], [S^1]\rangle
\]
only depends on the homotopy class of $g:S^1\to U(n)$, and these homotopy classes are parametrized by the additive group $\mathbb{Z}$ of the integers. Notice how the generator $[\tfrac{1}{2\pi i}\mathrm{tr}(\theta_{U(n)})]$ of $H^1(U(n);\mathbb{Z})$ has appeared. This shows how this proof is related to the one in the previous section via the transgression isomorphism $H^1(U(n);\mathbb{Z})\to H^2(BU(n);\mathbb{Z})$.
\par
It is useful to read the transgression isomorphism in terms of differential forms by passing to real coefficients and pretending
that  $BU(n)$ is a finite dimensional smooth manifold. This can be made completely rigorous in various ways, e.g., by 
looking at $BU(n)$ as an inductive limit of finite dimensional Grasmannians. Then a connection on the universal $U(n)$-bundle 
$EU(n)\to BU(n)$ is described \`a la Ehresmann by a $\mathfrak{u}_n$-valued $U(n)$-equivariant 1-form $A$ on $EU(n)$
 which gives the Maurer-Cartan form when restricted to the fibers. The $\mathbb{R}$-valued 1-form
$\frac{1}{2\pi i}\mathrm{tr}(A)$ restricted to the fibers gives the closed 1-form  $\frac{1}{2\pi i}\mathrm{tr}(\theta_{U(n)})$ which is the generator of $H^1(U(n),\mathbb{R})$; the differential $d\frac{1}{2\pi i}\mathrm{tr}(A)=\frac{1}{2\pi i}\mathrm{tr}(F_A)$ is an exact 2-form on $EU(n)$ which is $U(n)$-invariant and so is the pullback of a closed 2-form on $BU(n)$ which, since it represents the first Chern class, 
is the generator of $H^2(U(n),\mathbb{R})$.
\par
One sees that $\frac{1}{2\pi i}\mathrm{tr}$ plays a triple role in the above description, which might be 
initially confusing. To get a better understanding of what is going on, let us consider more generally an arbitrary compact connected Lie group $G$. Then the transgression isomorphism between $H^1(G,\mathbb{R})$ and $H^2(BG;\mathbb{R})$ is realized by a Chern-Simons element $\mathrm{CS}_1$ for the Lie algebra $\mathfrak{g}$. This element is characterized by the following property: for $A\in \Omega^1(EG;\mathfrak{g})$ the connection 1-form of a principal $G$-connection on $EG\to BG$, we have the following transgression diagram
\[
\xymatrix{
\langle F_A\rangle & \mathrm{CS}_1(A)\ar@{|->}[l]_{d}\ar@{|->}[rr]^{A=\theta_G}&&\mu_1(\theta_G)\;,
}
\] 
where on the left hand side $\langle\,- \rangle$ is a degree 2 invariant polynomial on $\mathfrak{g}$,
 and on the right hand side $\mu_1$ is 1-cocycle on $\mathfrak{g}$. One says that $\mathrm{CS}_1$ transgresses $\mu_1$ to $\langle\,-\rangle$. Via the identification of $H^1(G;\mathbb{R})$ with the degree one Lie algebra cohomology $H^1_{\mathrm{Lie}}(\mathfrak{g};\mathbb{R})$ and of $H^2(BG;\mathbb{R})$ with the vector space of degree 2 elements in the graded algebra $\mathrm{inv}(\mathfrak{g})$ (with elements of $\mathfrak{g}^*$ placed in degree 2), one sees that this indeed realizes the transgression isomorphism.

\subsection{The first Chern class as a morphism of stacks}\label{first-chern}
Note that,
by the end of the previous section, 
 the base manifold $S^1$ has completely disappeared. 
 This suggests that one should be able to describe 1-dimensional Chern-Simons theory with gauge group $U(n)$ more generally as a map
\[
\{\text{$U(n)$-bundles with connections on $X$}\}/\text{iso}\rightarrow \mathbf{??}\;,
\]
where now $X$ is an arbitrary manifold, and ``$\mathbf{??}$" is some natural target to be determined. To try to figure out what this natural target could be, let us look at something simpler and forget the connection. Then we know that the first Chern class gives a morphism of sets
\[
c_1:\{\text{$U(n)$-bundles on $X$}\}/\text{iso}\rightarrow H^2(X;\mathbb{Z})\;.
\]
Here the right hand side is much closer to the left hand side than it might appear at first sight. Indeed, the second integral cohomology group of $X$ precisely classifies principal $U(1)$-bundles on $X$ up to isomorphism, so that the first Chern class is actually a map
\[
c_1:\{\text{$U(n)$-bundles on $X$}\}/\text{iso}\rightarrow \{\text{$U(1)$-bundles over $X$}\}/\text{iso}\;.
\] 
Writing $\mathbf{B}U(n)(X)$ and $\mathbf{B}U(1)(X)$ for the groupoids 
of principal $U(n)$- and $U(1)$-bundles over $X$, respectively,
\footnote{That is, for the collections of all such
bundles, with bundle isomorphisms as morphisms.} 
one can further rewrite $c_1$ as a function
\[
c_1:\pi_0\mathbf{B}U(n)(X)\to \pi_0\mathbf{B}U(1)(X)
\]
between the connected components of these groupoids.
This immediately leads one to suspect that $c_1$ could actually be $\pi_0(\mathbf{c}_1(X))$ for some morphism of groupoids $\mathbf{c}_1(X):\mathbf{B}U(n)(X)\to \mathbf{B}U(1)(X)$. Moreover, naturality of the first Chern class suggests that, independently of $X$, there should actually be a morphism of stacks
\[
  \mathbf{c}_1:\mathbf{B}U(n)\to \mathbf{B}U(1)
\]
over the site of smooth manifolds.\footnote{The reader
unfamiliar with the language of higher stacks and simplicial presheaves in differential 
geometry can find an introduction in \cite{FSS}.} Since a smooth manifold is built by patching together,
in a smooth way, open balls of $\mathbb{R}^n$ for some $n$, this in turn is equivalent to saying that $\mathbf{c}_1:\mathbf{B}U(n)\to \mathbf{B}U(1)$ is a morphism of stacks over the full sub-site of Cartesian spaces, where by definition a Cartesian space is a smooth manifold diffeomorphic to $\mathbb{R}^n$ for some $n$. 
To see that $c_1$ is indeed induced by a morphism of stacks, notice that $\mathbf{B}U(n)$ can be obtained by stackification from the simplicial presheaf which to a Cartesian space $U$ associates the nerve of the action groupoid\footnote{Given a set $X$ with an action of a group $G$ on it, the action groupoid $X/\!/G$ is the (small) groupoid having $X$ as set of objects and with $\mathrm{Hom}_{X/\!/G}(x,y)=\{g\in G\text{ such that }g\cdot x=y\}$. The composition of morphisms is given by the product in $G$. }  $*/\!/C^\infty(U;U(n))$. This is nothing but saying, in a very compact way, that to give a principal $U(n)$-bundle on a smooth manifold $X$ one picks a good open cover $\mathcal{U}=\{U_\alpha\}$ of $X$ and local data given by smooth functions on the double intersection
\[
g_{\alpha\beta}:U_{\alpha\beta}\to U(n)
\]
such that $g_{\alpha\beta}g_{\beta\gamma}g_{\gamma\alpha}=1$ on the triple intersection,
$U_{\alpha\beta\gamma}$. The group homomorphism 
\[
\det: U(n)\to U(1)
\]
maps local data $\{g_{\alpha\beta}\}$ for a principal $U(n)$ bundle to local data $\{h_{\alpha\beta}=\det(g_{\alpha\beta})\}$ for a principal $U(1)$-bundle and, by the basic properties of the first Chern class, one sees that
\[
\mathbf{B}\!\det: \mathbf{B}U(n)\to \mathbf{B}U(1)
\] 
induces $c_1$ at the level of isomorphism classes, i.e., one can take $\mathbf{c}_1=\mathbf{B}\!\det$.

Note that there is a canonical notion of \emph{geometric realization} of stacks on smooth manifolds by topological 
spaces (see section 4.3.4.1 of \cite{survey}). Under this realization the morphism of stacks $\mathbf{B}\mathrm{det}$ becomes a 
continuous function of classifying spaces $B U(n) \to K(\mathbb{Z},2)$ which represents the
universal first Chern class.

\subsection{Adding connections to the picture}\label{adding-connections} 
The above discussion suggests that what should really lie behind 1-dimensional 
Chern-Simons theory with gauge group $U(n)$ is a morphism of stacks 
\[
\hat{\mathbf{c}}_1:\mathbf{B}U(n)_{\mathrm{conn}}\to \mathbf{B}U(1)_{\mathrm{conn}}
\]
from the stack of $U(n)$-principal bundles with connection to the stack of $U(1)$-principal bundles with connection, lifting the first Chern class. This morphism is easily described, as follows.
Local data for a $U(n)$-principal bundle with connection on a smooth manifold $X$ are
\begin{itemize}
\item smooth $\mathfrak{u}_n$-valued 1-forms $A_\alpha$ on $U_\alpha$;
\item smooth functions $g_{\alpha\beta}\colon U_{\alpha\beta}\to U(n)$,
\end{itemize}
such that
\begin{itemize}
\item $A_\beta=g_{\alpha\beta}^{-1}A_\alpha g_{\alpha\beta}+g_{\alpha\beta}^{-1}dg_{\alpha\beta}$ on $U_{\alpha\beta}$;
\item $g_{\alpha\beta}g_{\beta\gamma}g_{\gamma\alpha}=1$ on $U_{\alpha\beta\gamma}$,
\end{itemize}
and this is equivalent to saying that $\mathbf{B}U(n)_{\mathrm{conn}}$ is 
the stack of simplicial sets
\footnote{
It is noteworthy that this indeed is a stack on the site $\mathrm{CartSp}$. On the larger but equivalent site of 
all smooth manifolds it is just a prestack that needs to be further stackified.}
which to a Cartesian space $U$ assigns the nerve of the action groupoid
\[
\Omega^1(U;\mathfrak{u}_n)/\!/C^\infty(U;U(n))\;,
\]
where the action is given by
$
g:A\mapsto g^{-1}Ag+g^{-1}dg$.
To give a morphism $\hat{\mathbf{c}}_1:\mathbf{B}U(n)_{\mathrm{conn}}\to \mathbf{B}U(1)_{\mathrm{conn}}$ we therefore just need to give a morphism of simplicial prestacks
\[
\mathcal{N}(\Omega^1(-;\mathfrak{u}_n)/\!/C^\infty(-;U(n)))\longrightarrow \mathcal{N}(\Omega^1(-;\mathfrak{u}_1)/\!/C^\infty(-;U(1)))
\]
lifting
\[
\mathbf{B}\!\det:\mathcal{N}(*/\!/C^\infty(-;U(n)))\longrightarrow \mathcal{N}(*/\!/C^\infty(-;U(1)))\;,
\]
where $\mathcal{N}$ is the nerve of the indicated groupoid.
In more explicit terms, we have to give a natural linear morphism 
\[
\varphi:\Omega^1(U;\mathfrak{u}_n)\to \Omega^1(U;\mathfrak{u}_1)\;,
\]
 such that
\[
\varphi(g^{-1}Ag+g^{-1}dg)=\varphi(A)+\det(g)^{-1}d \det(g)\;,
\]
and it is immediate to check that the linear map 
\[
\mathrm{tr}: \mathfrak{u}_n\to \mathfrak{u}_1
\]
does indeed induce such a morphism $\varphi$. In the end we get a commutative diagram of stacks
\[
\xymatrix{
\mathbf{B}U(n)_{\mathrm{conn}}\ar[d]\ar[r]^-{\hat{\mathbf{c}}_1}& \mathbf{B}U(1)_{\mathrm{conn}}\ar[d]\\
\mathbf{B}U(n)\ar[r]^-{\mathbf{c}_1}& \mathbf{B}U(1)\;,
}
\]
where the vertical arrows forget the connections.

\subsection{Degree 2 differential cohomology}\label{degree-2}
If we now fix a base manifold $X$ and look at isomorphism classes 
of principal $U(n)$-bundles (with connection) on $X$, we get a commutative diagram of sets
\[
\xymatrix{
{\{\text{$U(n)$-bundles with connection on $X$}\}/\text{iso}}\ar[d]\ar[r]^{\phantom{mmmmmmmmmm}\hat{c}}& \hat{H}^2(X;\mathbb{Z})\ar[d]\\
{\{\text{$U(n)$-bundles on $X$}\}/\text{iso}}\ar[r]^{\phantom{mmmm}c}& {H}^2(X;\mathbb{Z})\;,
}
\]
where $\hat{H}^2(X;\mathbb{Z})$ is the second differential cohomology group of $X$. This is defined as the degree 0 hypercohomology group of $X$ with coefficients in the two-term Deligne complex, i.e., in the sheaf of complexes
\[
C^\infty(-;U(1))\xrightarrow{\frac{1}{2\pi i}d\mathrm{log}} \Omega^1(-;\mathbb{R})\;,
\] 
with $\Omega^1(-;\mathbb{R})$ in degree zero \cite{brylinski,gajer}.
That $\hat{H}^2(X;\mathbb{Z})$ classifies principal $U(1)$-bundles with connection is manifest by this description: via the Dold-Kan correspondence, the sheaf of complexes indicated above precisely gives a simplicial presheaf which produces $\mathbf{B}U(1)_{\mathrm{conn}}$ via stackification. Note that we have two natural morphisms  of complexes of sheaves
\[
\raisebox{20pt}{
\xymatrix{
C^\infty(-;U(1))\ar[d]\ar[r]^-{\frac{1}{2\pi i}d\mathrm{log}}&\Omega^1(-;\mathbb{R})\ar[d]\\
C^\infty(-;U(1))\ar[r]&0
}}
\qquad \quad
{\rm and}
\qquad \quad
\raisebox{20pt}{
\xymatrix{
C^\infty(-;U(1))\ar[d]\ar[r]^-{\frac{1}{2\pi i}d\mathrm{log}}&\Omega^1(-;\mathbb{R})\ar[d]\\
0\ar[r]&\Omega^2(-;\mathbb{R})_{\mathrm{cl}}\;.
}
}
\]
The first one induces the forgetful morphism $\mathbf{B}U(1)_{\mathrm{conn}}\to\mathbf{B}U(1)$, 
while the second one induces the curvature morphism 
$F_{(-)} : \mathbf{B}U(1)_{\mathrm{conn}}\to \Omega^2(-;\mathbb{R})_{\mathrm{cl}}$ mapping a $U(1)$-bundle with connection to its curvature 2-form. From this one sees that degree 2 differential cohomology implements in a natural geometric way the simple idea of having an integral cohomology class together with a closed 2-form representing it in de Rham cohomology. 

The last step that we need to recover the 1-dimensional Chern-Simons action functional 
from section \ref{TheBasicDefinition} 
is to give a natural morphism 
\[
\mathrm{hol}: \hat{H}^2(S^1;\mathbb{Z})\to U(1)
\]
so as to realize the 1-dimensional Chern-Simons action functional as the composition
\[
\xymatrix{
\mathrm{CS}_1:{\{\text{$U(n)$-bundles with connection on $X$}\}/\text{iso}}
  \ar[r]^-{\hat{c}}& \hat{H}^2(X;\mathbb{Z})\ar[r]^-{\mathrm{hol}}&U(1)}.
\]
As the notation ``$\mathrm{hol}$" suggests, this morphism is nothing but the holonomy morphism mapping a principal $U(1)$-bundle with connection on $S^1$ to its holonomy.
\par
An enlightening perspective from which to look at this situation 
is in terms of fiber integration and moduli stacks of principal $U(1)$-bundles with connections over a base manifold $X$. Namely, for a fixed $X$ we can consider the \emph{mapping stack}
\[
\mathbf{Maps}(X,\mathbf{B}U(1)_{\mathrm{conn}})
\,,
\] 
which is presented by the simplicial presheaf
that sends a Cartesian space $U$ to the nerve of the groupoid of principal $U(1)$-bundles with connection on $U\times X$. In other words, $\mathbf{Maps}(X,\mathbf{B}U(1)_{\mathrm{conn}})$ is the 
\emph{internal hom} space between $X$ and $\mathbf{B}U(1)_{\mathrm{conn}}$ in the category of simplicial sheaves over the site of smooth manifolds. Then, if $\Sigma_1$ is an oriented compact manifold of dimension one, the fiber integration formula from \cite{gomi-terashima1,gomi-terashima2}  can be naturally interpreted as a morphism of simplicial sheaves
\[
\mathrm{hol}_{\Sigma_1}: \mathbf{Maps}(\Sigma_1,\mathbf{B}U(1)_{\mathrm{conn}}) \to \underline{U}(1)\;,
\]
where on the right one has the sheaf of 
smooth
functions with values in $U(1)$. Taking global sections over the point one gets the morphism of simplicial sets
\[
\mathrm{hol}_{\Sigma_1}:\mathbf{H}(\Sigma_1,\mathbf{B}U(1)_{\mathrm{conn}})\to U(1)_{\mathrm{discr}}\;,
\]
where on the right the Lie group $U(1)$ is seen as a 0-truncated simplicial object and where $\mathbf{H}(\Sigma_1,\mathbf{B}U(1)_{\mathrm{conn}})$ is 
(the nerve of) the groupoid of principal $U(1)$-bundles with connection on $X$. Finally, 
passing to isomorphism classes/connected components one gets the morphism 
\[
\hat{H}^2(\Sigma_1;\mathbb{Z})\to U(1)\;.
\]
This morphism can also be described in purely algebraic terms by noticing that for any 1-dimensional oriented compact manifold $\Sigma_1$ the short exact sequence of complexes of sheaves
\[
\xymatrix{
0\ar[r] &0\ar[r]\ar[d]&C^\infty(-;U(1))\ar[r]\ar[d]^{\frac{1}{2\pi i}d\mathrm{log}}&C^\infty(-;U(1))\ar[r]\ar[d]&0\\
0\ar[r] &\Omega^1(-;\mathbb{R})\ar[r]&\Omega^1(-;\mathbb{R})\ar[r]&0\ar[r]&0
}
\]
induces an isomorphim
\[
\Omega^1(\Sigma_1)/\Omega^1_{\mathrm{cl},\mathbb{Z}}(\Sigma_1)\xrightarrow{\sim}\hat{H}^{2}(\Sigma_1;\mathbb{Z}) 
\]
in hypercohomology,
where $\Omega^1(\Sigma_1)/\Omega^1_{\mathrm{cl},\mathbb{Z}}(\Sigma_1)$ is the group of differential $1$-forms on $\Sigma_1$ modulo those $1$-forms which are closed and have integral periods. In terms of this isomorphism, the holonomy map is realized as the composition
\[
\hat{H}^{2}(\Sigma_1;\mathbb{Z})\xrightarrow{\sim}\Omega^1(\Sigma_1)/\Omega^1_{\mathrm{cl},\mathbb{Z}}(\Sigma_1)\xrightarrow{\exp\left(2\pi i \int_{\Sigma_1}-\right)}U(1).
\]

\subsection{The Brylinski-McLaughlin 2-cocycle}\label{bml-2}
It is natural to expect that the lift of the universal first Chern class $c_1$ to a morphism of stacks 
$\mathbf{c}_1 : \mathbf{B}U(n)_{\mathrm{conn}}\to \mathbf{B}U(1)_{\mathrm{conn}}$ is a particular case of a more general construction 
that holds for the generator $c$ of the second integral cohomology group of an arbitrary compact connected Lie group $G$ with $\pi_1(G)\cong \mathbb{Z}$. Namely, if $\langle\,-\rangle$ is the degree 2 invariant polynomial on $\mathfrak{g}[2]$ corresponding to the characteristic class $c$, then for any $G$-connection $\nabla$ on a principal $G$-bundle $P\to X$ one has that $\langle F_\nabla\rangle$ is a closed 2-form on $X$ representing the integral class $c$. This precisely suggests that $(P,\nabla)$ defines an element in degree 2 differential cohomology, giving a map
\[
\{\text{$G$-bundles with connection on $X$}\}/\text{iso}\to \hat{H}^2(X;\mathbb{Z}).
\]
That this is indeed so can be seen following Brylinski and McLaughlin 
\cite{brylinski-mclaughlin}
(see \cite{bm:pont} for an exposition an \cite{bm:geom4-I,bm:geom4-II} for related discussion). 
Let $\{A_\alpha,g_{\alpha\beta}\}$ the local data for a $G$-connection on $P\to X$, 
relative to a trivializing good open cover $\mathcal{U}$ of $X$. Then, since $G$ is connected and the open sets $U_{\alpha\beta}$ are contractible, we can smoothly extend the transition functions 
$g_{\alpha\beta}:U_{\alpha\beta}\to G$ to 
functions $\hat{g}_{\alpha\beta}:[0,1]\times U_{\alpha\beta}\to G$ with $\hat{g}_{\alpha\beta}(0)=e$, the identity element of $G$, and $\hat{g}_{\alpha\beta}(1)=g_{\alpha\beta}$. Using the functions $\hat{g}_{\alpha\beta}$ one can interpolate from $A_\alpha\bigr\vert_{U_{\alpha\beta}}$ to $A_\beta\vert_{U_{\alpha\beta}}$ by defining the $\mathfrak{g}$-valued 1-form
\[
\hat{A}_{\alpha\beta}=\hat{g}_{\alpha\beta}^{-1}A_\alpha\vert_{U_{\alpha\beta}}\hat{g}_{\alpha\beta}+\hat{g}_{\alpha\beta}^{-1}d\hat{g}_{\alpha\beta}
\]
on $U_{\alpha\beta}$. 
Now pick a real-valued 1-cocycle $\mu_1$ on the Lie algebra $\mathfrak{g}$ representing the cohomology class $c$ and a Chern-Simons element $\mathrm{CS}_1$ realizing the transgression from $\mu_1$ to $\langle\,-\rangle$. Then the element 
\[
        (\mathrm{CS}_1(A_\alpha), \int_{\Delta^1} \mathrm{CS}_1(\hat A_{\alpha\beta}) ~\,\mathrm{mod}\,\mathbb{Z})
\]
is a degree 2 cocycle in the  {\Cech}-Deligne total complex lifting the cohomology class $c\in H^2(BG,\mathbb{Z})$ to a differential cohomology class $\hat{c}$. Notice how modding out by $\mathbb{Z}$ in  the integral  $\int_{\Delta^1} \mathrm{CS}_1(\hat A_{\alpha\beta})$ precisely takes care of $G$ being connected but not simply connected, with $H^1(G;\mathbb{Z})\cong \pi_1(G)\cong \mathbb{Z}$. That is, choosing two different extensions $\hat{g}_{\alpha\beta}$ of $g_{\alpha\beta}$ will produce two different values for that integral, but their difference will lie in the rank 1 lattice of 1-dimensional periods of $G$,  and with the correct normalization this will be a copy of $\mathbb{Z}$.
\par
A close look at the construction of  Brylinski and McLaughlin, see \cite{FSS},
 reveals that it actually provides a refinement of the characteristic class $c\in H^2(BG;\mathbb{Z})$ to a commutative diagram of stacks
\[
\xymatrix{
\mathbf{B}G_{\mathrm{conn}}\ar[d]\ar[r]^-{\hat{\mathbf{c}}}& \mathbf{B}U(1)_{\mathrm{conn}}\ar[d]\\
\mathbf{B}G\ar[r]^-{{\mathbf{c}}}& \mathbf{B}U(1)\;.
}
\]

\subsection{The presymplectic form on $\mathbf{B}U(n)_{\mathrm{conn}}$}
\label{presymplectic}

In geometric quantization it is customary
 to call \emph{pre-quantization} of a symplectic manifold $(M,\omega)$ the datum of a 
 $U(1)$-principal bundle with connection on $M$ whose curvature 
form is $\omega$.\footnote{See for instance \cite{Kostant} for an original reference 
on geometric quantization, \cite{Woodhouse} for a comprehensive account, and \cite{Rogers} for further pointers.}
Furthermore, it is shown that most of the good features of symplectic manifolds continue to hold under the weaker hypothesis that the 2-form $\omega$ is only closed; this leads to introducing the term 
\emph{pre-symplectic manifold} to denote a smooth manifold equipped with a  closed $2$-form $\omega$ and 
to  speak of \emph{prequantum line bundles} for these. In terms of the morphisms of stacks described in the previous sections, a prequantization of a presymplectic manifold is a lift of the morphism $\omega: M\to \Omega^2(-\mathbb{R})_{\mathrm{cl}}$ to a map $\nabla$ fitting into a commuting diagram
\[
 \raisebox{20pt}{
\xymatrix{
& \mathbf{B}U(1)_{\mathrm{conn}}\ar[d]^{F_{(-)}}
  \\
  M\ar[r]^-{\omega}\ar[ur]^{\nabla}&\Omega^2(-;\mathbb{R})_{\mathrm{cl}}\;,
}
}
\]
where the vertical arrow is the curvature morphism. 
From this perspective there is no reason to restrict $M$ to being a manifold. 
By taking $M$ to be the universal moduli stack $\mathbf{B}U(n)_{\mathrm{conn}}$, we see that 
the morphism $\hat{\mathbf{c}}_1$ can be naturally interpreted as giving 
a canonical prequantum line bundle over $\mathbf{B}U(n)_{\mathrm{conn}}$, whose curvature 2-form
\[
\omega_{\mathbf{B}U(n)_{\mathrm{conn}}}:\mathbf{B}U(n)_{\mathrm{conn}}\xrightarrow{\hat{\mathbf{c}}_1}\mathbf{B}U(1)_{\mathrm{conn}}\xrightarrow{F} \Omega^2(-;\mathbb{R})_{\mathrm{cl}}
\]
is the natural presymplectic 2-form on the stack $\mathbf{B}U(n)_{\mathrm{conn}}$: the
invariant polynomial $\langle -\rangle$ viewed in the context of stacks. 
The datum of a principal $U(n)$-bundle with connection $(P,\nabla)$ on a manifold $X$ is equivalent to the datum of a morphism $\varphi:X\to \mathbf{B}U(n)_{\mathrm{conn}}$, and the pullback $\varphi^*\omega_{\mathbf{B}U(n)_{\mathrm{conn}}}$ of the canonical 2-form on $\mathbf{B}U(n)_{\mathrm{conn}}$ is the curvature 2-form $\frac{1}{2\pi i}\mathrm{tr}(F_\nabla)$ on $X$.   If $(P,\nabla)$ is a principal $U(n)$-bundle with connection over a compact closed oriented 1-dimensional manifold  $\Sigma_1$ and the morphism $\varphi:\Sigma_1\to \mathbf{B}U(n)_{\mathrm{conn}}$ defining it can be extended to a morphism $\tilde{\varphi}:\Sigma_2\to \mathbf{B}U(n)_{\mathrm{conn}}$ for some 2-dimensional oriented manifold $\Sigma_2$ with $\partial\Sigma_2=\Sigma_1$, then
\[
CS_1(\nabla)=\exp\int_{\Sigma_2} \tilde{\varphi}^*\omega_{\mathbf{B}U(n)_{\mathrm{conn}}}\;,
\]
and the right hand side is independent of the extension $\tilde{\varphi}$. In other words, 
\[
CS_1(\nabla)=\exp\int_{\Sigma_2} \mathrm{tr}(F_{\tilde{\nabla}})\;,
\]
for any extension $(\tilde{P},\tilde{\nabla})$ of $(P,\nabla)$ to $\Sigma_2$. This way we recover the definition of the Chern-Simons action functional for $U(n)$-principal connections on $S^1$ given in section \ref{TheBasicDefinition}.
\par
More generally, the differential refinement $\hat{\mathbf{c}}$ of a characteristic class $c$ of a compact connected Lie group $G$ with $H^1(G;\mathbb{Z})\cong \mathbb{Z}$, endows the stack $\mathbf{B}G_{\mathrm{conn}}$ with a canonical presymplectic structure with a prequantum line bundle given by $\hat{\mathbf{c}}$ itself, and the same considerations apply.

\subsection{The determinant as a holonomy map}\label{determinant}
We have so far met two natural maps with target the sheaf $\underline{U}(1)$ of smooth functions with values in 
the group $U(1)$. The first 
one was the determinant 
\[
\det:\underline{U}(n)\to \underline{U}(1)\;,
\]
and the second one was the holonomy map
\[
\mathrm{hol}_X:\mathbf{Maps}(X;\mathbf{B}U(1)_{\mathrm{conn}})\to \underline{U}(1)\;,
\]
defined on the moduli stack of principal $U(1)$-bundles with connection on a 1-dimensional compact oriented manifold $X$. To see how these two are related, take $X=S^1$ and notice that, by definition, a morphism from a smooth manifold $X$ to the stack $\mathbf{Maps}(S^1;\mathbf{B}U(n)_{\mathrm{conn}})$ is the datum of a a principal $U(n)$-bundle with connection over the product manifold $X\times S^1$. Taking the holonomy of the $U(n)$-connection along the fibers of $X\times S^1\to X$ locally defines a smooth $U(n)$-valued function on $X$ which is well defined up to conjugation. In other words, holonomy along $S^1$ defines a morphism from $X$ to the stack $\underline{U}(n)/\!/_{\mathrm{Ad}}\underline{U}(n)$, where Ad indicates the adjoint action. 
Since this construction is natural in $X$ we have defined a natural $U(n)$-holonomy morphism
\[
\mathrm{hol}^{U(n)}:\mathbf{Maps}(S^1;\mathbf{B}U(n)_{\mathrm{conn}})\to \underline{U}(n)/\!/_{\mathrm{Ad}}\underline{U}(n)\;.
\]
For $n=1$, due to the fact that $U(1)$ is abelian, 
we also have a natural morphism $U(1)/\!/_{\mathrm{Ad}}U(1)\to U(1)$, and the holonomy map $\mathrm{hol}_{S^1}$ factors as 
\[
\mathrm{hol}_{S^1}:\mathbf{Maps}(S^1;\mathbf{B}U(1)_{\mathrm{conn}})\xrightarrow{\mathrm{hol}^{U(1)}} \underline{U}(1)/\!/_{\mathrm{Ad}}\underline{U}(1)\to \underline{U}(1)\;.
\]
Therefore, by naturality of $\mathbf{Maps}$ we obtain the following commutative diagram
\[
\xymatrix{
&\mathbf{Maps}(S^1;\mathbf{B}U(n)_{\mathrm{conn}})\ar[d]_{\mathrm{hol}^{U(n)}}
\ar[rr]^{\hspace{-1mm}\mathbf{Maps}(S^1,\hat{\mathbf{c}}_1)}
&& \mathbf{Maps}(S^1;\mathbf{B}U(1)_{\mathrm{conn}})\ar[d]^{\mathrm{hol}^{U(1)}}&\\
\underline{U}(n)\ar[r]&\underline{U}(n)/\!/_{\mathrm{Ad}}\underline{U}(n)\ar[rr]^{\det}&&\underline{U}(1)/\!/_{\mathrm{Ad}}\underline{U}(1)\ar[r]&\underline{U}(1)\;,}
\]
where the leftmost bottom arrow is the natural quotient projection $\underline{U}(n)\to  \underline{U}(n)/\!/_{\mathrm{Ad}}\underline{U}(n)$. In the language of \cite{survey} (3.9.6.4)
one says that the determinant map is the ``concretification'' 
of the morphism $\mathbf{Maps}(S^1,\hat{\mathbf{c}}_1)$, we come back to this in 
section \ref{DifferentialModuli} below. 
This construction immediately generalizes to the case of an arbitrary compact connected Lie group $G$ with $H^1(G;\mathbb{Z})\cong \mathbb{Z}$: the Lie group morphism $\rho:G\to U(1)$ integrating the Lie algebra cocycle $\mu_1$ corresponding to the characteristic class $c\in H^2(BG;\mathbb{Z})$ is the concretification of $\mathbf{Maps}(S^1,\hat{\mathbf{c}})$.

\subsection{Killing the first Chern class: $SU(n)$-bundles}\label{killing}
  Recall from the theory of characteristic classes 
  (see \cite{milnor-stasheff})
  that the first Chern class is the obstruction to reducing the structure group of a principal $U(n)$-bundle to $SU(n)$. In the stacky perspective that 
  we have been adopting so far this amounts to saying that the stack $\mathbf{B}SU(n)$ of principal $SU(n)$-bundles is the \emph{homotopy fiber} of $\mathbf{c}_1$, hence the object fitting into the
  homotopy pullback diagram of stacks of the form
\[
  \raisebox{20pt}{
  \xymatrix{
    \mathbf{B}SU(n)\ar[r]\ar[d]&{*}\ar[d]
	\\
    \mathbf{B}U(n)\ar[r]^-{\mathbf{c}_1}&\mathbf{B}U(1)\;.
	}}
	\]
By the universal property of the homotopy pullback, this says that an $SU(n)$-principal
bundle over a smooth manifold $X$  is equivalently a $U(n)$-principal bundle $P$, together with a choice of trivialization of the associated determinant $U(1)$-principal bundle.
Moreover, the whole groupoid of $SU(n)$-principal bundles 
on  $X$ is equivalent to the groupoid of $U(n)$-principal bundles 
on $X$ equipped with a trivialization of their associated 
determinant bundle. To explicitly see this equivalence, let us write the local data for a morphism from a smooth manifold $X$ to the homotopy pullback above. In terms of a fixed good open cover $\mathcal{U}$ of $X$, these are:
\begin{itemize}
\item smooth functions $\rho_\alpha:U_\alpha\to U(1)$;
\item smooth functions $g_{\alpha\beta}:U_{\alpha\beta}\to U(n)$,
\end{itemize}
subject to the constraints
\begin{itemize}
\item $\det(g_{\alpha\beta})\rho_\beta=\rho_\alpha$ on $U_{\alpha\beta}$;
\item $g_{\alpha\beta} g_{\beta\gamma} g_{\gamma\alpha}=1$ on  $U_{\alpha\beta\gamma}$.
\end{itemize}
Morphisms between $\{\rho_\alpha,g_{\alpha\beta}\}$ and $\{\rho_\alpha',g_{\alpha\beta}'\}$ are the gauge transformations locally given by $U(n)$-valued functions $h_{\alpha}$ on $U_{\alpha}$ such that $h_\alpha g_{\alpha\beta}=g_{\alpha\beta}' h_\beta$ and $\rho_\alpha\det(h_\alpha)=\rho_\alpha'$. The classical description of objects in $\mathbf{B}SU(n)$ corresponds to the gauge fixing $\rho_\alpha\equiv 1$; at the level of morphisms, imposing this gauge fixing constrains the gauge transformation $h_\alpha$ to satisfy $\det(h_\alpha)=1$, i.e. to take values in $SU(n)$. From a categorical point of view, this amounts to saying that the embedding of the groupoid of $SU(n)$-principal bundles over $X$ into the groupoid of morphisms from $X$ to the homotopy fiber of $\mathbf{c}_1$ given by $\{g_{\alpha\beta}\}\mapsto \{1,g_{\alpha\beta}\}$ is fully faithful. It is also essentially surjective: use the embedding $U(1)\to U(n)$ given by $e^{it}\mapsto (e^{it},1,1,\dots, 1)$ to lift $\rho_\alpha^{-1}$ to a $U(n)$-valued function $h_\alpha$ with $\det(h_\alpha)={\rho_\alpha}^{-1}$; then $\{h_\alpha\}$ is an isomorphism between $\{\rho_\alpha,g_{\alpha\beta}\}$ and $\{1,h_\alpha g_{\alpha\beta}{h_\beta}^{-1}\}$.

\vspace{3mm}
Similarly, the stack of $SU(n)$-principal bundles 
with $\mathfrak{su}_n$-connections is the homotopy pullback
\[
  \raisebox{20pt}{
  \xymatrix{
    \mathbf{B}SU(n)_{\mathrm{conn}}
	\ar[r]
	\ar[d]
	&
	{*}
	\ar[d]
	\\
    \mathbf{B}U(n)_{\mathrm{conn}}\ar[r]^-{\hat{\mathbf{c}}_1}
	&
	\mathbf{B}U(1)_{\mathrm{conn}}~\;.
 }
 }
\]
Details on this homotopy pullback description of $\mathbf{B}SU(n)_{\mathrm{conn}}$ can be found in \cite{FiorenzaSatiSchreiberI}. 

\vspace{3mm}
In summary, what we have discussed means that the map $\hat {\mathbf{c}}_1$ 
between universal moduli stacks equivalently plays the following different roles:
\begin{enumerate}
  \item it is a smooth and differential refinement of the universal first Chern class;
  \item it induces a 1-dimensional Chern-Simons action functional by \emph{transgression} to maps from the circle;
  \item it represents the obstruction to
   lifting a smooth unitary structure to a smooth special unitary structure.   
\end{enumerate}
In the following we will consider higher analogs of $\hat {\mathbf{c}}_1$ 
and will see these different but equivalent
 roles of universal differential characteristic maps amplified further. 
\par
As a concluding remark, let us notice that if $X$ is a smooth manifold and $G$ a Lie group, then the homotopy fiber of a morphism $f:X\to \mathbf{B}G$, i.e., the homotopy pullback
\[
  \raisebox{20pt}{
  \xymatrix{
    P
	\ar[r]
	\ar[d]
	&
	{*}
	\ar[d]
	\\
    X\ar[r]^-{f}
	&
	\mathbf{B}G
 }
 }
\]
is a principal $G$-bundle $P\to X$. Since the principal $G$-bundle $P\to X$ is induced by the morphism $f$ to the moduli stack of principal $G$-bundles, one says that $P\to X$ is \emph{modulated} by $f:X\to \mathbf{B}G$. Under topological realization, this reproduces the familiar construction of principal $G$-bundles over $X$ as pullbacks of the universal principal $G$-bundle $EG\to BG$ via a morphism $f:X\to BG$. This terminology extends to the case of $X$ being an arbitrary stack and $G$ an arbitrary (higher) smooth group, so, for instance, one can say that the stack $\mathbf{B}SU(n)$ is the principal $U(1)$-bundle over $\mathbf{B}U(n)$ modulated by the morphism $\mathbf{c}_1$. Similarly, if $f$ is a morphism from a smooth manifold $X$ to the moduli stack $\mathbf{B}G_{\mathrm{conn}}$ of principal $G$-connections, by composing $f$ with the forgetful morphism $\mathbf{B}G_{\mathrm{conn}}\to \mathbf{B}G$ and taking the homotopy fiber, we get a homotopy commutative diagram
\[
  \raisebox{20pt}{
  \xymatrix{
    P
	\ar[r]^-{\omega_P}
	\ar[d]
	&
	\Omega^1(-,\mathfrak{g})\ar[r]\ar[d]&
	{*}
	\ar[d]
	\\
    X\ar[r]^-{f}
    &\mathbf{B}G_{\mathrm{conn}}\ar[r]
	&
	\mathbf{B}G
 }
 }
\]
which shows how the principal $G$-bundle $P$ gets canonically endowed by a $\mathfrak{g}$-valued 1-form $\omega_P$, where $\mathfrak{g}$ is the Lie algebra of $G$. The pair $(P,\omega_P)$ is the principal $G$-connection on $X$ modulated by $f:X\to \mathbf{B}G_{\mathrm{conn}}$. Again, this terminology extends to stacks and smooth higher groups.

\section{The archetypical example: 3d Chern-Simons theory}
\label{3d}

We now pass from the toy example of 1-dimensional Chern-Simons theory to the 
archetypical example of 3-dimensional Chern-Simons theory, and in fact to its
extended (or ``multi-tiered'') geometric prequantization.
 
While this is a big step as far as the content of the theory goes, a
pleasant consequence of the higher geometric formulation of the 1d theory above is that 
\emph{conceptually} essentially nothing new happens when we move from 1-dimensional theory to 3-dimensional theory (and further). 
For the 3d theory we only need to restrict our attention to simply connected compact simple Lie groups, so as to have $\pi_3(G)\cong \mathbb{Z}$ as the first nontrivial homotopy group, and to move from stacks 
to higher stacks, or more precisely, to 3-stacks.\footnote{For non-simply connected groups one needs a little
bit more structure, as we briefly indicate in section \ref{CupProductCS}.} By definition, a higher stack is a \emph{simplicial sheaf} (or $\infty$-sheaf) on some site of definition. In particular, a \emph{smooth higher stack} is a higher stack on the site of smooth manifolds. Since in this text the site of definition will always be the site of smooth manifolds,\footnote{Actually, in section \ref{super-SC} also the more general site of smooth supermanifolds will be considered.} we will often just say ``higher stack'', or even just ``stack'',  to mean ``smooth higher stack''. Notice that an ordinary (i.e., set-valued) sheaf is precisely a 0-truncated simplicial sheaf, and that an ordinary (i.e., groupoid-valued) stack is precisely a 1-truncated simplicial sheaf. Therefore, if one calls $n$-stack an $n$-truncated simplical sheaf, we have that from the higher stacks perspective sheaves and stacks are 0- and 1-stacks, respectively. As for ordinary stacks, when a higher stack represents some moduli problem, we will call it a higher moduli stack. 
\par
We will denote by $\mathbf{H}$ the $(\infty,1)$-category of smooth higher stacks. Since it is an $\infty$-category of $\infty$-sheaves on a site, it is an example of an $\infty$-topos \cite{lurie}.

\subsection{Higher $U(1)$-bundles with connections and differential cohomology}

The basic 3-stack naturally appearing in ordinary 3d Chern-Simons theory is the 3-stack $\mathbf{B}^3U(1)_{\mathrm{conn}}$ of principal $U(1)$-3-bundles with connection (also known as $U(1)$-bundle-2-gerbes with connection). It is convenient to introduce in general the $n$-stack $\mathbf{B}^nU(1)_{\mathrm{conn}}$ and to describe its relation to differential cohomology.
\par
By definition, $\mathbf{B}^n U(1)_{\mathrm{conn}}$ is the $n$-stack obtained by stackifying the prestack on Cartesian spaces which corresponds, via the Dold-Kan correspondence, 
to the $(n+1)$-term Deligne complex 
\[
\underline{U}(1)[n]_D^\infty=\biggl( \underline{U}(1)\xrightarrow{\frac{1}{2\pi i}d\mathrm{log}} \Omega^1(-;\mathbb{R})\xrightarrow{d}\cdots\xrightarrow{d}\Omega^n(-;\mathbb{R})\biggr)\;,
\]
where $\underline{U}(1)$ is the sheaf of smooth functions with values in $U(1)$, and with $\Omega^n(-;\mathbb{R})$ in degree zero. It is immediate from the definition that the equivalence classes of $U(1)$-$n$-bundles with connection on a smooth manifold $X$ are classified by the $(n+1)$-st differential cohomology group of $X$,
\[
\hat{H}^{n+1}(X;\mathbb{Z})\cong \mathbb{H}^0(X;\underline{U}(1)[n]_D^\infty)\cong
\pi_0 \mathbf{H}(X;\mathbf{B}^n U(1)_{\mathrm{conn}})\;,
\]
where in the middle we have degree zero hypercohomology of $X$ with coefficients in $\underline{U}(1)[n]_D^\infty$.
Similarly, the $n$-stack of $U(1)$-$n$-bundles (without connection) $\mathbf{B}^nU(n)$ is obtained via Dold-Kan and stackification from the sheaf of chain complexes
\[
\underline{U}(1)[n]=\biggl(\underline{U}(1)\to 0\to\cdots\to0\biggr)\;,
\]
with $C^\infty(-;U(1))$ in degree $n$. Equivalence classes of $U(1)$-$n$-bundles on $X$ are in natural bijection with
\[
{H}^{n+1}(X;\mathbb{Z})\cong H^n(X;\underline{U}(1))\cong \mathbb{H}^0(X;\underline{U}(1)[n]) \cong
\pi_0 \mathbf{H}(X;\mathbf{B}^n U(1))
\;.
\]
The obvious morphism of chain complexes of sheaves $\underline{U}(1)[n]_D^\infty\to \underline{U}(1)[n]$ induces the ``forget the connection'' morphism $\mathbf{B}^nU(1)_{\mathrm{conn}}\to \mathbf{B}^nU(1)$ and, at the level of equivalence classes, the natural morphism
\[
\hat{H}^{n+1}(X;\mathbb{Z})\to {H}^{n+1}(X;\mathbb{Z})
\]
from differential cohomology to integral cohomology. If we denote by $\Omega^{n+1}(-;\mathbb{R})_{\mathrm{cl}}$ the sheaf (a 0-stack) of closed $n$-forms, then the morphism of complexes $\underline{U}(1)[n]_D^\infty\to \Omega^{n+1}(-;\mathbb{R})_{\mathrm{cl}}$ given by
\[
\xymatrix{
\underline{U}(1)\ar[d]\ar[rr]^-{\tfrac{1}{2\pi i}d\mathrm{log}}&& \Omega^1(-;\mathbb{R})\ar[d]\ar[r]^-{d}&\cdots\ar[r]^-{d}\ar[d]&\Omega^n(-;\mathbb{R})\ar[d]^{d}\\
0\ar[rr]&& 0\ar[r]&\cdots\ar[r]&\Omega^{n+1}(-;\mathbb{R})_{\mathrm{cl}}
}
\]
induces the morphism of stacks $\mathbf{B}^nU(1)_{\mathrm{conn}}\xrightarrow{F_{(-)}} \Omega^{n+1}(-;\mathbb{R})_{\mathrm{cl}}$ mapping a circle $n$-bundle ($(n-1)$-bundle gerbe) with connection to the curvature $(n+1)$-form of its connection. At the level of differential cohomology, 
this is the morphism 
\[
\hat{H}^{n+1}(X;\mathbb{Z})\to \Omega^{n+1}(X;\mathbb{R})_{\mathrm{cl}}\;.
\]
The last $n$-stack we need to introduce to complete this sketchy picture of differential cohomology 
formulated on universal moduli stacks 
is the $n$-stack $\flat\mathbf{B}^{n+1}\mathbb{R}$ associated with the chain complex of sheaves
\[
\flat\underline{\mathbb{R}}[n+1]^\infty=\biggl(  \Omega^1(-;\mathbb{R})\xrightarrow{d}\cdots\xrightarrow{d}\Omega^{n}(-;\mathbb{R})\xrightarrow{d}\Omega^{n+1}(-;\mathbb{R})_{\mathrm{cl}}\biggr)\;,
\]
with $\Omega^{n+1}(-;\mathbb{R})_{\mathrm{cl}}$ in degree zero. The obvious morphism of complexes of sheaves $\Omega^{n+1}(-;\mathbb{R})_{\mathrm{cl}}\to \flat\underline{\mathbb{R}}[n+1]^\infty$ induces a morphism of stacks $\Omega^{n+1}(-;\mathbb{R})_{\mathrm{cl}}\to\flat\mathbf{B}^{n+1}\mathbb{R}$. Moreover one can show (see, e.g., \cite{FSS,survey}) that there is a ``universal curvature characteristic'' morphism
$\mathrm{curv} :  \mathbf{B}^nU(1)\to \flat\mathbf{B}^{n+1}\mathbb{R}$ and a homotopy pullback diagram
\[
\raisebox{20pt}{
 \xymatrix{
  \mathbf{B}^nU(1)_{\mathrm{conn}}\ar[r]^-{F} \ar[d]
  &
  \Omega^{n+1}(-;\mathbb{R})_{\mathrm{cl}}\ar[d]
  \\
  \mathbf{B}^nU(1)\ar[r]^-{\mathrm{curv}} &\flat\mathbf{B}^{n+1}\mathbb{R}\;,
}}
\]
of higher moduli stacks, 
which induces in cohomology the commutative diagram
\[
\raisebox{20pt}{
\xymatrix{
  \hat{H}^{n+1}(X;\mathbb{Z})\ar[r]^-{F} \ar[d]
  &
  \Omega^{n+1}(X;\mathbb{R})_{\mathrm{cl}}\ar[d]
  \\
  {H}^{n+1}(X;\mathbb{Z})
  \ar[r]^{\mathrm{}} &{H}^{n+1}_{\mathrm{dR}}(X;\mathbb{R})\;.
}}
\]
This generalizes to any degree $n\geq 1$ what we remarked in section \ref{degree-2} for the degree 2 case: differential cohomology encodes in a systematic and geometric way the simple idea of having an integral cohomology class together with a closed differential form form representing it in de Rham cohomology. For $n=0$ we have $\hat{H}^1(X;\mathbb{Z})\equiv H^0(X;\underline{U}(1))=C^\infty(X;U(1))$ and the map $\hat{H}^1(X;\mathbb{Z})\to {H}^1(X;\mathbb{Z})$ is the morphism induced in cohomology by the short exact sequence of sheaves
\[
0\to \underline{\mathbb{Z}}\to \underline{\mathbb{R}}\to \underline{U}(1)\to 1\;.
\]
At the level of stacks, this corresponds to the morphism
\[
U(1) \to \mathbf{B}\mathbb{Z}
\]
induced by the canonical principal $\mathbb{Z}$-bundle $\mathbb{R}\to U(1)$. 

\subsection{Compact simple and simply connected Lie groups}
From a cohomological point of view, a compact simple and simply connected Lie group $G$ is the degree 3 analogue of the group $U(n)$ considered in our 1-dimensional toy model. That is, the homotopy (hence the homology) of $G$ is trivial up to degree 3, 
and $\pi_3(G)\cong H^3(G;\mathbb{Z})\cong \mathbb{Z}$, by the Hurewicz isomorphism,. Passing from $G$ to its classifying space $BG$ we find $H^4(BG;\mathbb{Z})\cong\mathbb{Z}$, so that the fourth integral cohomology group of $BG$ is generated by a fundamental characteristic class $c\in  H^4(BG;\mathbb{Z})$. All other elements in $H^4(BG;\mathbb{Z})$ are of the form $kc$ for some integer $k$, usually called the ``level'' in the physics literature. For $P$ a $G$-principal bundle over a smooth manifold $X$, we will write $c(P)$ for the cohomology class $f^*c\in H^4(X,\mathbb{Z})$, where $f:X\to BG$ is any classifying map for $P$. This way we realize $c$ a map
\[
c:\{\text{principal $G$-bundles on $X$}\}/\text{iso}\to H^4(X;\mathbb{Z})\;.
\]
\par
Moving to real coefficients, the fundamental characteristic class $c$ is represented, via the isomorphism $H^4(BG;\mathbb{R})\cong H^3(G;\mathbb{R})\cong H^3_{\mathrm{Lie}}(\mathfrak{g},\mathbb{R})$ by the canonical 3-cocycle $\mu_3$ on the Lie algebra $\mathfrak{g}$ of $G$, i.e., up to normalization, to the 3-cocycle $\langle[-,-],-\rangle$, where $\langle-,-\rangle$ is the Killing form of $\mathfrak{g}$ and 
$[-,-]$ is the Lie bracket. On the other hand, via the Chern-Weil isomorphism
\[
H^*(BG;\mathbb{R})\cong \mathrm{inv}(\mathfrak{g}[2])\;,
\]
the characteristic class $c$ corresponds to the Killing form, seen as a degree four invariant polynomial on $\mathfrak{g}$ (with elements of $\mathfrak{g}^*$ placed in degree 2). The transgression between $\mu_3$ and $\langle-,-\rangle$ is witnessed by the canonical degree 3 Chern-Simons element $\mathrm{CS}_3$ of $\mathfrak{g}$. That is, for a $\mathfrak{g}$-valued 1-form $A$ on some manifold, let 
\[
\mathrm{CS}_3(A)=\langle A,dA\rangle+\tfrac{1}{3}\langle A,[A,A]\rangle\;.
\]
Then, for $A\in \Omega^1(EG;\mathfrak{g})$ the connection 1-form of a principal $G$-connection on $EG\to BG$, we have the following transgression diagram
\[
\xymatrix{
\langle F_A,F_A\rangle & \mathrm{CS}_3(A)\ar@{|->}[l]_-{d}\ar@{|->}[rr]^{A=\theta_G\phantom{mmm}}&&\mu_3(\theta_G,\theta_G,\theta_G)\;,
}
\] 
where $\theta_G$ is the Maurer-Cartan form of $G$ (i.e., the restriction of $A$ to the fibers of $EG\to BG$) and $F_A=dA+\frac{1}{2}[A,A]$ is the curvature 2-form of $A$. Notice how both the invariance of the Killing form and the Maurer-Cartan equation $d\theta_G+\frac{1}{2}[\theta_G,\theta_G]=0$ play a r\^ole in the above transgression diagram.

\subsection{The differential refinement of degree 4 characteristic classes}\label{refinement}
The description of the Brylinski-McLaughlin 2-cocycle from section \ref{bml-2} has an evident generalization to degree four. Indeed, let $\{A_\alpha,g_{\alpha\beta}\}$ the local data for a $G$-connection $\nabla$ on $P\to X$, 
relative to a trivializing good open cover $\mathcal{U}$ of $X$, with $G$ a compact simple and simply connected Lie group. Then, since $G$ is connected and the open sets $U_{\alpha\beta}$ are contractible, we can smoothly extend the transition functions $g_{\alpha\beta}:U_{\alpha\beta}\to G$ to 
functions $\hat{g}_{\alpha\beta}:[0,1]\times U_{\alpha\beta}\to G$ with $\hat{g}_{\alpha\beta}(0)=e$, the identity element of $G$, and $\hat{g}_{\alpha\beta}(1)=g_{\alpha\beta}$, and using the functions $\hat{g}_{\alpha\beta}$ one can interpolate from $A_\alpha\bigr\vert_{U_{\alpha\beta}}$ to $A_\beta\bigr\vert_{U_{\alpha\beta}}$ as in section \ref{bml-2}, defining a $\mathfrak{g}$-valued 1-form $\hat{A}_{\alpha\beta}=\hat{g}_{\alpha\beta}^{-1}A_\alpha\bigr\vert_{U_{\alpha\beta}}\hat{g}_{\alpha\beta}+\hat{g}_{\alpha\beta}^{-1}d\hat{g}_{\alpha\beta}$. On the triple intersection $U_{\alpha\beta\gamma}$ we have the paths in $G$
  $$
      \xymatrix{
        & g_{\alpha\beta} \ar[dr]^{\hat g_{\alpha\beta} \cdot \hat g_{\beta\gamma}}
        \\
        e 
          \ar[rr]_{\hat g_{\alpha\gamma}}^{{\ }}="t" 
          \ar[ur]^{\hat g_{\alpha\beta}}
          && 
         {g}_{\alpha\gamma}\;.
      }
    $$
Since $G$ is simply connected we can find smooth functions
\[
\hat{g}_{\alpha\beta\gamma}:U_{\alpha\beta\gamma}\times \Delta^2\to G
\]
filling these 2-simplices, and we can use these to extend the interpolation between $\hat{A}_{\alpha\beta}$, $\hat{A}_{\beta\gamma}$ and $\hat{A}_{\gamma\alpha}$ over the 2-simplex. Let us denote this interpolation by $\hat{A}_{\alpha\beta\gamma}$. Finally, since $G$ is 2-connected, on the quadruple  intersections we can find smooth functions 
\[
\hat g_{\alpha\beta\gamma\delta} : U_{\alpha\beta\gamma\delta}\times \Delta^3 \to G
\]
cobounding the union of the 2-simplices corresponding to the $\hat{g}_{\alpha\beta\gamma}$'s on the triple intersections. 
We can again use the $\hat g_{\alpha\beta\gamma\delta}$'s to interpolate between the $\hat{A}_{\alpha\beta\gamma}$'s over the 3-simplex. Finally, one considers the 
degree zero {\Cech}-Deligne  cochain with coefficients in $\underline{U}(1)[3]^\infty_D$
     \begin{equation}
        \left( \mathrm{CS}_3(A_\alpha), \int_{\Delta^1} \mathrm{CS}_3(\hat A_{\alpha\beta }) , \int_{\Delta^2} \mathrm{CS}_3(\hat A_{\alpha\beta\gamma}),
        \int_{\Delta^3} \mathrm{CS}_3(\hat A_{\alpha\beta\gamma\delta}) ~\,\mathrm{mod}\,\mathbb{Z}\right)\;.
\end{equation}
Brylinski and McLaughlin \cite{brylinski-mclaughlin} show 
(see also \cite{bm:pont} for an exposition and \cite{bm:geom4-I,bm:geom4-II} for related discussion)
that this is indeed a degree zero {\Cech}-Deligne cocycle, and thus defines an element in $\hat{H}^4(X;\mathbb{Z})$. 
Moreover, they show that this cohomology class only depends on the isomorphism class of $(P,\nabla)$, 
inducing therefore a well-defined map
\[
\hat{c}:\{\text{$G$-bundles with connection on $X$}\}/\text{iso}\to \hat{H}^4(X;\mathbb{Z})\;.
\]
Notice how modding out by $\mathbb{Z}$ in the rightmost integral in 
the above cochain
 precisely takes care of $\pi_3(G)\cong H^3(G;\mathbb{Z})\cong \mathbb{Z}$. Notice also that, by construction, 
\[
  \int_{\Delta^3} \mathrm{CS}_3(\hat A_{\alpha\beta\gamma\delta})= \int_{\Delta^3} \hat g_{\alpha\beta\gamma\delta}^* \; \mu_3(\theta_G \wedge \theta_G \wedge \theta_G)\;, 
\]
where $\theta_G$ is the Maurer-Cartan form of $G$. Hence the Brylinski-McLaughlin cocycle lifts the degree 3 cocycle with coefficients in $\underline{U}(1)$ 
\[
\int_{\Delta^3} \hat g_{\alpha\beta\gamma\delta}^* \; \mu_3(\theta_G \wedge \theta_G \wedge \theta_G)\,\mod\,\mathbb{Z}\;,
\]
which represents the characteristic class $c(P)$ in $H^3(X;\underline{U}(1))\cong H^4(X;\mathbb{Z})$. 
As a result, the differential characteristic class $\hat{c}$ lifts the characteristic class $c$, i.e., we have  a natural commutative diagram
\[
\xymatrix{
\{\text{$G$-bundles with connection on $X$}\}/\text{iso}\ar[d]\ar[r]^{\phantom{mmmmmmmmm}\hat{c}}& \hat{H}^4(X;\mathbb{Z})\ar[d]\\
\{\text{$G$-bundles on $X$}\}/\text{iso}\ar[r]^{\phantom{mmmm}c}& {H}^4(X;\mathbb{Z})\;.
}
\]
By looking at the Brylinski-McLaughlin construction 
through the eyes of simplicial integration of $\infty$-Lie algebras
one sees \cite{FSS} that the above commutative diagram is naturally enhanced to a commutative diagram of stacks
\[
\raisebox{20pt}{
\xymatrix{
\mathbf{B}G_{\mathrm{conn}}\ar[d]\ar[r]^-{\hat{\mathbf{c}}}& \mathbf{B}^3U(1)_{\mathrm{conn}}\ar[d]\\
\mathbf{B}G\ar[r]^-{{\mathbf{c}}}& \mathbf{B}^3U(1)\;.
}}
\]
As we are going to show, the morphism $\hat{\mathbf{c}}:\mathbf{B}G_{\mathrm{conn}}\to \mathbf{B}^3U(1)_{\mathrm{conn}}$ 
that refines
 the characteristic class $c$ to a morphism of stacks is the morphism secretly governing all basic features of level 1 three-dimensional Chern-Simons theory with gauge group $G$. Similarly, for any $k\in \mathbb{Z}$, one has a morphism of stacks 
\[
k\hat{\mathbf{c}}:\mathbf{B}G_{\mathrm{conn}}\to \mathbf{B}^3U(1)_{\mathrm{conn}}
\]
governing level $k$ 3d Chern-Simons theory with gauge group $G$. 
Indeed, this map may be regarded as the very Lagrangian of 3d Chern-Simons theory extended (ÒlocalizedÓ, Òmulti-tieredÓ) to codimension 3. This means that we have data assigned to $k$-dimensional manifolds with corners for any $0\leq k\leq 3$, giving a representation of the symmetric monoidal $(\infty,3)$-category of fully extended cobordism in dimension 3 \cite{LurieTQFT}. In the next section we describe the \emph{closed manifolds sector} of this extended field theory. A description of the full field theory, including manifolds with boundaries and corners, can be obtained along the same lines by extending the Gomi and Terashima formulas for integration of Deligne cocycles on manifolds with boundaries \cite{gomi-terashima1} to manifolds with corners.

\subsection{Prequantum $n$-bundles on moduli stacks of $G$-connections on a fixed manifold}
\label{PrequantumLineBundlesOnModuliStacks}

We discuss now how the differential refinement $\hat {\mathbf{c}}$ of the universal characteristic
map $c$ constructed above serves as the \emph{extended} Lagrangian for 3d Chern-Simons theory
in that its \emph{transgression} to mapping stacks out of $k$-dimensional manifolds yields
all the ``geometric prequantum'' data of Chern-Simons theory in the corresponding dimension, 
in the sense of geometric quantization. For the purpose of this exposition 
we use terms such as ``prequantum $n$-bundle'' freely without formal definition. We expect the 
reader can naturally see at least vaguely the higher prequantum picture alluded to here. 
A more formal survey of these notions is in section \ref{PrequantumInHigherCodimension}.

\medskip

If $X$ is  a compact oriented manifold without boundary, then there is a fiber integration in differential cohomology lifting fiber integration in integral cohomology \cite{HopkinsSinger}:
\[
\xymatrix{
\hat{H}^{n+\dim X}(X\times Y;\mathbb{Z})\ar[d]\ar[r]^{\phantom{mmm}\int_X}& \hat{H}^{n}(Y;\mathbb{Z})\ar[d]\\
{H}^{n+\dim X}(X\times Y;\mathbb{Z})\ar[r]^{\phantom{mmm}\int_X}& {H}^{n}(Y;\mathbb{Z})\;.
}
\]
In \cite{gomi-terashima1} Gomi and Terashima describe an explicit lift of this to the level of \v{C}ech-Deligne cocycles; see also \cite{dupont-ljungmann}. One observes \cite{FiorenzaSatiSchreiberIII} that
such a lift has a natural interpretation  as a morphism of moduli stacks
\[
\mathrm{hol}_X: \mathbf{Maps}(X,\mathbf{B}^{n+\dim X}U(1)_{\mathrm{conn}})\to \mathbf{B}^{n}U(1)_{\mathrm{conn}}
\]
from the $(n+\dim X)$-stack of moduli of $U(1)$-$(n+\dim X)$-bundles with connection over $X$ to the $n$-stack of  $U(1)$-$n$-bundles with connection (section 2.4 of \cite{FiorenzaSatiSchreiberIII}).
Therefore, if $\Sigma_k$ is a compact 
oriented manifold of dimension $k$ with $0\leq k\leq 3$, we have a composition
\[
\mathbf{Maps}(\Sigma_k,\mathbf{B}G_{\mathrm{conn}})\xrightarrow{\mathbf{Maps}(\Sigma_k,\hat{\mathbf{c}})}\mathbf{Maps}(\Sigma_k,\mathbf{B}^3U(1)_{\mathrm{conn}})\xrightarrow{\mathrm{hol}_{\Sigma_k}} \mathbf{B}^{3-k}U(1)_{\mathrm{conn}}\;.
\]
This is the canonical $U(1)$-$(3-k)$-bundle with connection over the moduli space of principal $G$-bundles with connection over $\Sigma_k$ induced by $\hat {\mathbf{c}}$: the
\emph{transgression} of $\hat {\mathbf{c}}$ to the mapping space. 
Composing on the right with the curvature morphism
we get the underlying canonical closed $(4-k)$-form 
\[
\mathbf{Maps}(\Sigma_k,\mathbf{B}G_{\mathrm{conn}})\to \Omega^{4-k}(-;\mathbb{R})_{\mathrm{cl}}
\] 
on this moduli space. 
In other words, the moduli stack of principal $G$-bundles with connection over $\Sigma_k$ carries a canonical 
\emph{pre-$(3-k)$-plectic structure} 
(the higher order generalization of a symplectic structure, \cite{Rogers}) 
and, moreover, this is equipped with a canonical geometric prequantization:
the above $U(1)$-$(3-k)$-bundle with connection. 

Let us now investigate in more  detail the cases $k=0,1,2,3$.

\subsubsection{$k=0$: the universal Chern-Simons 3-connection $\hat{\mathbf{c}}$}

The connected 0-manifold $\Sigma_0$ is the point and, by definition of $\mathbf{Maps}$,
 one has a canonical identification
\[
\mathbf{Maps}(*,\mathbf{S})\cong \mathbf{S}
\]
for any (higher) stack $\bf S$. Hence the morphism 
\[
\mathbf{Maps}(*,\mathbf{B}G_{\mathrm{conn}})\xrightarrow{\mathbf{Maps}(*,\hat{\mathbf{c}})}\mathbf{Maps}(*,\mathbf{B}^3U(1)_{\mathrm{conn}})
\]
is nothing but the 
universal differential characteristic map $\hat{\mathbf{c}}:\mathbf{B}G_{\mathrm{conn}}\to \mathbf{B}^3U(1)_{\mathrm{conn}}$ that refines the universal characteristic class $c$. 
This map modulates a circle 3-bundle with connection (bundle 2-gerbe) on the universal moduli stack
of $G$-principal connections. For $\nabla : X \longrightarrow \mathbf{B}G_{\mathrm{conn}}$
any given $G$-principal connection on some $X$, the pullback 
$$
  \hat {\mathbf{c}}(\nabla) : 
  \xymatrix{
    X \ar[r]^-\nabla & \mathbf{B}G_{\mathrm{conn}} \ar[r]^-{\hat {\mathbf{c}}} & \mathbf{B}^3 U(1)_{\mathrm{conn}}
  }
$$
is a 3-bundle (bundle 2-gerbe) on $X$ which is sometimes in the literature called the 
\emph{Chern-Simons 2-gerbe} of the given connection $\nabla$. Accordingly, $\hat {\mathbf{c}}$ modulates
the \emph{universal} Chern-Simons bundle 2-gerbe with universal 3-connection.
From the point of view of higher geometric quantization, this is the \emph{prequantum 3-bundle}
of extended prequantum Chern-Simons theory.

This means that the prequantum $U(1)$-$(3-k)$-bundles associated with $k$-dimensional manifolds are all determined by  the the prequantum $U(1)$-3-bundle associated with the point, in agreement with the formulation of fully extended topological field theories \cite{FHLT}.
We will denote by the symbol $\omega^{(4)}_{\mathbf{B}G_{\mathrm{conn}}}$ the pre-3-plectic 4-form induced on $\mathbf{B}G_{\mathrm{conn}}$ by the curvature morphism.

\subsubsection{$k=1$: the Wess-Zumino-Witten bundle gerbe}
\label{TheWZWGerbe}

We now come to the transgression of the extended Chern-Simons Lagrangian to the 
closed connected 1-manifold, the circle $\Sigma_1 = S^1$. 
Here we find a higher analog of the construction described in section \ref{determinant}. 
Notice that, on the one hand, we can think  of the mapping stack 
$\mathbf{Maps}(\Sigma_1, \mathbf{B}G_{\mathrm{conn}}) \simeq  \mathbf{Maps}(S^1 , \mathbf{B}G_{\mathrm{conn}})$
 as a kind of moduli stack of $G$-connections on the circle
-- up to a slight subtlety, which we explain in more detail below in section \ref{DifferentialModuli}.
On the other hand, we can think of that mapping stack 
 as the \emph{free loop space} of the universal moduli stack $\mathbf{B}G_{\mathrm{conn}}$.

The subtlety here is related to the differential refinement, 
so it is instructive to first discard the differential refinement and consider just the 
smooth characteristic map $\mathbf{c} : {\mathbf{B}G \to \mathbf{B}^3 U(1)}$
which underlies the extended Chern-Simons Lagrangian and which  modulates the universal circle 3-bundle on $\mathbf{B}G$ (without connection). Now, for every pointed stack 
$* \to \mathbf{S}$ we have the corresponding (categorical) \emph{loop space}
$
  \Omega \mathbf{S} := * \times_{\mathbf{S}} *
$,
which is the homotopy pullback of the point inclusion along itself. Applied to 
the moduli stack $\mathbf{B}G$ this recovers the Lie group $G$, identified with the sheaf (i.e, the $0$-stack) of smooth functions with target $G$:
$
  \Omega \mathbf{B}G \simeq \underline{G}
  $.
This kind of looping/delooping equivalence is familiar from the homotopy theory of classifying spaces;
but notice that since we are working with smooth (higher) stacks, the loop space
 $\Omega \mathbf{B}G$ also knows the smooth structure of the group $G$, i.e. it knows $G$ as a Lie group. Similarly,
we have
$$
  \Omega \mathbf{B}^3 U(1)\simeq \mathbf{B}^2 U(1)
$$
and so forth in higher degrees.
Since the looping operation is functorial, we may also apply it to the characteristic map $\mathbf{c}$
itself to obtain a map
$$
  \Omega \mathbf{c} :    \underline{G} \to \mathbf{B}^2 U(1)
$$
which modulates a $\mathbf{B}U(1)$-principal 2-bundle on the Lie group $G$.
This is also known as the \emph{WZW-bundle gerbe}; see \cite{GR, SchweigertWaldorf}. 
The reason, as discussed there and as we will see in a moment, is that
this is the 2-bundle that underlies the 2-connection with surface holonomy over a worldsheet 
given by the Wess-Zumino-Witten action functional.  
However, notice first that there is more structure implied here: for any pointed stack $\mathbf{S}$ there is a natural equivalence  $
  \Omega {\mathbf{S}} \simeq \mathbf{Maps}_*(\mathbf{\Pi}(S^1), {\mathbf{S}})
  $,
 between the loop space object $ \Omega {\mathbf{S}}$ and the moduli stack of \emph{pointed maps} from the categorical circle  $\mathbf{\Pi}(S^1)\simeq \mathbf{B}\mathbb{Z}$ to $\mathbf{S}$. 
Here $\mathbf{\Pi}$ denotes the \emph{path $\infty$-groupoid} of a given (higher) stack.\footnote{The existence and functoriality of the path $\infty$-groupoids is one of the 
features characterizing the higher topos of higher smooth stacks as being \emph{cohesive}, see \cite{survey}.}
On the other hand, if we do not fix the base point then we obtain the \emph{free loop space object}
$
  \mathcal{L} {\mathbf{S}} \simeq \mathbf{Maps}(\mathbf{\Pi}(S^1),  {\mathbf{S}})
 $. Since a map $\mathbf{\Pi}(\Sigma)\to \mathbf{B}G$ is equivalently 
a  map $\Sigma \to \flat \mathbf{B}G$, i.e., a flat $G$-principal connection on $\Sigma$,
the free loop space $\mathcal{L}\mathbf{B}G$ is equivalently the moduli stack of (necessarily flat) $G$-principal connections on $S^1$.
We will come back to this perspective in section \ref{DifferentialModuli} below.
The homotopies that do not fix the base point act by conjugation on loops, hence
we have, for any smooth  (higher) group, that 
$$
  \mathcal{L}\mathbf{B}G \simeq \underline{G}/\!/_{\mathrm{Ad}}\underline{G}
$$
is the (homotopy) quotient of the adjoint action of $G$ on itself; see \cite{NSSa} for details
on homotopy actions of smooth  higher groups. For $G$ a Lie group this is the familiar adjoint
action quotient stack. But the expression holds fully generally. Notably, we also have
$$
  \mathcal{L}\mathbf{B}^3 U(1) \simeq \mathbf{B}^2 U(1)/\!/_{\mathrm{Ad}}\mathbf{B}^2 U(1)
$$
and so forth in higher degrees.
However, in this case, since the smooth 3-group $\mathbf{B}^2 U(1)$ is abelian (it is a groupal $E_\infty$-algebra)
the adjoint action splits off in a direct factor and we have a projection
$$
  \xymatrix{
    \mathcal{L}\mathbf{B}^3 U(1)
\simeq
	\mathbf{B}^2 U(1) \times ({*}/\!/\mathbf{B}^2 U(1))
	\ar[r]^-{p_1}
	&
	\mathbf{B}^2 U(1)
  }\;.
$$
In summary, this means that the map $\Omega \mathbf{c}$ modulating the WZW 2-bundle over $G$ descends
to the adjoint quotient to the map
$$
  p_1 \circ \mathcal{L}\mathbf{c} 
  :
    \underline{G}/\!/_{\mathrm{Ad}}\underline{G}
	\to
	\mathbf{B}^2 U(1)
  \,,
$$
and this means that the WZW 2-bundle is canonically equipped with the structure of an 
\emph{$\mathrm{ad}_G$-equivariant} bundle gerbe, a crucial feature of the WZW bundle gerbe
\cite{GR, GSW}.

We emphasize that the derivation here is fully general and holds for any smooth  (higher) group $G$
and any smooth characteristic map $\mathbf{c} : {\mathbf{B}G \to \mathbf{B}^n U(1)}$.
Each such pair induces a WZW-type $(n-1)$-bundle on the smooth  (higher) group $G$ modulated by
$\Omega \mathbf{c}$ and equipped with $G$-equivariant structure exhibited by
$p_1 \circ \mathcal{L}\mathbf{c}$. We discuss such higher examples of higher Chern-Simons-type theories
with their higher WZW-type functionals further below in section \ref{CupProductCS}.

We now turn to the differential refinement of this situation. In analogy to the 
above construction, but taking care of the connection data in 
the extended Lagrangian $\hat {\mathbf{c}}$, we find a
homotopy commutative diagram in $\mathbf{H}$ of the form
\[
\hspace{-2mm}
\xymatrix{
&\mathbf{Maps}(S^1;\mathbf{B}G_{\mathrm{conn}})\ar[d]_{\mathrm{hol}}\ar[rr]^{\hspace{-3mm}\mathbf{Maps}(S^1,\hat{\mathbf{c}})}
&& \mathbf{Maps}(S^1;\mathbf{B}^3U(1)_{\mathrm{conn}})\ar[d]^{\mathrm{hol}}&\\
\underline{G}\ar[r]&\underline{G}/\!/_{\mathrm{Ad}}\underline{G}\ar[rr]^{\hspace{-6mm}\mathbf{wzw}\phantom{mmm}}&&\mathbf{B}^2U(1)_{\mathrm{conn}}/\!/_{\mathrm{Ad}}\mathbf{B}^2U(1)_{\mathrm{conn}}\ar[r]&\mathbf{B}^2U(1)_{\mathrm{conn}}\;,
}
\]
where the vertical maps are obtained by forming holonomies of (higher) connections along
the circle.
The lower horizontal row is the differential refinement of 
$\Omega \mathbf{c}$: it modulates the Wess-Zumino-Witten $U(1)$-bundle gerbe with connection
\[
\mathbf{wzw}:\underline{G}\to \mathbf{B}^2U(1)_{\mathrm{conn}}\;.
\]
That $\mathbf{wzw}$ is indeed the correct differential refinement can be seen, 
for instance, by interpreting the construction by Carey-Johnson-Murray-Stevenson-Wang in \cite{CJMSW} in terms of the above diagram. 
That is, choosing a basepoint $x_0$ in $S^1$ one obtains a canonical lift of the leftmost vertical arrow:
\[
\xymatrix{
&\mathbf{Maps}(S^1;\mathbf{B}G_{\mathrm{conn}})\ar[d]^{\mathrm{hol}}\\
\underline{G}\ar[r]\ar[ru]^-{(P_{x_0},\nabla_{x_0})}&\underline{G}/\!/_{\mathrm{Ad}}\underline{G}\;,
}
\]
where $(P_{x_0}\nabla_{x_0})$ is the principal $G$-bundle with connection on the product $G\times S^1$ characterized by the property that the holonomy of $\nabla_{x_0}$ along $\{g\}\times S^1$ 
with starting point $(g,x_0)$ is  the element $g$ of $G$. Correspondingly, we have a homotopy commutative diagram
\[
\hspace{-2mm}
\xymatrix{
&\mathbf{Maps}(S^1;\mathbf{B}G_{\mathrm{conn}})\ar[d]_{\mathrm{hol}}\ar[rr]^{\hspace{-2mm}\mathbf{Maps}(S^1,\hat{\mathbf{c}})}
&& \mathbf{Maps}(S^1;\mathbf{B}^3U(1)_{\mathrm{conn}})\ar[d]^{\mathrm{hol}}\ar[dr]^{~~~\mathrm{hol}_{S^1}}&\\
\underline{G}\ar[ru]^-{(P_{x_0},\nabla_{x_0})}\ar[r]&\underline{G}/\!/_{\mathrm{Ad}}\underline{G}
\ar[rr]^-{\mathbf{wzw}}&&\mathbf{B}^2U(1)_{\mathrm{conn}}/\!/_{\mathrm{Ad}}\mathbf{B}^2U(1)_{\mathrm{conn}}\ar[r]&\mathbf{B}^2U(1)_{\mathrm{conn}}\;.
}
\]
Then Proposition 3.4 from \cite{CJMSW} identifies the upper path (hence also the lower path) from $\underline{G}$ to $\mathbf{B}^2U(1)_{\mathrm{conn}}$ with the Wess-Zumino-Witten bundle gerbe. 
\par
Passing to equivalence classes of global sections, we see that $\mathbf{wzw}$ induces, for any smooth manifold $X$,
 a natural map $C^\infty(X;G)\to \hat{H}^3(X;\mathbb{Z})$.
In particular, if $X=\Sigma_2$ is a compact Riemann surface, we can further integrate over $X$ to get
\[
wzw:C^\infty(\Sigma_2;G)\to  \hat{H}^3(\Sigma_2;\mathbb{Z})\xrightarrow{\int_{\Sigma_2}} U(1)\;.
\]
This is the \emph{topological term} in the Wess-Zumino-Witten model; see \cite{Ga, FW, CJM}.
Notice how the fact that $\mathbf{wzw}$ factors through $\underline{G}/\!/_{\mathrm{Ad}}\underline{G}$ gives the conjugation invariance of the Wess-Zumino-Witten bundle gerbe, hence of the topological term in the Wess-Zumino-Witten model.

\subsubsection{$k=2$: the symplectic structure on the moduli space of flat connections on Riemann surfaces}
\label{TheSymplecticStructureOnModuliSpaceOfConnections}

For $\Sigma_2$ a compact Riemann surface, the transgression of the extended Lagrangian
$\hat {\mathbf{c}}$ yields a map
\[
\mathbf{Maps}(\Sigma_2;\mathbf{B}G_{\mathrm{conn}})\xrightarrow{\mathbf{Maps}(\Sigma_2,\hat{\mathbf{c}})}
 \mathbf{Maps}(\Sigma_2;\mathbf{B}^3U(1)_{\mathrm{conn}})\xrightarrow{\mathrm{hol}_{\Sigma_2}} \mathbf{B}U(1)_{\mathrm{conn}}
 \,,
\]
modulating a circle-bundle with connection on the moduli space of gauge fields on $\Sigma_2$.
The underlying curvature of this connection is the map obtained by composing this with
\[
  \xymatrix{
    \mathbf{B}U(1)_{\mathrm{conn}}
    \ar[r]^{F_{(-)}}
	&
    \Omega^2(-;\mathbb{R})_{\mathrm{cl}}
  }\;,
\]
which gives the canonical presymplectic 2-form
\[
 \omega 
 :
 \xymatrix{
   \mathbf{Maps}(\Sigma_2;\mathbf{B}G_{\mathrm{conn}})
   \ar[r]
   &
   \Omega^2(-;\mathbb{R})_{\mathrm{cl}}
  }
\]
on the moduli stack of principal $G$-bundles with connection on $\Sigma_2$. 
Equivalently, this is the transgression of the invariant polynomial 
$\langle -\rangle : \xymatrix{\mathbf{B}G_{\mathrm{conn}} \ar[r] &  \Omega^4_{\mathrm{cl}}}$
to the mapping stack out of $\Sigma_2$. The restriction of this 2-form to the moduli stack $\mathbf{Maps}(\Sigma_2;\flat\mathbf{B}G_{\mathrm{conn}})$ of flat principal $G$-bundles on $\Sigma_2$ induces a canonical symplectic structure  on the moduli space
\[
\mathrm{Hom}(\pi_1(\Sigma_2),G)/_{\mathrm{Ad}}G
\]
of flat $G$-bundles on $\Sigma_2$. Such a symplectic structure seems to have been first made explicit
in \cite{AB} and then identified as the phase space structure of Chern-Simons theory in \cite{Witten}.
Observing that differential forms on the moduli stack, and
hence de Rham cocycles $\mathbf{B}G \to \flat_{\mathrm{dR}}\mathbf{B}^{n+1}U(1)$,
may equivalently be expressed by simplicial forms on the bar complex of $G$, one recognizes
in the above transgression construction a stacky refinement of the construction of \cite{We}. Here
$\flat_{\mathrm{dR}}\mathbf{B}^{n+1}U(1)$ is the $n$-stack of flat de Rham coefficients, obtained via the Dold-Kan correspondence by the truncated de Rham complex 
\[
 \Omega^1(-;\mathbb{R})\xrightarrow{d}\cdots\xrightarrow{d}\Omega^n(-;\mathbb{R})\xrightarrow{d}\Omega^{n+1}_{\mathrm{cl}}(-;\mathbb{R})\;.
\]
\par
To see more explicitly what this form $\omega$ is, consider any test manifold $U \in \mathrm{CartSp}$. 
Over  this the map of stacks $\omega$ is a function which sends a $G$-principal connection 
$A \in \Omega^1(U \times \Sigma_2)$ (using that every $G$-principal bundle over $U \times \Sigma_2$
is trivializable) to the 2-form 
$$
  \int_{\Sigma_2} \langle F_{A} \wedge F_{A}\rangle
  \in
  \Omega^2(U)
  \,.
$$
Now if $A$ represents a field in the phase space, hence an element in the concretification of 
the mapping stack, then it has no ``leg''
\footnote{That is, when written in local coordinates $(u, \sigma)$ on $U \times \Sigma_2$,
then  
$A=A_i(u, \sigma) du^i + A_j (u, \sigma) d\sigma^j$ reduces to the second summand.}
 along $U$, and so it is a 1-form on $\Sigma_2$ that depends
smoothly on the parameter $U$: it is a $U$-parameterized \emph{variation} of such a 1-form.
Accordingly, its curvature 2-form splits as
$$
  F_{A} = F_A^{\Sigma_2} + d_U A
  \,,
$$
where $F_A^{\Sigma_2} := d_{\Sigma_2} A + \tfrac{1}{2}[A \wedge A]$ 
is the $U$-parameterized collection of curvature forms on $\Sigma_2$. 
The other term is the \emph{variational differential} of the $U$-collection of forms.
Since the fiber integration map $\int_{\Sigma_2} : \Omega^4(U \times \Sigma_2) \to \Omega^2(U)$ 
picks out the component of $\langle F_A \wedge F_A\rangle$ with two legs along $\Sigma_2$
and two along $U$, integrating over the former we have that
$$
  \omega|_U
  = 
  \int_{\Sigma_2} \langle F_A \wedge F_A\rangle
  = 
  \int_{\Sigma_2} \langle d_U A \wedge d_U A \rangle
  \in
  \Omega^2_{\mathrm{cl}}(U)
  \,.
$$
In particular if we consider, without loss of generality,
$(U = \mathbb{R}^2)$-parameterized variations and expand 
$$
  d_U A = (\delta_1 A) du^1 + (\delta_2 A) du^2 \in \Omega^2(\Sigma_2 \times U)
  \,,
$$
then 
$$
  \omega|_U = \int_{\Sigma_2} \langle \delta_1 A, \delta_2 A \rangle
  \,.
$$
In this form the symplectic structure appears, for instance, in prop. 3.17 of \cite{FreedCS1}
(in \cite{Witten} this corresponds to (3.2)).

In summary, this means that the circle bundle with connection obtained by transgression of the 
extended Lagrangian $\hat {\mathbf{c}}$ is a \emph{geometric prequantization} of the phase space of
3d Chern-Simons theory. Observe that traditionally prequantization involves an
arbitrary \emph{choice}: the choice of prequantum bundle with connection whose 
curvature is the given symplectic form. 
Here we see that in \emph{extended} prequantization this 
choice is eliminated, or at least reduced: while there may be many differential cocycles
lifting a given curvature form, only few of them arise by transgression from 
higher differential cocycles in top codimension. In other words, the restrictive choice
of the single geometric prequantization of the invariant polynomial 
$\langle -,-\rangle : \mathbf{B}G_{\mathrm{conn}} \to \Omega^4_{\mathrm{cl}}$
by $\hat{\mathbf{c}} : \mathbf{B}G_{\mathrm{conn}}\to \mathbf{B}^3 U(1)_{\mathrm{conn}}$
down in top codimension induces canonical choices of prequantization 
over all $\Sigma_k$ in all lower codimensions $(n-k)$.

\subsubsection{$k=3$: the Chern-Simons action functional}

Finally, for $\Sigma_3$ a compact oriented 3-manifold without boundary,
transgression of the extended Lagrangian $\hat {\mathbf{c}}$ produces the morphism
\[
\mathbf{Maps}(\Sigma_3;\mathbf{B}G_{\mathrm{conn}})\xrightarrow{\mathbf{Maps}(\Sigma_3,\hat{\mathbf{c}})}
 \mathbf{Maps}(\Sigma_3;\mathbf{B}^3U(1)_{\mathrm{conn}})\xrightarrow{\mathrm{hol}_{\Sigma_3}} 
 \underline{U}(1)\;.
\]
Since the morphisms in $\mathbf{Maps}(\Sigma_3; \mathbf{B}G_{\mathrm{conn}})$ are 
\emph{gauge transformations} between field configurations, while $\underline{U}(1)$ has
no non-trivial morphisms, this map necessarily gives a \emph{gauge invariant} $U(1)$-valued function
on field configurations. 
Indeed, evaluating over the point and passing to isomorphism classes
(hence to gauge equivalence classes), this induces the \emph{Chern-Simons action functional}
\[
S_{\hat{\mathbf{c}}}:\{\text{$G$-bundles with connection on~} \Sigma_3\}/\text{iso}\to U(1)\;.
\]
It follows from the description of $\hat{\mathbf{c}}$ given in section \ref{refinement} that 
if the principal $G$-bundle $P\to \Sigma_3$ is trivializable then
\[
S_{\hat{\mathbf{c}}}(P,\nabla)=\exp 2\pi i\int_{\Sigma_3} \mathrm{CS}_3(A)\;,
\]
where $A\in \Omega^1(\Sigma_3,\mathfrak{g})$ is the $\mathfrak{g}$-valued 1-form on $\Sigma_3$ representing 
the connection $\nabla$ in a chosen trivialization of $P$. This is actually always the case, but notice two things: first, in the stacky description one does not need to know a priori that every principal $G$-bundle on a 3-manifold is trivializable; second, the independence of $S_{\hat{\mathbf{c}}}(P,\nabla)$ of the trivialization chosen is automatic from the fact that $S_{\hat{\mathbf{c}}}$ is a morphism of stacks read at the level of equivalence classes.\par

Furthermore, if $(P,\nabla)$ can be extended to a principal $G$-bundle with connection $(\tilde{P},\tilde{\nabla})$ over a compact 4-manifold $\Sigma_4$ bounding $\Sigma_3$, one has
\[
S_{\hat{\mathbf{c}}}(P,\nabla)=\exp 2\pi i\int_{\Sigma_4}\tilde{\varphi}^*\omega^{(4)}_{\mathbf{B}G_{\mathrm{conn}}}=\exp 2\pi i\int_{\Sigma_4}\langle F_{\tilde{\nabla}},F_{\tilde{\nabla}}\rangle\;,
\]
where $\tilde{\varphi}:\Sigma_4\to \mathbf{B}G_{\mathrm{conn}}$ is the morphism corresponding to the extended bundle $(\tilde{P},\tilde{\nabla})$. Notice that the right hand side is independent of the extension chosen. Again, this is always the case, so one can actually take the above equation as a definition of the Chern-Simons action functional, see, e.g., \cite{FreedCS1, FreedCS2}. However, notice how in the stacky approach we do not need a priori to know that the oriented cobordism ring is trivial in dimension 3. Even more remarkably, the stacky point of view tells us that there would be a natural and well-defined 3d Chern-Simons action functional even if the oriented cobordism ring were nontrivial in dimension 3 or that not every $G$-principal bundle on a 3-manifold were trivializable.
An instance of checking that a nontrivial 
higher cobordism group vanishes can be found in \cite{KS2}, allowing for the 
application of the construction of Hopkins-Singer \cite{HopkinsSinger}.

\subsubsection{
The Chern-Simons action functional with Wilson loops}
\label{WithWilsonLoops}

To conclude our exposition of the examples of 1d and 3d Chern-Simons theory in higher geometry,
we now briefly discuss how both unify into the theory of 3d Chern-Simons gauge fields
with Wilson line defects. Namely,
 for every embedded knot
$$\iota : S^1 \hookrightarrow \Sigma_3$$ in the closed 3d worldvolume and every complex linear representation
$R : G \to \mathrm{Aut}(V)$ one can consider the  \emph{Wilson loop observable} $W_{\iota,R}$ mapping a gauge field 
$A : {\Sigma \to \mathbf{B}G_{\mathrm{conn}}}$, to the corresponding
``Wilson loop holonomy''
$$
  W_{\iota,R}:A \mapsto \mathrm{tr}_{R}( \mathrm{hol}(\iota^*A)) \in \mathbb{C}
  \,.
$$
This is the trace, in the given representation, of the parallel transport defined by the connection $A$ around the loop $\iota$
(for any choice of base point).
It is an old observation\footnote{This can be traced back to \cite{BalachandranBorchardtStern};
a nice modern review can be found in section 4 of 
\cite{Beasley}.} 
that this Wilson loop $W(C,A,R)$ is itself the \emph{partition function}
of a 1-dimensional topological $\sigma$-model quantum field theory that describes the topological
sector of a particle charged under the nonabelian background gauge field $A$.
In section 3.3 of \cite{Witten} it was therefore emphasized that  Chern-Simons theory
with Wilson loops should really be thought of as given by a single Lagrangian
which is the sum of the 3d Chern-Simons Lagrangian for the gauge field as above, 
plus that for this topologically charged particle.

We now briefly indicate how this picture is naturally captured by
higher geometry and refined to a single \emph{extended} Lagrangian for coupled 1d and 3d Chern-Simons theory,
given by maps on higher moduli stacks. In doing this, we will also see how
the ingredients of Kirillov's orbit method  and the Borel-Weil-Bott theorem
find a natural rephrasing in the context of smooth differential moduli stacks.
The key observation is that for 
$
  \langle \lambda , -\rangle 
$
an integral weight for our simple, connected, simply connected and compact  Lie group $G$, the
contraction of $\mathfrak{g}$-valued differential forms with $\lambda$ extends
to a morphism of smooth moduli stacks of the form
$$
  \langle \mathbf{\lambda} , - \rangle
  \;:\, 
    \Omega^1(-,\mathfrak{g})//\underline{T}_\lambda
	\to
	\mathbf{B}U(1)_{\mathrm{conn}}
  \,,
$$
where  $T_\lambda \hookrightarrow G$ is the maximal torus of $G$ which is the stabilizer subgroup of
$\langle \lambda, -\rangle$ under the coadjoint action of $G$ on $\mathfrak{g}^*$. 
Indeed, 
this is just the classical statement that exponentiation of $\langle \lambda,-\rangle$ 
induces an isomorphism between the integral weight lattice $\Gamma_{\mathrm{wt}}(\lambda)$ relative to the maximal torus $T_\lambda$ and the $\mathbb{Z}$-module $\mathrm{Hom}_{\mathrm{Grp}}(T_\lambda,U(1))$ and that under this isomorphism a gauge transformation
of a $\mathfrak{g}$-valued 1-form $A$ turns into that of the $\mathfrak{u}(1)$-valued 1-form
$\langle \lambda, A\rangle$.

Comparison with the discussion in section \ref{toy-example} shows that this is 
the extended Lagrangian of a 1-dimensional Chern-Simons theory. In fact it is just a slight
variant of the trace-theory discussed there: if we realize $\mathfrak{g}$ as a matrix Lie algebra
and write $\langle \alpha, \beta\rangle = \mathrm{tr}(\alpha \cdot \beta)$ as the matrix trace,
then the above Chern-Simons 1-form is 
given 
by the ``$\lambda$-shifted trace''
$$
  \mathrm{CS}_\lambda(A) := \mathrm{tr}(\lambda \cdot A) \in \Omega^1(-;\mathbb{R})
  \,.
$$
Then, clearly, while the ``plain'' trace is invariant under the adjoint action of all of $G$, 
the $\lambda$-shifted trace is invariant only under the subgroup $T_\lambda$ of $G$ that fixes $\lambda$. 

Notice that the domain of $\langle \mathbf{\lambda}, -\rangle$ 
naturally sits 
inside $\mathbf{B}G_{\mathrm{conn}}$
by the canonical map
$$
    \Omega^1(-,\mathfrak{g})/\!/ \underline{T}_\lambda
	\to
    \Omega^1(-,\mathfrak{g})/\!/ \underline{G}
	\simeq
	\mathbf{B}G_{\mathrm{conn}}
  \,.
$$
One sees that the homotopy fiber of this map is the \emph{coadjoint orbit} 
$\mathcal{O}_\lambda \hookrightarrow \mathfrak{g}^*$ of $\langle \lambda, -\rangle$,
equipped with the map of stacks
$$
  \mathbf{\theta}
   :
    \mathcal{O}_{\lambda} \simeq \underline{G}/\!/\underline{T}_\lambda 
	\to
	\Omega^1(-,\mathfrak{g})/\!/\underline{T}_\lambda
$$
which over a test manifold $U$ sends $g \in C^\infty(U,G)$ to 
the pullback $g^* \theta_G$ of the Maurer-Cartan form.
Composing this with the above extended Lagrangian $\langle \mathbf{\lambda}, -\rangle$
yields a map
$$
  \langle \mathbf{\lambda}, \mathbf{\theta}\rangle
  : 
  \xymatrix{
    \mathcal{O}_\lambda
	\ar[r]^-{\mathbf{\theta}}
	&
	\Omega^1(-,\mathfrak{g})/\!/\underline{T}_\lambda
	\ar[r]^-{\mathbf{\langle \lambda, -\rangle}}
	&
	\mathbf{B}U(1)_{\mathrm{conn}}
  }
$$
which modulates a canonical $U(1)$-principal bundle with connection on the coadjoint orbit.
One finds that this is the canonical prequantum bundle used in the orbit method 
 \cite{Kirillov}.
 In particular its curvature is 
 the canonical symplectic form on the coadjoint orbit.
%

So far this shows how the ingredients of the orbit method are incarnated in smooth moduli 
stacks. This now immediately induces Chern-Simons theory with Wilson loops by
considering the map $ \Omega^1(-,\mathfrak{g})/\!/ \underline{T}_\lambda\to	\mathbf{B}G_{\mathrm{conn}}$
itself as the target\footnote{This means that here we are secretely moving from the topos of (higher) stacks on smooth manifolds to its \emph{arrow topos}, see section \ref{FieldsInSlices} below.} for a field theory defined on knot inclusions $\iota: S^1 \hookrightarrow \Sigma_3$.
%
%
This means that a field configuration
is a diagram of smooth stacks of the form
\[
 \raisebox{20pt}{
  \xymatrix{
     S^1 \ar[rr]^-{(\iota^*A)^g}_>{\ }="s" \ar[d]_\iota^>{\ }="t" 
	 && \Omega^1(-,\mathfrak{g})/\!/\underline{T}_\lambda
	 \ar[d]
	 \\
	 \Sigma_3
	 \ar[rr]_-{A}
	 &&
	~ \mathbf{B}G_{\mathrm{conn}}\;,
	 \ar@{=>}^g "s"; "t"
  }
  }
  \label{FieldsForCSWithWilsonLoop}
\]
i.e., that a field configuration consists of 
\begin{itemize}
  \item a gauge field $A$ in the ``bulk'' $\Sigma_3$;
  \item a $G$-valued function $g$ on the embedded knot
\end{itemize}
such that 
 the
restriction of the ambient gauge field $A$ to the knot is equivalent, via the gauge transformation $g$, to a $\mathfrak{g}$-valued connection on $S^1$ whose local $\mathfrak{g}$-valued 1-forms are related to each other by local gauge transformations taking values in the torus $T_\lambda$. 
Moreover, a gauge transformation between two such field configurations $(A,g)$ and $(A',g')$ is a pair $(t_{\Sigma_3},t_{S^1})$ consisting of a 
$G$-gauge transformation $t_{\Sigma_3}$ on $\Sigma_3$ and a $T_\lambda$-gauge transformation
$t_{S^1}$ on $S^1$, intertwining the 
gauge transformations $g$ and $g'$. In particular if the bulk gauge field on $\Sigma_3$ is held
fixed, i.e., if $A=A'$, then $t_{S^1}$ satisfies the equation $g' = g\, t_{S^1}$. This means that 
the Wilson-line components of gauge-equivalence classes
of field configurations are naturally identified with smooth functions $S^1 \to G/T_{\lambda}$, i.e., with smooth functions on the Wilson loop with values in the coadjoint orbit.
This is essentially a rephrasing of the above statement that $G/{T_\lambda}$ is the homotopy
fiber of the inclusion of the moduli stack of Wilson line field configurations into the  moduli stack of bulk field configurations.

We may postcompose the two horizontal maps in this square with our two extended
Lagrangians, that for 1d and that for 3d Chern-Simons theory, to get the diagram
$$
 \raisebox{20pt}{
  \xymatrix{
     S^1 \ar[rr]^-{(\iota^*A)^g}_>{\ }="s" \ar[d]_\iota^>{\ }="t" 
	 && \Omega^1(-,\mathfrak{g})/\!/T
	 \ar[d]
	 \ar[r]^-{\langle \mathbf{\lambda}, -\rangle}
	 &
	 \mathbf{B}U(1)_{\mathrm{conn}}
	 \\
	 \Sigma_3
	 \ar[rr]^-{A}
	 &&
	 \mathbf{B}G_{\mathrm{conn}}
	 \ar[r]^-{\hat {\mathbf{c}}}
	 &
	 \mathbf{B}^3 U(1)_{\mathrm{conn}}\;.
	 \ar@{=>}_g "s"; "t"
  }
  }
$$
Therefore, writing $\mathbf{Fields}_{\mathrm{CS}+ \mathrm{W}}\left(S^1 \stackrel{\iota}{\hookrightarrow} \Sigma_3\right)$ for the moduli stack of field configurations for Chern-Simons theory with Wilson lines, we find
two action functionals as the composite top and left morphisms in the diagram
$$
 \raisebox{40pt}{
  \xymatrix{
    \mathbf{Fields}_{\mathrm{CS}+ \mathrm{W}}\left(S^1 \stackrel{\iota}{\hookrightarrow} \Sigma_3\right)
	\ar[r]
	\ar[d]
	&
	\mathbf{Maps}(\Sigma_3, \mathbf{B}G_{\mathrm{conn}})
	\ar[d]
	\ar[rrr]^-{\mathrm{hol}_{\Sigma_3}\mathbf{Maps}(\Sigma_3, \hat {\mathbf{c}})}
	&&&
	\underline{U}(1)
	\\
	\mathbf{Maps}(S^1, \Omega^1(-,\mathfrak{g})/\!/T_\lambda)
	\ar[r]
	\ar[dd]|{\mathrm{hol}_{S^1} \mathbf{Maps}(S^1, \langle \mathbf{\lambda},-\rangle)}
	&
	\mathbf{Maps}(S^1, \mathbf{B}G_{\mathrm{con}})
	\\
	\\
	\underline{U}(1)
  }
  }
$$
in $\mathbf{H}$, where the top left square is the homotopy pullback that characterizes
maps in $\mathbf{H}^{(\Delta^1)}$ in terms of maps in $\mathbf{H}$.
The product of these is the
action functional
$$
\xymatrix{
\mathbf{Fields}_{\mathrm{CS}+ \mathrm{W}}\left(S^1 \stackrel{\iota}{\hookrightarrow} \Sigma_3\right)\ar[r]& \mathbf{Maps}(\Sigma_3, \mathbf{B}^3 U(1)_{\mathrm{conn}})\times  \mathbf{Maps}(S^1, \mathbf{B} U(1)_{\mathrm{conn}})
\ar[d]\\
&\underline{U}(1)\times \underline{U}(1)\ar[r]^-{\cdot}
&\underline{U}(1)\;.
&
}
$$
where the rightmost arrow is the multiplication in $U(1)$.
Evaluated on a field configuration with components $(A,g)$ as just discussed, this 
is 
$$
  \exp\left(
    2 \pi i 
	\left(
	   \int_{\Sigma_3} \mathrm{CS}_3(A)
	   + 
       \int_{S^1} \langle \lambda, (\iota^*A)^g\rangle 
	\right)
  \right)
  \,.
$$
This is indeed the action functional for Chern-Simons theory with Wilson loop $\iota$
in the representation $R$ correspponding to the integral weight $\langle \lambda, -\rangle$ by
the Borel-Weil-Bott theorem, 
as reviewed for instance 
in Section 4 of \cite{Beasley}.

\medskip

Apart from being an elegant and concise repackaging of this well-known action functional
and the quantization conditions that go into it, the above reformulation in terms of stacks immediately 
leads to prequantum line bundles in Chern-Simons theory with Wilson loops. Namely, by considering the codimension 1 case, one finds the
the symplectic structure and
the canonical prequantization for the moduli stack of field configurations on surfaces with specified singularities at
specified punctures \cite{Witten}.
Moreover, this is just the first example in a general mechanism of (extended) action functionals
with defect and/or boundary insertions. Another example of the same mechanism is the 
gauge coupling action functional of the open string. This we discuss in section \ref{OpenStringGaugeCoupling}
below.

\section{Extension to more general examples}
\label{CupProductCS}

The way we presented the two examples of the previous sections indicates that they
are clearly just the beginning of a 
rather general pattern of extended prequantized higher gauge theories of Chern-Simons type:
for every smooth higher group $G$ with universal differential higher moduli stack 
$\mathbf{B}G_{\mathrm{conn}}$ (and in fact for any higher moduli stack at all, 
as further discussed in section \ref{Sigma-Models} below) every differentially
refined universal characteristic map of stacks
$$
  \mathbf{L}
    :
  \xymatrix{
    \mathbf{B}G_{\mathrm{conn}}
	\ar[r]
	&
	\mathbf{B}^n  U(1)_{\mathrm{conn}}
  }
$$
constitutes an extended Lagrangian -- hence, by iterated transgression, the action functional, 
prequantum theory and WZW-type 
action functional -- of an $n$-dimensional  Chern-Simons type gauge field theory with (higher) gauge group $G$. Moreover, just moving 
from higher stacks on the site of smooth manifolds to higher stacks on the site of smooth supermanifolds one has an immediate and natural generalization to  super-Chern-Simons theories. Here we briefly
survey some examples of interest, which were introduced in detail in \cite{SSSIII}
and \cite{FiorenzaSatiSchreiberIII}. Further examples and further details can be found in 
section 5.7 of \cite{survey}. 

\subsection{String connections and twisted String structures}
\label{string}

Notice how we have moved from the 1d Chern-Simons theory of section \ref{toy-example} to the 3d Chern-Simon theory of section \ref{3d} by replacing the connected but not 1-connected compact Lie group $U(n)$ with a compact 2-connected but not 3-connected Lie group $G$. The natural further step towards a higher dimensional Chern-Simons theory would then be to consider a compact Lie group which is (at least) 3-connected. Unfortunately, there exists no such Lie group: if $G$ is compact and simply connected then its third homotopy group will be nontrivial, see e.g. \cite{Mi}. However, a solution to this problem does exist if we move from compact Lie groups to the more general context of smooth higher groups, i.e. if we focus on the stacks of principal bundles rather than on their gauge groups. As a basic example, think of 
how we obtained the stacks $\mathbf{B}SU(n)$ and $\mathbf{B}SU(n)_{\mathrm{conn}}$ out of $\mathbf{B}U(n)$ and $\mathbf{B}U(n)_{\mathrm{conn}}$ in section \ref{killing}. There we first obtained these stacks as homotopy fibers of the morphisms of stacks
\[
\mathbf{c}_1:\mathbf{B}U(n)\to \mathbf{B}U(1)\;;\qquad \qquad \hat{\mathbf{c}}_1:\mathbf{B}U(n)_{\mathrm{conn}}\to \mathbf{B}U(1)_{\mathrm{conn}}
\]
refining the first Chern class. Then, in a second step, we identified these homotopy fibers with the stack of principal bundles (with and without connection) for a certain compact Lie group, which turned out to be $SU(n)$. However,  
the homotopy fiber definition would have been meaningful even in case we would have been unable to show that there was a compact Lie group behind it, or even in case there would have been none 
such. This may seem too far a generalization, but actually Milnor's theorem \cite{Mi1} would have assured us 
in any case that there existed a \emph{topological} group $SU(n)$ whose classifying space is homotopy equivalent to the topological realization of the homotopy fiber $\mathbf{B}SU(n)$, that is, equivalently, to the homotopy fiber of the topological realization of the morphism $\mathbf{c}_1$. This is nothing but the topological characteristic map
\[
c_1: BU(n)\to BU(1)\simeq K(\mathbb{Z},2)
\]
defining the first Chern class. In other words, one defines the space $BSU(n)$ as the homotopy pullback
\[
\xymatrix{BSU(n)\ar[d]\ar[r]&{*}\ar[d]\\
BU(n)\ar[r]^-{c_1}&K(\mathbb{Z},2)\;;
}
\]
the based loop space 
$\Omega BSU(n)$  has a natural structure of topological group ``up to homotopy'', and Milnor's theorem precisely tells us that we can strictify it, i.e. we can find a topological group $SU(n)$ (unique up to homotopy) such that
$
SU(n)\simeq \Omega BSU(n)$.
Moreover, $BSU(n)$, defined as a homotopy fiber, will be a classifying space for this ``homotopy-$SU(n)$'' group. From this perspective, we see that having a model for the homotopy-$SU(n)$ which is a compact Lie group is surely something nice to have, but that we would have nevertheless been able to  speak in a rigorous and well-defined way of the groupoid of smooth $SU(n)$-bundles over a smooth manifold $X$ even in case such a compact Lie model did not exist. The same considerations apply to the stack of principal $SU(n)$-bundles with connections.
\par
These considerations may look redundant, since one is well aware that there is indeed a compact Lie group $SU(n)$ with all the required features. However, this way of reasoning becomes prominent and indeed essential when we move to higher characteristic classes. The fundamental example is probably the following. For $n\geq 3$ the spin group $\mathrm{Spin}(n)$ is compact and simply connected with $\pi_3(\mathrm{Spin}(n))\cong \mathbb{Z}$. The generator of $H^4(B\mathrm{Spin}(n);\mathbb{Z})$ is the first fractional Pontrjagin class $\tfrac{1}{2}p_1$, which can be equivalently seen as a characteristic map
\[
\tfrac{1}{2}p_1\colon B\mathrm{Spin}(n)\to K(\mathbb{Z};4)\;.
\]
The String group $\mathrm{String}(n)$ is then defined as the topological group whose classifying space is the homotopy fiber of  $\frac{1}{2}p_1$, i.e., the homotopy pullback
\[
\xymatrix{B\mathrm{String}(n)\ar[d]\ar[r]&{*}\ar[d]\\
B\mathrm{Spin}(n)\ar[r]^{\frac{1}{2}p_1}&K(\mathbb{Z},4)\;;
}
\]
this defines $\mathrm{String}(n)$ uniquely up to homotopy. The topological group $\mathrm{String}(n)$ is 6-connected with $\pi_7(\mathrm{String}(n))\cong \mathbb{Z}$. The generator of $H^8(B\mathrm{String}(n);\mathbb{Z})$ is the second 
fractional Pontrjagin class $\frac{1}{6}p_2$, see \cite{SSSII}. One can then define the 3-stack of smooth $\mathrm{String}(n)$-principal bundles as the homotopy pullback
\[
\xymatrix{\mathbf{B}\mathrm{String}(n)\ar[d]\ar[r]&{*}\ar[d]\\
\mathbf{B}\mathrm{Spin}(n)\ar[r]^{\frac{1}{2}\mathbf{p}_1}&\mathbf{B}^3U(1)\;,
}
\]
where $\tfrac{1}{2}\mathbf{p}_1$ is the morphism of stacks whose topological realization is $\tfrac{1}{2}p_1$. In other words, a String$(n)$-principal bundle over a smooth manifold $X$ is the datum of a 
Spin$(n)$-principal bundle over $X$ together with a trivialization of the associated 
principal $U(1)$-3-bundle. The characteristic map 
\[
\tfrac{1}{6}p_2: B\mathrm{String}(n)\to K(\mathbb{Z};8)
\]
is the topological realization of a morphism of stacks
\[
\tfrac{1}{6}\mathbf{p}_2: \mathbf{B}\mathrm{String}(n)\to \mathbf{B}^7U(1)\;,
\]
see \cite{SSSIII, FSS}. Similarly, one can define the 3-stack of smooth 
String bundles with connections as the homotopy pullback
\[
\xymatrix{\mathbf{B}\mathrm{String}(n)_{\mathrm{conn}}\ar[d]\ar[r]&{*}\ar[d]\\
\mathbf{B}\mathrm{Spin}(n)_{\mathrm{conn}}\ar[r]^{\frac{1}{2}\hat{\mathbf{p}}_1}&\mathbf{B}^3U(1)_{\mathrm{conn}}\;,
}
\]
where $\tfrac{1}{2}\hat{\mathbf{p}}_1$ is the lift of $\tfrac{1}{2}{\mathbf{p}}_1$ to the stack of Spin$(n)$-bundles with connections. Again, this means that a String$(n)$-bundle with connection over a smooth manifold $X$ is the datum of a Spin$(n)$-bundle with connection over $X$ together with a trivialization of the associated $U(1)$-3-bundle with connection. The morphism $\frac{1}{6}\mathbf{p}_2$ lifts to a morphism
 \[
\tfrac{1}{6}\hat{\mathbf{p}}_2: \mathbf{B}\mathrm{String}(n)_{\mathrm{conn}}\to \mathbf{B}^7U(1)_{\mathrm{conn}}\;,
\]
see \cite{FSS}, and this defines a 7d Chern-Simons theory with gauge group the String$(n)$-group.
\par
In the physics literature one usually considers also a more flexible notion of String connection, in which one requires that the underlying $U(1)$-3-bundle of a Spin$(n)$-bundle with connection is trivialized, but does not require the underlying 3-connection to be trivialized. In terms of stacks, this corresponds to considering the homotopy pullback
\[
\xymatrix{\mathbf{B}\mathrm{String}(n)_{\mathrm{conn}'}\ar[d]\ar[r]&{*}\ar[d]\\
\mathbf{B}\mathrm{Spin}(n)_{\mathrm{conn}}\ar[r]^{~~\frac{1}{2}\mathbf{p}_1}&\mathbf{B}^3U(1)\;;
}
\]
see, e.g., \cite{waldorf}. Furthermore, it is customary to consider not only the case where the underlying $U(1)$-3-bundle (with or without connection) is trivial, but also the case when it is equivalent to a fixed \emph{background}  $U(1)$-3-bundle (again, eventually with connection). Notably, the connection 3-form of this fixed background is the C-field of the M-theory literature
(cf. \cite{tcu, tcu2}). 
The moduli stacks of Spin$(n)$-bundles on a smooth manifold $X$ with possibly nontrivial fixed $U(1)$-3-bundle background are called \cite{SSSIII} moduli stacks of \emph{twisted} String bundles on $X$. A particular intersting case is when the twist is independent of $X$, hence is itself 
given by a universal characteristic class, hence by a twisting morphism 
$\mathbf{c}:\xymatrix{\mathbf{S} \ar[r] & \mathbf{B}^3U(1)}$, where $\mathbf{S}$ is some 
(higher moduli) stack. In this case, indeed, one can define the stack $\mathbf{B}\mathrm{String}(n)^{\mathbf{c}}$ of $\mathbf{c}$-twisted String$(n)$-structures as the homotopy pullback
 \[
  \xymatrix{\mathbf{B}\mathrm{String}(n)^{\mathbf{c}}\ar[d]\ar[r]&{\mathbf{S}}\ar[d]^{\mathbf{c}}\\
\mathbf{B}\mathrm{Spin}(n)\ar[r]^-{\frac{1}{2}\hat{\mathbf{p}}_1}&\mathbf{B}^3U(1)\;,
}
\]
and similarly for the stack of $\mathbf{c}$-twisted String$(n)$-connections. 
This is a higher analog of $\mathrm{Spin}^c$-structures, whose universal moduli stack
sits in the analogous homotopy pullback diagram
$$
  \raisebox{20pt}{
  \xymatrix{
    \mathbf{B}\mathrm{Spin}^c(n) \ar[r] \ar[d] & \mathbf{B}U(1) \ar[d]^{\mathbf{c}_1 \, \mathrm{mod}\, 2}
	\\
	\mathbf{B}SO(n) \ar[r]^-{\mathbf{w}_1} & \mathbf{B}^2 \mathbb{Z}_2
  }}
  \,.
$$
(For more on higher $\mathrm{Spin}^c$-structures see also \cite{tw, tw2} and section 5.2 of \cite{survey}.)
By a little abuse of terminology, when the twisting morphism $\mathbf{a}$ is the refinement of a characteristic class for a compact simply connected simple Lie group $G$ to a morphism of stacks 
$\mathbf{a}:\mathbf{B}G\to \mathbf{B}^3U(1)$, one may speak of \emph{$G$-twisted structures} rather than of 
$\mathbf{a}$-twisted structures. For instance, in heterotic string theory $G$ is the group $E_8\times E_8$ and $\mathbf{a}$ is a stacky refinement of the second Chern class.
\par
By the discussion in section \ref{3d} the differential twisting maps $\tfrac{1}{2}\hat {\mathbf{p}}_1$ and
$\hat {\mathbf{a}}$ appearing here are at the same time extended Lagrangians of Chern-Simons theories.
Together with the nature of 
homotopy pullback, it follows \cite{FSS}
that a field $\phi : {X \to \mathbf{B}\mathrm{String}^{\mathbf{a}}}_{\mathrm{conn}}$ consists of 
pairs of fields and homotopy between their Chern-Simons data, namely of
\begin{enumerate}
  \item a $\mathrm{Spin}$-connection $\nabla_{\mathfrak{so}}$;
  \item a $G$-connection $\nabla_{\mathfrak{g}}$;
  \item a twisted 2-form connection $B$ whose curvature 3-form $H$ is locally given by
    $H = d B + \mathrm{CS}(\nabla_{\mathfrak{so}}) - \mathrm{CS}(\nabla_{\mathfrak{g}})$.
\end{enumerate}
This the the data for (Green-Schwarz-)anomaly-free background gauge fields 
(gravity, gauge field, Kalb-Ramond field) for the heterotic string \cite{SSSIII}.
A further refinement of this construction yields the universal moduli stack for the
 \emph{supergravity C-field} configurations in terms of $E_8$-twisted String connections
 \cite{FiorenzaSatiSchreiberII}. Here the presence of the differential 
 characteristic maps $\hat {\mathbf{c}}$ induces the Chern-Simons gauge-coupling
 piece of the supergravity 2-brane (the \emph{M2-brane}) action functional.

\subsection{Cup-product Chern-Simons theories}

In section \ref{3d} we had restricted attention to 3d Chern-Simons theory with simply connected
gauge groups. Another important special case of 3d Chern-Simons theory is that for 
gauge group the circle group $U(1)$, which is of course not simply connected. In this case the
univsersal characteristic map that controls the theory is the differential refinement of the
\emph{cup product class} $c_1 \cup c_1$. Here we briefly indicate this case and the 
analogous higher dimensional Chern-Simons theories obtained from cup products of higher 
classes and from higher order cup products.

\medskip

The cup product $\cup$ in integral cohomology can be lifted to a cup product 
$\hat{\cup}$ in differential cohomology, i.e., for any smooth manifold $X$ we have a natural commutative diagram 
\[
\xymatrix{
\hat{H}^p(X;\mathbb{Z})\otimes \hat{H}^q(X;\mathbb{Z})\ar[d]\ar[r]^{\phantom{mmm}\hat{\cup}} & \hat{H}^{p+q}(X;\mathbb{Z})\ar[d]\\
{H}^p(X;\mathbb{Z})\otimes {H}^q(X;\mathbb{Z})\ar[r]^{\phantom{mm}{\cup}} & {H}^{p+q}(X;\mathbb{Z})\;,
}
\]
for any $p,q\geq 0$. Moreover, this cup product is induced by a cup product defined at the level of \v{C}ech-Deligne cocycles, the so called  \emph{Beilinson-Drinfeld cup product}, see \cite{brylinski}. 
This, in turn, may be seen \cite{FiorenzaSatiSchreiberIII} 
to come from a morphism of higher universal moduli stacks
$$
\widehat{\mathbf{\cup}}
:
\mathbf{B}^{n_1}U(1)_{\mathrm{conn}}\times \mathbf{B}^{n_2}U(1)_{\mathrm{conn}}\to \mathbf{B}^{n_1+n_2+1}U(1)_{\mathrm{conn}}\;.
$$
Moreover, since the Beilinson-Deligne cup product is associative up to homotopy, this induces a well-defined morphism
$$
\mathbf{B}^{n_1}U(1)_{\mathrm{conn}}\times \mathbf{B}^{n_2}U(1)_{\mathrm{conn}}\times \cdots \times \mathbf{B}^{n_{k+1}}U(1)_{\mathrm{conn}}\to \mathbf{B}^{n_1+\cdots+n_{k+1}+k}U(1)_{\mathrm{conn}}\;.
$$
In particular, for $n_1=\cdots=n_{k+1}=3$, one finds a cup product morphism
$$
\left(\mathbf{B}^{3}U(1)_{\mathrm{conn}}\right)^{k+1}\to \mathbf{B}^{4k+3}U(1)_{\mathrm{conn}}\;.
$$
Furthermore, one sees from the explicit expression of the Beilinson-Deligne cup product that, on a local chart $U_\alpha$, if the 3-form datum of a connection on a $U(1)$-3-bundle is the 3-form $C_\alpha$, then the $(4k+3)$-form local datum for the corresponding connection on the associated $U(1)$-$(4k+3)$-bundle is
\[
C_\alpha\wedge \underbrace{dC_\alpha \wedge \cdots \wedge{dC_\alpha}}_{k\text{ times}}\;.
\]
\par
Now let $G$ be a compact and simply connected simple Lie group and let $\hat{\mathbf{c}}:\mathbf{B}G_{\mathrm{conn}}\to \mathbf{B}^3U(1)_{\mathrm{conn}}$ be the morphism of stacks underlying the fundamental characteristic class $c\in H^4(BG,\mathbb{Z})$. Then we can consider the $(k+1)$-fold product of $\hat{\mathbf{c}}$ with itself:
  \[
    \hat {\mathbf{c}}~ \hat {\mathbf{\cup}}~ \hat {\mathbf{c}}~ \hat {\mathbf{\cup}}\cdots ~ \hat {\mathbf{\cup}}~ \hat {\mathbf{c}} 
	: 
	\xymatrix{
	  \mathbf{B}G_{\mathrm{conn}}
	  \ar[rr]^{\hspace{-7mm}(\hat {\mathbf{c}},\dots,\hat {\mathbf{c}})}
	  &&
	 \left( \mathbf{B}^3 {{U}(1)}_{\mathrm{conn}}\right)^{k+1}
	  \ar[r]^<<<<{~\hat {\mathbf{\cup}}}
	  &
	  \mathbf{B}^{4k+3} U(1)_{\mathrm{conn}}\;.
	}
  \]
 If $X$  is a compact oriented smooth manifold, fiber integration along $X$ gives the  morphism
 \[
 \mathbf{Maps}(X,\mathbf{B}G_{\mathrm{conn}})\xrightarrow{}
 \mathbf{Maps}(X,\mathbf{B}^{4k+3} U(1)_{\mathrm{conn}})\xrightarrow{\mathrm{hol}_X}\mathbf{B}^{4k+3-\dim X} U(1)_{\mathrm{conn}}\;.
 \]
In particular, if $\dim X=4k+3$, by evaluating over the point and taking equivalence classes we get a canonical morphism
\[
\{\text{$G$-bundles with connections on $X$}\}/\text{iso}\to U(1)\;.
\] 
This is the action functional of the $(k+1)$-fold \emph{cup product Chern-Simons theory} induced by the $(k+1)$-fold cup product of $c$ with itself \cite{FiorenzaSatiSchreiberIII}. This way one obtains, for every $k\geq 0$,  a $(4k+3)$-dimensional theory starting with a 3d Chern-Simons theory.
Moreover, in the special case that the principal $G$-bundle on $X$ is topologically trivial, this action functional has a particularly simple expression: it is given by
\[
\exp 2\pi i \int_X \mathrm{CS}_3(A)\wedge \langle F_A, F_A\rangle\wedge \cdots \wedge \langle F_A, F_A\rangle\;,
\]
where $A\in \Omega^1(X;\mathfrak{g})$ is the $\mathfrak{g}$-valued 1-form on $X$ representing the connection in the chosen trivialization of the $G$-bundle. But notice that in this more general situation now 
not \emph{every} gauge field configuration will have an underlying trivializable (higher) bundle
anymore, the way it was true for the 3d Chern-Simons theory of a simply connected Lie group in
section \ref{3d}. 
\par
More generally, one can consider an arbitrary smooth (higher) group $G$, e.g. $U(n)\times \mathrm{Spin}(m)\times \mathrm{String}(l)$,  together with $k+1$ characteristic maps
 $
  \hat {\mathbf{c}}_i
   :
  \mathbf{B}G_{\mathrm{conn}}
  \to 
  \mathbf{B}^{n_i} {{U}(1)}_{\mathrm{conn}}$
and
one can form the $(k+1)$-fold product
  \[
    \hat {\mathbf{c}}_1~ \hat {\mathbf{\cup}}\cdots ~ \hat {\mathbf{\cup}}~ \hat {\mathbf{c}}_{k+1} 
	: 
	\mathbf{B}G_{\mathrm{conn}}
	  \rightarrow
	  \mathbf{B}^{n_1+\cdots n_{k+1}+k} {{U}(1)}_{\mathrm{conn}}
	\,,
	  \]
inducing a $(n_1+\cdots n_{k+1}+k)$-dimensional Chern-Simons-type theory. For instance, if $G_1$ and $G_2$ are two compact simply connected simple Lie groups, then we have a 7d cup product Chern-Simons theory associated with the cup product $ \hat {\mathbf{c}}_1 \hat {\mathbf{\cup}}  \hat {\mathbf{c}}_2$. If  $(P_1,\nabla_1)$ and $(P_2,\nabla_2)$ are a pair of topologically trivial principal $G_1$- and $G_2$-bundles with connections  over a 7-dimensional oriented compact manifold without boundary $X$, the action functional of this Chern-Simons theory on this pair is given by
\[
\exp 2\pi i \int _X \mathrm{CS}_3(A_1)\wedge \langle F_{A_2}, F_{A_2}\rangle=\exp 2\pi i \int _X \mathrm{CS}_3(A_2)\wedge \langle F_{A_1}, F_{A_1}\rangle\;,
\]
where $A_i$ is the connection 1-forms of $\nabla_i$, for $i=1,2$.
Notice how in general a $G_i$-principal bundle on a 7-dimensional manifold is not topologically trivial, but still we have a well defined cup-product Chern-Simons action $S_{ \hat {\mathbf{c}}_1 \hat {\mathbf{\cup}}  \hat {\mathbf{c}}_2}$. In the topologically nontrivial situation, however, there will not be such a simple global expression for the action.
\par

Let us briefly mention a few representative important examples from string theory and M-theory 
which admit a natural interpretation as cup-product Chern-Simons theories,
the details of which can be found in \cite{FiorenzaSatiSchreiberIII}. 
For all examples presented below we write the Chern-Simons action for the topologically trivial sector.

\begin{itemize}
 \item {\it Abelian higher dimensional CS theory and self-dual higher gauge theory.}
  For every $k \in \mathbb{N}$ the differential cup product yields the extended Lagrangian
  $$
    \mathbf{L}
	:
	\xymatrix{
      \mathbf{B}^{2k+1}U(1)_{\mathrm{conn}}
	  \ar[r]
	  & 
	  \mathbf{B}^{2k+1}U(1)_{\mathrm{conn}} \times \mathbf{B}^{2k+1}U(1)_{\mathrm{conn}}
	  \ar[r]^-{\widehat{\mathbf{\cup}}}
	  &
	  \mathbf{B}^{4k+3}U(1)_{\mathrm{conn}}
	}
  $$
  for a $4k+3$-dimensional Chern-Simons theory of $(2k+1)$-form connections on higher circle bundles
 (higher bundle gerbes). Over a $4k+3$-dimensional manifold $\Sigma$ the corresponding 
  action functional applied to gauge fields $A$ whose underlying bundle is trivial is given by 
$$
   \exp 2\pi i\int_{\Sigma}\mathrm{CS}_1(A) \cup d \mathrm{CS}_1(A)=\exp 2\pi i\int_{\Sigma}A \wedge F_A
   \,,
$$
where $F_A=dA$ is the curvature of a $U(1)$-connection $A$. 
Similarly, the transgression of $\mathbf{L}$ to codimension 1 over
a manifold $\Sigma$ of dimenion $4k+2$ yields the prequantization
of a symplectic form on $(2k+1)$-form connections which, by a derivation analogous to that in
section \ref{TheSymplecticStructureOnModuliSpaceOfConnections}, is given by 
$$
  \omega(\delta A_1 , \delta A_1)
  =
  \int_\Sigma \delta A_1 \wedge \delta A_1
  \,.
$$
A complex polarization of this symplectic structure is given by a choice of conformal metric on $\Sigma$
and the corresponding canonical coordinates are complex Hodge self-dual forms on $\Sigma$. 
This yields the famous holographic relation between higher abelian Chern-Simons theory and 
self-dual higher abelian gauge theory in one dimension lower.

\item {\it The M5-brane self-dual theory:}
In particular, for $k = 1$ it was argued in \cite{Witten5BraneEffective} that the 7-dimensional
Chern-Simons theory which we refine to an extended prequantum theory by the extended Lagrangian 
  $$
    \mathbf{L}
	:
	\xymatrix{
      \mathbf{B}^{3}U(1)_{\mathrm{conn}}
	  \ar[r]
	  & 
	  \mathbf{B}^{3}U(1)_{\mathrm{conn}} \times \mathbf{B}^{3}U(1)_{\mathrm{conn}}
	  \ar[r]^-{\widehat{\mathbf{\cup}}}
	  &
	  \mathbf{B}^{7}U(1)_{\mathrm{conn}}
	}
  $$
  describes, in this holographic manner, 
  the quantum theory of the self-dual 2-form in the 6-dimensional worldvolume theory of a single M5-brane.
  Since moreover in \cite{WittenAdSCFT} it was argued that this abelian 7-dimensional Chern-Simons theory
  is to be thought of as the abelian piece in the Chern-Simons term of 11-dimensional supergravity
  compactified on a 4-sphere, and since this term in general receives non-abelian corrections
  from ``flux quantization'' (see \cite{FiorenzaSatiSchreiberII} for a review of these and for discussion
  in the present context of higher moduli stacks), 
  we discussed in \cite{FiorenzaSatiSchreiberI} 
  the appropriate non-abelian refinement of this 
  7d Chern-Simons term, which contains also cup product terms of the form
  $\hat {\mathbf{a}}_1\widehat {\mathbf{\cup}} \hat {\mathbf{a}}_2$ as well we the term 
  $\tfrac{1}{6}\widehat{\mathbf{p}}_2$ from section \ref{string}.

\item {\it Five-dimensional and eleven-dimensional supergravity:}
The topological part of the five-dimensional supergravity action is 
$\exp 2\pi i\int_{Y^5} A \wedge F_A \wedge F_A\;,
\)
where $A$ is a $U(1)$-connection. 
Writing the action as
$\exp 2\pi i\int_{Y^5} \mathrm{CS}_1(A) \cup d\mathrm{CS}_1(A) \cup d\mathrm{CS}_1(A)$, one sees this is  
 a 3-fold Chern-Simons theory. 
Next, in eleven dimensions, 
the C-field $C_3$ with can be viewed as a 3-connection on 
a 2-gerbe with 4-curvature $G_4$.  By identifying the C-field with 
the Chern-Simons 3-form $\mathrm{CS}_3(A)$ of a $U(1)$-3-connection $A$, 
the topological action $\exp 2\pi i
\int_{Y^{11}} C_3 \wedge G_4 \wedge G_4$, is seen to be of the form
 $\exp 2\pi i
 \int_{Y^{11}} \mathrm{CS}_3(A) \cup d\mathrm{CS}_3(A) \cup d\mathrm{CS}_3(A)$. This realizes the 11d supergravity C-field action as the action for a 3-tier cup-product abelian Chern-Simons theory induced by a morphism of 3-stacks \cite{FiorenzaSatiSchreiberII}.

\end{itemize}

\subsection{Super-Chern-Simons theories}\label{super-SC}

The (higher) topos $\mathbf{H}$ of (higher) stacks on the smooth site of manifolds which we have been 
considering for most of this paper has an important property common to various similar toposes such as
that on supermanifolds: it  satisfies a small set of axioms called (differential) 
\emph{cohesion}, see \cite{survey}. Moreover, essentially every
construction described in the above sections  makes sense in an arbitrary cohesive topos. 
For constructions like homotopy pullbacks, mapping spaces, adjoint actions etc., this is true 
for every topos, while the differential cohesion in addition guarantees the existence of 
differential geometric structures such as de Rham coefficients, connections, differential cohomology, etc.
This setting allows to transport all considerations based on the cohesion axioms across
various kinds of geometries. Notably, one can speak of higher \emph{supergeometry}, and 
hence of fermionic quantum fields,
simply by declaring the site of definition to be that of supermanifolds: 
indeed, the higher topos of (higher) stacks on supermanifolds is differentially cohesive
(\cite{survey}, section 4.6). 
This leads to a natural notion of \emph{super-Chern-Simons theories}. 

In order to introduce these notions, we need a digression on higher complex line bundles.
Namely, we have been using the $n$-stacks $\mathbf{B}^n U(1)$, but without any substantial
change in the theory we could also use the $n$-stacks $\mathbf{B}^n \mathbb{C}^\times$
with the multiplicative group $U(1)$ of norm 1 complex numbers replaced by the full mutliplicative group of non-zero complex numbers. Since we have a fiber sequence
\[
\mathbb{R}_{>0}\to \mathbb{C}^\times \to U(1)
\]
with topologically contractible fiber, under geometric realization ${\vert -\vert}$
the canonical map $\mathbf{B}^n U(1) \to \mathbf{B}^n \mathbb{C}^\times$ becomes an equivalence. 
Nevertheless, some constructions are more naturally expressed in terms of $U(1)$-principal
$n$-bundles, while others are more naturally expressed in terms of $\mathbb{C}^\times$-principal 
$n$-bundles, and so it is useful to be able to switch from one description to the other.  For $n=1$ this is the familiar fact that the classifying space of principal $U(1)$-bundles is homotopy equivalent to the classifying space of complex line bundles. For $n=2$ we still have a noteworthy (higher) linear algebra interpretation: $\mathbf{B}^2 \mathbb{C}^\times$ is naturally identified with the 2-stack 
$2 \mathbf{Line}_{\mathbb{C}}$ of \emph{complex line 2-bundles}.
Namely, for $R$ a commutative ring (or more generally an $E_\infty$-ring), one considers the 2-category of $R$-algebras, bimodules and 
bimodule homomorphisms (e.g. \cite{De}). 
We may think of this as the 2-category of \emph{2-vector spaces} over $R$
(appendix A of \cite{Schreiber08}, section 4.4. of \cite{SchreiberWaldorf}, section 7 of \cite{FHLT}).
Notice that this 2-category is naturally braided monoidal. 
We then write 
$$
  \xymatrix{
    2 \mathrm{Line}_R~
    \ar@{^{(}->}[r]
    &
    2 \mathrm{Vect}_R	
  }
$$
for the full sub-2-groupoid on those objects which are invertible under this tensor product:
the \emph{2-lines} over $R$. This is the \emph{Picard 2-groupoid}
over $R$, and with the inherited monoidal structure it is a 3-group, 
the \emph{Brauer 3-group} of $R$. Its homotopy groups have
a familiar algebraic interpretation:
\begin{itemize}
  \item $\pi_0(2 \mathrm{Line}_R)$ is the \emph{Brauer group} of $R$;
  \item $\pi_1(2 \mathrm{Line}_R)$ is the ordinary \emph{Picard group} of $R$ (of ordinary $R$-lines);
  \item $\pi_2(2\mathrm{Line}_R) \simeq R^\times $ is the \emph{group of units}.
\end{itemize}
(This is the generalization to $n=2$ of the  familiar Picard 1-groupoid $1 \mathrm{Line}_R$ of invertible $R$-modules.)
Since the construction is natural in $R$ and naturality respects 2-lines, by taking $R$ to be a sheaf of $k$-algebras, with $k$ a fixed field, one defines the 2-stacks $2\mathbf{Vect}_{k}$ of
$k$-\emph{2-vector bundles} and $2\mathbf{Line}_{k}$ of 2-line bundles over $k$. If $k$ is algebraically closed, then there is, up to equivalence, only a single 2-line and only a single
invertible bimodule, hence 
$2 \mathbf{Line}_k \simeq \mathbf{B}^2 k^\times$.
In particular, we have that 
$$
  2\mathbf{Line}_{\mathbb{C}} \simeq \mathbf{B}^2 \mathbb{C}^\times
  \,.
$$

The background $B$-field of the bosonic string has a natural interpretation as a section of the differential refinement $\mathbf{B}^2 \mathbb{C}^\times_{\mathrm{conn}}$ of the 2-stack $\mathbf{B}^2 \mathbb{C}^\times$. Hence, by the above discussion, it is identified with a   
2-connection on a complex 2-line bundle. However, a careful analysis,
due to \cite{DFM} and made more explicit in \cite{FreedLecture}, shows that
for the superstring the background $B$-field is more refined. Expressed in the language of 
higher stacks the statement is that the superstring $B$-field is a 
connection on a complex \emph{super}-2-line bundle. This means that 
one has to move from the (higher) topos of (higher) stacks on the site of smooth manifolds to 
that of stacks on the site of smooth supermanifolds (section 4.6 of \cite{survey}). The 2-stack of complex 2-line bundles is then replaced by the 2-stack 
$2\mathbf{sLine}_{\mathbb{C}}$ of super-2-line bundles, whose global points are complex Azumaya superalgebras. Of these there are, up to equivalence, not just one but two: the canonical super 2-line and its 
``superpartner'' \cite{Wall}. Moreover, there are now, up to equivalence, two different invertible 2-linear
maps from each of these super-lines to itself. In summary, the homotopy sheaves of 
the super 2-stack of super line 2-bundles are
\begin{itemize}
  \item $ \pi_0(2\mathbf{sLine}_{\mathbb{C}}) \simeq \mathbb{Z}_2$,
  \item $\pi_1(2\mathbf{sLine}_{\mathbb{C}}) \simeq \mathbb{Z}_2$,
  \item $\pi_2(2\mathbf{sLine}_{\mathbb{C}}) \simeq \mathbb{C}^\times$.
\end{itemize}
Since the homotopy groups of the group $\mathbb{C}^{\times}$ are 
$\pi_0(\mathbb{C}^\times)=0$
and $\pi_1(\mathbb{C}^\times)=\mathbb{Z}$,
it follows that the geometric realization of this 2-stack has homotopy groups
\begin{itemize}
  \item $ \pi_0(\vert2\mathbf{sLine}_{\mathbb{C}}\vert) \simeq \mathbb{Z}_2$,
  \item $\pi_1(\vert2\mathbf{sLine}_{\mathbb{C}}\vert) \simeq \mathbb{Z}_2$,
  \item $\pi_2(\vert2\mathbf{sLine}_{\mathbb{C}}\vert) \simeq 0$,
  \item $\pi_3(\vert2\mathbf{sLine}_{\mathbb{C}}\vert) \simeq \mathbb{Z}$.
\end{itemize}
These are precisely the correct coefficients for the twists of complex K-theory \cite{DK}, witnessing the
fact that the $B$-field background of the superstring twists the Chan-Paton bundles on the 
D-branes \cite{DFM,  FreedLecture}.

The braided monoidal structure of the 2-category of complex super 2-vector spaces induces on 
$2\mathbf{sLine}_{\mathbb{C}}$ the structure of a \emph{braided 3-group}. Therefore, 
one has a naturally defined 3-stack $\mathbf{B}(2\mathbf{sLine}_{\mathbb{C}})_{\mathrm{conn}}$ which is the 
supergeometric refinement of the coefficient object $\mathbf{B}^3 \mathbb{C}^{\times}_{\mathrm{conn}}$
for the extended Lagrangian of bosonic 3-dimensional Chern-Simons theory.  
Therefore, for $G$ a super-Lie group a super-Chern-Simons theory, inducing a
super-WZW action functional on $G$, is naturally given by an extended Lagrangian which is a map
of higher moduli stacks of the form
$$
  \mathbf{L}
    : 
    \mathbf{B}G_{\mathrm{conn}}
   \to
    \mathbf{B}(2\mathbf{sLine}_{\mathbb{C}})_{\mathrm{conn}} 
  \,.
$$
Notice that, by the canonical inclusion 
$\mathbf{B}^3 \mathbb{C}^\times_{\mathrm{conn}}
\to \mathbf{B}(2\mathbf{sLine}_{\mathbb{C}})_{\mathrm{conn}}$, every bosonic 
extended Lagrangian of 3d Chern-Simons type induces such a supergeometric theory
with trivial super-grading part. 


\section{Outlook: Higher prequantum theory}

The discussion in sections \ref{toy-example} and \ref{3d}
of low dimensional Chern-Simons theories and the survey on higher
dimensional Chern-Simons theories in section \ref{CupProductCS}, formulated and extended 
in terms of higher stacks, is a first indication of a fairly comprehensive theory
of higher and extended prequantum gauge field theory that is naturally incarnated in a 
suitable context of higher stacks. In this last section we give a brief glimpse of some 
further aspects. Additional, more comprehensive expositions and further pointers
are collected for instance in \cite{survey,TwistedStructuresLecture}.

\subsection{$\sigma$-models} 
\label{Sigma-Models}

The Chern-Simons theories presented in the previous sections are manifestly special examples of the following general construction: one has a universal (higher) stack $\mathbf{Fields}$ of field configurations for a certain field theory, equipped with an \emph{extended} Lagrangian, namely 
with a map of higher stacks
$$
  \mathbf{L} \;:\; \mathbf{Fields} \to \mathbf{B}^n U(1)_{\mathrm{conn}}
$$
to the $n$-stack of $U(1)$-principal $n$-bundles with connections.
The Lagrangian $\mathbf{L}$ induces  Lagrangian data in arbitrary codimension: for every 
closed oriented worldvolume $\Sigma_k$ of dimension $k \leq n$ there is 
a \emph{transgressed} Lagrangian
$$
 \xymatrix{
    \mathbf{Maps}(\Sigma_k;\mathbf{Fields})
	\ar[rr]^-{\hspace{-1mm} \mathbf{Maps}(\Sigma_k;\mathbf{L})}
	&&
	 \mathbf{Maps}(\Sigma_k; \mathbf{B}^n U(1)_{\mathrm{conn}})	
	 \ar[r]^-{\hspace{-1.5mm}\mathrm{hol}_{\Sigma_k}}
	&
	\mathbf{B}^{n-k} U(1)_{\mathrm{conn}}
  }
$$
defining the (off-shell) prequantum $U(1)$-$(n-k)$-bundle of the given field theory. In particular, the curvature forms of these bundles induce the canonical pre-$(n-k)$-plectic structure on the moduli stack of field configurations on $\Sigma_k$.
\par
In codimension $0$, i.e., for $k=n$ one has the morphism of stacks
\[
\exp (2\pi i \int_{\Sigma_n} -~): \mathbf{Maps}(\Sigma_n;\mathbf{Fields})\to \underline{U}(1)
\]
and so taking global sections over the point and passing to equivalence classes one finds the \emph{action functional}
\[
\exp (2\pi i \int_{\Sigma_n}-~): \{\text{Field configurations}\}/\text{equiv}\to U(1)\;.
\]
Notice how the stacky origin of the action functional automatically implies that its value only depends on the gauge equivalence class of a given field configuration. 
Moreover, the action functional of an extended Lagrangian field theory as above is manifestly a $\sigma$-model action functional: the target ``space'' is the 
universal moduli stack of field configurations itself. Furthermore, the composition
\[
\omega : 
\mathbf{Fields} \xrightarrow{ \mathbf{L}} \mathbf{B}^n U(1)_{\mathrm{conn}}\xrightarrow{F_{(-)}}\Omega^{n+1}(-;\mathbb{R})_{\mathrm{cl}}
\]
shows that 
the stack of field configurations is naturally equipped with a 
pre-$n$-plectic structure \cite{Rogers}, which means that 
actions of extended Lagrangian field theories in the above sense are examples of $\sigma$-models with (pre)-$n$-plectic targets. 
For \emph{binary} dependence of the $n$-plectic form on the fields this includes the AKSZ $\sigma$-models
\cite{AKSZ,CattaneoFelder, CattaneoFelder01, Ikeda93, Ikeda02, RoytenbergAKSZ, ScSt94, ScSt94b}. 
Namely, the target space of the AKSZ $\sigma$-model is a symplectic dg-manifold, and this can be equivalently seen as an $L_\infty$-algebroid $\mathfrak{P}$ endowed with a quadratic and non-degenerate 
invariant polynomial. Moreover, this symplectic dg-manifold is equipped with a canonical Hamiltonian, which can be seen as a cocycle on the $L_\infty$-algebroid $\mathfrak{P}$, and with a Lagrangian density, which can be seen as a Chern-Simons element transgressing the Hamiltonian cocycle to the invariant polynomial on $\mathfrak{P}$. As shown in \cite{KoSt07, KoSt10}, a field configuration in the $n$-dimensional AKSZ $\sigma$-model can be identified with a $\mathfrak{P}$-connection on a trivial dg-bundle on the worldsheet $\Sigma_{n}$, and the Chern-Simons action functional for such a connection is then seen to be the AKSZ action:
\[
\{\text{trivial bundles with $\mathfrak{P}$-connections on $\Sigma_{n}$}\}\xrightarrow{\int_{\Sigma_{n}}L_{AKSZ}} \mathbb{R}.
\]
Notice how we are in a situation completely similar to that of our toy example in section \ref{TheBasicDefinition}. It is therefore clear how to globalize the constructions from \cite{KoSt07, KoSt10} to the case of nontrivial bundles: the space of fields will be the moduli stack $\mathbf{P}_{\mathrm{conn}}$ of principal $\mathfrak{P}$-connections, and the Chern-Simons element for $\mathfrak{P}$ 
will be promoted to a morphism of higher stacks
\[
\mathbf{P}_{\mathrm{conn}}\xrightarrow{\mathbf{L}_{AKSZ}}\mathbf{B}^nU(1)_{\mathrm{conn}},
\]
as shown in \cite{CW-AKSZ}. In codimension 0, one finds the exponentiated AKSZ action functional
\[
\exp (2\pi i \int_{\Sigma_n} \mathbf{L}_{AKSZ}):\mathbf{Maps}(\Sigma_n,\mathbf{P}_{\mathrm{conn}})\to 
\underline{U}(1)\]
promoted to a morphism of stacks; in codimension 1, by composing with the curvature morphism, one finds the pre-symplecic structure
\[
\mathbf{Maps}(\Sigma_{n-1},\mathbf{P}_{\mathrm{conn}})\to 
\mathbf{B}^1U(1)_{\mathrm{conn}}\to \Omega^2_{\mathrm{cl}}
\]
on the (extended) phase space of the AKSZ $\sigma$-model, see, e.g., \cite{Cattaneo-Mnev-ReshetikhinI,Cattaneo-Mnev-ReshetikhinII}.

For instance, from this perspective, the action functional of classical 3d Chern-Simons theory is the $\sigma$-model action functional with target the stack $\mathbf{B}G_{\mathrm{conn}}$
equipped with the pre-3-plectic 
form $\langle -,-\rangle : \mathbf{B}G_{\mathrm{conn}} \to \Omega^4_{\mathrm{cl}}$
(the Killing form invariant polynomial)
as discussed in section \ref{3d}. 
If we consider binary invariant polynomials in \emph{derived} geometry, hence on objects with components
also in negative degree, then also closed bosonic string field theory as in \cite{Z}
is an example (see 5.7.10 of \cite{survey}) as are constructions such as \cite{Co}.
Examples of $n$-plectic structures of higher arity 
on moduli stacks of higher gauge fields are in 
\cite{FiorenzaSatiSchreiberI, FiorenzaSatiSchreiberIII}.

More generally, we have transgression of the extended Lagrangian over manifolds $\Sigma_k$
with boundary $\partial \Sigma_k$. Again by inspection of the constructions in 
\cite{gomi-terashima1} in terms of Deligne complexes, one finds that 
under the Dold-Kan correspondence these induce the corresponding constructions on higher
moduli stacks: the \emph{higher parallel transport} of $\mathbf{L}$ over $\Sigma_k$
yields a \emph{section} of the $(n-k+1)$-bundle which is modulated over the boundary by
$\mathbf{Maps}(\partial \Sigma_k, \mathbf{B}G_{\mathrm{conn}}) \to \mathbf{B}^{n-k+1}U(1)_{\mathrm{conn}}$.
This is the incarnation at the prequantum level of the \emph{propagator} of the full
extended TQFT in the sense of \cite{LurieTQFT} of the propagator over $\Sigma_k$, as 
indicated in \cite{Lipsky}.
Further discussion of this full prequantum field theory obtained this way is well beyond the scope of
the present article. However, below in section \ref{PrequantumInHigherCodimension} we indicate how familiar 
\emph{anomaly cancellation} constructions in open string theory naturally arise as examples
of such transgression of extended Lagrangians over worldvolumes with boundary.

\subsection{Fields in slices: twisted differential structures}
\label{FieldsInSlices}

Our discussion of $\sigma$-model-type actions in the previous section 
might seem to suggest that all the fields that
one encounters in field theory have moduli that form (higher) stacks on the site of 
smooth manifolds. However, this is actually not the case
and one need not look too far in order to find a counterexample: 
the field of gravity in general relativity is a (pseudo-)Riemannian metric on spacetime, 
and there is no such thing as a stack of (pseudo-)Riemannian metrics on the smooth site. 
This is nothing but the elementary fact that a 
(pseudo-)Riemannian metric cannot be pulled back along an arbitrary smooth morphism between manifolds, 
but only along local diffeomorphisms. Translated into the language of stacks, this tells us that (pseudo-)Riemannian metrics is a stack on the \'etale site of smooth manifolds, but not on the smooth site.\footnote{See \cite{carchedi} for a comprehensive treatment of the \'etale site of smooth manifolds and of the 
higher topos of higher stacks over it.}
 Yet we can still look at (pseudo-)Riemannian metrics on a smooth $n$-dimensional manifold $X$ from the perspective of the topos $\mathbf{H}$ of stacks over the smooth site, and indeed this is
the more comprehensive point of view. Namely, 
working in $\mathbf{H}$ also means to work with all its \emph{slice toposes} 
(or \emph{over-toposes}) $\mathbf{H}/_{\mathbf{S}}$ over the various objects $\mathbf{S}$ in $\mathbf{H}$. For the field of 
gravity this means working in the slice $\mathbf{H}/_{\mathbf{B}GL(n;\mathbb{R})}$ over the stack $\mathbf{B}GL(n;\mathbb{R})$.\footnote{More detailed discussion 
of of how (quantum) fields generally are maps in slices of cohesive toposes has been given in the 
lecture notes \cite{TwistedStructuresLecture} and in sections 1.2.16, 5.4 of \cite{survey}.} 
\par
Once again, this seemingly frightening terminology is just a concise and rigorous way of expressing a familiar fact from Riemannian geometry: endowing a smooth $n$-manifold $X$ with a pseudo-Riemannian metric of signature $(p,n-p)$ is equivalent to performing a
 reduction of the structure group of the tangent bundle of $X$ to $O(p,n-p)$. Indeed, one can look at the tangent bundle (or, more precisely, at the associated frame bundle) as a morphism $\tau_X: X \to \mathbf{B}GL(n;\mathbb{R})$. 
 
 \paragraph{Example: Orthogonal structures.}
 The above reduction is then the datum of a homotopy lift of $\tau_X$
 \[
  \xymatrix{
    & \mathbf{B}O(p,n-p) \ar[d]
    \\
    X \ar[r]_-{\tau_X}^{\ }="t" \ar[ur]^{o_X}_{\ }="s" & \mathbf{B} \mathrm{GL}(n;\mathbb{R})\;,
	\ar@{=>}^{e} "s"; "t"
  }
\]
where the vertical arrow 
$$
  \mathbf{OrthStruc}_n
  :
  \xymatrix{
    \mathbf{B}O(p,n-p)
	\ar[r]
	& 
	\mathbf{B} \mathrm{GL}(n;\mathbb{R})
  }
$$
is induced by the inclusion of groups 
$O(p,n-p)\hookrightarrow GL(n;\mathbb{R})$. 
Such a commutative diagram is precisely a map 
$$
  (o_X, e)
   : 
  \xymatrix{
    \tau_X 
	\ar[r]
	&
	\mathbf{OrthStruc}_n
  }
$$
in the slice $\mathbf{H}/_{\mathbf{B}GL(n;\mathbb{R})}$. The homotopy $e$ appearing in the 
above diagram is
precisely the \emph{vielbein field} (frame field) which exhibits the reduction, hence 
which induces the Riemannian metric. So the moduli stack of Riemannian metrics
in $n$ dimensions is $\mathbf{OrthStruc}_n$, not as an object of the ambient cohesive topos
$\mathbf{H}$, but of the slice $\mathbf{H}_{/\mathbf{B}\mathrm{GL}(n)}$. Indeed, a 
map between manifolds regarded in this slice, namely a map 
$(\phi, \eta) : \tau_Y  \to \tau_X$, is equivalently a smooth map 
$\phi : Y \to X$ in $\mathbf{H}$, but equipped with an equivalence
$\eta : \phi^* \tau_X \to \tau_Y$. This precisely exhibits $\phi$ as a local
diffeomorphism. In this way the slicing formalism automatically knows along which kinds of maps 
metrics may be pulled back.

\paragraph{Example: (Exceptional) generalized geometry.} If we replace in the above
 example the map 
$\mathbf{OrthStruc}_n$ with inclusions of other maximal
compact subgroups, we similarly obtain the moduli stacks for \emph{generalized} geometry
(metric and B-field)
as appearing in type II superstring backgrounds (see, e.g., \cite{Hi}), given by
$$
  \mathbf{typeII} 
    : 
  \xymatrix{ 
     \mathbf{B}(O(n)\times O(n)) \ar[r] & \mathbf{B}O(n,n)
  }
  \;\;\in \mathbf{H}_{/\mathbf{B}O(n,n)}
$$
and of \emph{exceptional generalized geometry} appearing in compactifications of 11-dimensional 
supergravity \cite{Hu}, given by
$$
  \mathbf{ExcSugra}_n : \xymatrix{
    \mathbf{B}K_{n}
	\ar[r]
	&
	\mathbf{B}E_{n(n)}
  }
  \;\;
  \in 
  \mathbf{H}_{/\mathbf{B}E_{n(n)}},
$$
where $E_{n(n)}$ is the maximally non-compact real
form of the Lie group of rank $n$
with $E$-type Dynkin diagram, and $K_n\subseteq E_{n(n)}$ is a maximal compact subgroup.
For instance, a manifold $X$ in type II-geometry is represented by 
$\tau_X^{\mathrm{gen}} : X \to \mathbf{B}O(n,n)$ 
in the slice $\mathbf{H}_{/\mathbf{B}O(n,n)}$,
which is the map modulating what is called the \emph{generalized tangent bundle}, and a field
of generalized type II gravity is a map 
$(o_X^{\mathrm{gen}}, e) : \tau_X^{\mathrm{gen}} \to \mathbf{typeII}$
to the moduli stack in the slice. One checks that the homotopy $e$ is now precisely
what is called the \emph{generalized vielbein field} in type II geometry. 
We read off the kind of maps along which such fields may be pulled back: a map 
$(\phi,\eta) : \tau_Y^{\mathrm{gen}}\to \tau_X^{\mathrm{gen}}$
is a \emph{generalized} local diffeomorphism: a smooth map 
$\phi : Y \to X$
equipped with an equivalence of generalized tangent bundles 
$\eta : \phi^* \tau_X^{\mathrm{gen}} \to \tau_Y^{\mathrm{gen}}$.
A directly analogous discussion applies to the exceptional generalized geometry.

Furthermore,
 various topological structures are generalized fields in this sense, and become
fields in the more traditional sense after differential refinement. 

\paragraph{Example: Spin structures.}
The map
$\mathbf{SpinStruc} :  \mathbf{B}\mathrm{Spin} \to \mathbf{B}\mathrm{GL}$
is, when regarded as an object of $\mathbf{H}_{/\mathbf{B}\mathrm{GL}}$, the moduli 
stack of spin structures. Its differential refinement
$\mathbf{SpinStruc}_{\mathrm{conn}} :  \mathbf{B}\mathrm{Spin}_{\mathrm{conn}} \to 
 \mathbf{B}\mathrm{GL}_{\mathrm{conn}}$ is such that a domain object 
$\tau_X^{\nabla} \in \mathbf{H}_{/\mathbf{GL}_{\mathrm{conn}}}$ is given by an affine connection,
and a map $(\nabla_{\mathrm{Spin}} ,e) : \tau_X^{\nabla} \to \mathbf{SpinStruc}_{\mathrm{conn}}$
is precisely a \emph{Spin connection} and a Lorentz frame/vielbein which identifies $\nabla$
with the corresponding Levi-Civita connection.  

\par 
This example is the first in a whole tower of \emph{higher Spin structure} fields
\cite{SSSI, SSSII, SSSIII}, each of 
which is directly related to a corresponding higher Chern-Simons theory. 
The next higher example in this tower is the following.

\paragraph{Example: Heterotic fields.}
For $n\geq 3$, let $\mathbf{Heterotic}$ be the map
$$
  \mathbf{Heterotic} 
    : 
  \xymatrix{
    \mathbf{B}\mathrm{Spin}(n) \ar[rr]^-{(p, \tfrac{1}{2}\mathbf{p}_1{})}
    &&
    \mathbf{B}\mathrm{GL}(n;\mathbb{R}) \times \mathbf{B}^3 U(1)
  }
$$
regarded as an object in the slice $\mathbf{H}_{/\mathbf{B}\mathrm{GL}(n;\mathbb{R}) \times \mathbf{B}^3 U(1)}$. Here $p$ is the morphism induced by
\[
\mathrm{Spin}(n)\to O(n)\hookrightarrow GL(n;\mathbb{R})
\]
while $\frac{1}{2}\mathbf{p}_1:\mathbf{B}\mathrm{Spin}(n)\to \mathbf{B}^3 U(1)$ is the morphism of stacks underlying the first fractional Pontrjagin class which we met in section \ref{string}.
To regard a smooth manifold $X$ as an object in the  slice $\mathbf{H}_{/\mathbf{B}\mathrm{GL}(n;\mathbb{R}) \times \mathbf{B}^3 U(1)}$ means to equip it with a 
$U(1)$-3-bundle  $\mathbf{a}_X : X \to \mathbf{B}^3 U(1)$ in addition to the tangent bundle $\tau_X: X \to \mathbf{B}GL(n;\mathbb{R})$.  A Green-Schwarz anomaly-free background field configuration in heterotic string theory 
is (the differential refinement of) 
a map $(s_X,\phi) : (\tau_X, \mathbf{a}_X) \to \mathbf{Heterotic}$, i.e., a homotopy commutative diagram
$$
  \xymatrix{
    X 
	\ar[dr]_{\hspace{-4mm}(\tau_X, \mathbf{a}_X)}^{\ }="t"
	\ar[rr]^{s_X}_{\ }="s" 
	 && 
	\mathbf{B}\mathrm{Spin}
	\ar[dl]^{\mathbf{\phantom{mm}Heterotic}}
	\\
	&
	\mathbf{B}\mathrm{GL}(n)
	\times
	\mathbf{B}^3 U(1)\;.
	\ar@{=>}^{\phi} "s"; "t"
  }
$$
The 3-bundle  $\mathbf{a}_X$ serves as a twist: when $\mathbf{a}_X$ is trivial then we are in presence 
of a String structure on $X$; so it is customary to refer to $(s_X,\phi)$ as to an 
$\mathbf{a}_X$-\emph{twisted String structure} on $X$, in the sense of \cite{Wang, SSSIII}. The Green-Schwarz
anomaly cancellation condition is then imposed by requiring that  $\mathbf{a}_X$ 
(or rather its differential refinement) factors as
\[
  \xymatrix{
    X \longrightarrow
	\mathbf{B}SU
	\ar[r]^-{\mathbf{c}_2} 
	&
	\mathbf{B}^3 U(1)
  }\;,
\]
where $\mathbf{c}_2(E)$ is the morphism of stacks underlying the second Chern class. 
Notice that this says that the extended Lagrangians of $\mathrm{Spin}$- and $\mathrm{SU}$-Chern-Simons
theory in 3-dimensions, as discussed above, at the same time serve as the twists that control the
higher background gauge field structure in heterotic supergravity backgrounds.

\paragraph{Example: Dual heterotic fields.} 
Similarly, the morphism
$$
  \mathbf{DualHeterotic}
  :
  \xymatrix{
    \mathbf{B}\mathrm{String}(n)
    \ar[rr]^{(p, \tfrac{1}{6}\mathbf{p}_2)\phantom{mmm}}
    &&
	\mathbf{B}\mathrm{GL}(n;\mathbb{R}) \times \mathbf{B}^7 U(1)
  }
$$
governs field configurations for the dual heterotic string.
These examples, in their differentially refined version, have been discussed in \cite{SSSIII}.
The last example above is governed by the extended Lagrangian of the 7-dimensional Chern-Simons-type
higher gauge field theory of $\mathrm{String}$-2-connections. This has been discussed in 
\cite{FiorenzaSatiSchreiberI}.

There are many more examples of (quantum) fields modulated by objects in slices of a cohesive higher topos.
To close this brief discussion, notice that the previous example has an evident analog
in one lower degree: a central extension of Lie groups $A \to \hat G \to G$
induces a long fiber sequence 
$$
  \xymatrix{\underline{A} \longrightarrow \hat{\underline{G}} \longrightarrow \underline{G} \longrightarrow \mathbf{B}A \longrightarrow \mathbf{B}\hat G 
  \longrightarrow 
  \mathbf{B} G \ar[r]^-{\mathbf{c}} & \mathbf{B}^2 A}
$$
in $\mathbf{H}$, where $\mathbf{c}$ is the group 2-cocycle that classifies the extension. If 
we regard this as a coefficient object in the slice $\mathbf{H}_{/\mathbf{B}^2 A}$, then 
regarding a manifold $X$ in this slice means to equip it with an $(\mathbf{B}A)$-principal 2-bundle 
(an $A$-bundle gerbe) modulated by a map $\tau_X^{A} : {X \to \mathbf{B}^2A}$; 
and a field
$(\phi, \eta) : { \tau_X^A \to \mathbf{c} }$
is equivalently a $G$-principal bundle $P \to  X$ equipped with an equivalence 
$\eta : \mathbf{c}(E) \simeq \tau_X^A$ with the 2-bundle which obstructs
its lift to a $\hat G$-principal bundle (the ``lifting gerbe''). 
The differential refinement of this setup similarly yields $G$-gauge fields equipped with such 
an equivalence. A concrete example for this is discussed below in section \ref{PrequantumInHigherCodimension}.


This special case of fields in a slice is called a 
\emph{twisted (differential) $\hat G$-structure} in \cite{SSSIII}
and a \emph{relative field} in \cite{FT}. In more generality, 
the terminology \emph{twisted (differential) $\mathbf{c}$-structures} 
is used in \cite{SSSIII} 
to denote spaces of fields of the form $\mathbf{H}/_{\mathbf{S}}(\sigma_X,\mathbf{c})$ for some slice topos 
$\mathbf{H}/_{\mathbf{S}}$ and some coefficient object (or ``twisting object'') $\mathbf{c}$; see also  the  exposition in \cite{TwistedStructuresLecture}. In fact in full generality (quantum) fields in slice toposes
are equivalent to cocycles in (generalized and parameterized and possibly non-abelian and differential) 
\emph{twisted cohomology}. The constructions on which the above discussion is 
built is given in some generality in \cite{NSSa}.

\par
In many examples of twisted (differential) structures/fields in slices the twist 
is constrained to have a certain factorization. For instance the twist of the (differential) String-structure
in a heterotic background is constrained to be the (differential) second Chern-class of a 
(differential) $E_8 \times E_8$-cocycle, as mentioned in section \ref{string}; or
for instance  the gauging of the 1d Chern-Simons fields
on a knot in a 3d Chern-Simons theory bulk is constrained to be the restriction of the bulk gauge field,
as discussed in section \ref{WithWilsonLoops}. Another example is the twist of the Chan-Paton bundles
on D-branes, discussed below in section \ref{PrequantumInHigherCodimension}, which is constrained to be the
restriction of the ambient Kalb-Ramond field to the D-brane.
In all these cases the fields may be thought of as being maps in the slice topos
that arise 
from maps in the \emph{arrow topos} $\mathbf{H}^{\Delta^{1}}$.
A moduli stack here is a map of moduli stacks 
$$
  \mathbf{Fields}_{\mathrm{bulk}+\mathrm{def}} 
  : 
  \xymatrix{
    \mathbf{Fields}_{\mathrm{def}} \ar[r] & \mathbf{Fields}_{\mathrm{bulk}} 
  }
$$
in $\mathbf{H}$; and a domain on which such fields may be defined is an object 
$\Sigma_{\mathrm{bulk}} \in \mathbf{H}$ equipped  with a map (often, but not necessarily, an inclusion) 
${\Sigma_{\mathrm{def}} \to \Sigma_{\mathrm{bulk}}} $, and a field configuration is
a square of the form
$$
  \raisebox{20pt}{
  \xymatrix{
    \Sigma_{\mathrm{def}} \ar[rr]^-{\phi_{\mathrm{def}}}_{\ }="s"
	\ar[d]^>{\ }="t"
	&& \mathbf{Fields}_{\mathrm{def}}
	\ar[d]^{\mathbf{Fields}}
	\\
	\Sigma_{\mathrm{bulk}} \ar[rr]_-{\phi_{\mathrm{bulk}}} 
	&& \mathbf{Fields}_{\mathrm{bulk}}
	\ar@{=>}^\simeq "s"; "t"
  }
  }
$$
in $\mathbf{H}$. If we now fix $\phi_{\mathrm{bulk}}$ then $(\phi_{\mathrm{bulk}})|_{\Sigma_{\mathrm{def}}}$
serves as the twist, in the above sense, for $\phi_{\mathrm{def}}$.
If  $\mathbf{Fields}_{\mathrm{def}}$ is trivial (the point/terminal object), 
then such a field is a cocycle in \emph{relative cohomology}: a cocycle $\phi_{\mathrm{bulk}}$ on 
$\Sigma_{\mathrm{bulk}}$ equipped with a trivialization $(\phi_{\mathrm{bulk}})|_{\Sigma_{\mathrm{def}}}$
of its restriction to $\Sigma_{\mathrm{def}}$.

The fields in Chern-Simons theory with Wilson loops displayed in section \ref{FieldsForCSWithWilsonLoop}
clearly constitute an example of this phenomenon. Another example is the field content of type II string theory on 
a 10-dimensional spacetime $X$ with D-brane $Q \hookrightarrow X$,  
for which the above diagram reads
$$
  \raisebox{20pt}{
  \xymatrix{
    Q \ar[rr]_{\ }="s" \ar@{^{(}->}[d]^>{\ }="t" 
    && \mathbf{B}\mathrm{PU}_{\mathrm{conn}}
	\ar[d]^{\mathbf{dd}_{\mathrm{conn}}}
	\\
	X
	\ar[rr]^-B
	&&
	\mathbf{B}^2 U(1)_{\mathrm{conn}}\;,
	\ar@{=>} "s"; "t"
  }}
$$
discussed further below in section \ref{PrequantumInHigherCodimension}. In
\cite{FiorenzaSatiSchreiberII} we discussed how the supergravity C-field over
an 11-dimensional Ho{\v r}ava-Witten background with 10-dimensional boundary $X \hookrightarrow Y$
is similarly a relative cocyle, with the coefficients controled, once more, by the
extended Chern-Simons Lagrangian 
$$
  \hat {\mathbf{c}} : \xymatrix{ \mathbf{B}(E_8 \times E_8)_{\mathrm{conn}} \ar[r] & \mathbf{B}^3 U(1)_{\mathrm{conn}} }
  \,,
$$
now regarded in $\mathbf{H}^{(\Delta^1)}$.

\subsection{Differential moduli stacks}
\label{DifferentialModuli}

In the exposition in sections \ref{toy-example} and \ref{3d} above we referred, for
ease of discussion, to the mapping stacks of the form
$\mathbf{Maps}(\Sigma_k, \mathbf{B}G_{\mathrm{conn}})$ as moduli stacks of $G$-gauge fields on $\Sigma_k$.
From a more refined perspective this is not quite true. While certainly the global
points of these mapping stacks are equivalently the $G$-gauge field configurations on $\Sigma_k$, 
for $U$ a parameter space, the $U$-parameterized collections in the mapping stack are not
quite those of the intended moduli stack: for the former these are 
gauge fields and gauge transformations on $U \times \Sigma_k$, while for the latter these
are genuine cohesively $U$-parameterized collections of gauge fields on $\Sigma_k$.

In the exposition above we saw this difference briefly in 
section \ref{TheSymplecticStructureOnModuliSpaceOfConnections}, where we constrained a 1-form
$A \in \Omega^1(U \times \Sigma, \mathfrak{g})$ (a $U$-plot of the mapping stack) to vanish
on vector fields tangent to $U$; this makes it a smooth function on $U$ with values in 
connections on $\Sigma$. More precisely, for $G$ a Lie group and $\Sigma$ a smooth manifold, let 
$$
  G \mathbf{Conn}(\Sigma) \in \mathbf{H}
$$
be the stack which assigns to any $U \in \mathrm{CartSp}$ the groupoid of smoothly $U$-parameterized
collections of smooth $G$-principal connections on $\Sigma$, and of smoothly $U$-parameterized collections
of smooth gauge transformations between these connections. This is the actual moduli stack of $G$-connections.
In this form, but over a different site of definition, it appears for instance in geometric 
Langlands duality. In physics this stack is best known in the guise of its infinitesimal approximation:
the corresponding Lie algebroid is dually the (off-shell) \emph{BRST-complex} of the gauge theory, and
the BRST ghosts are the cotangents to the morphisms in $G\mathbf{Conn}(\Sigma)$ at the identity.

Notice that while the mapping stack is itself not quite the right answer, there is a canonical map
that comes to the rescue
$$
  \xymatrix{
    \mathbf{Maps}(\Sigma, \mathbf{B}G_{\mathrm{conn}})
	\ar[r]
	&
	G \mathbf{Conn}(\Sigma)
  }
  \,.
$$
We call this the \emph{concretification} map. We secretly already saw an example of this in 
section \ref{TheWZWGerbe}, where this was the map 
${\mathbf{Maps}(S^1, \mathbf{B}G_{\mathrm{conn}}) \longrightarrow \underline{G}/\!/_{\mathrm{Ad}} \underline{G}}$.

In more complicated examples, such as for higher groups $G$ and base spaces $\Sigma$ which are not
plain manifolds, it is in general less evident what $G\mathbf{Conn}(\Sigma)$ should be. But
if the ambient higher topos is cohesive, then there is a general abstract procedure that 
produces the differential moduli stack. This is discussed in sections 3.9.6.4 and 4.4.15.3
of \cite{survey} and in \cite{Nuiten}.

\subsection{Prequantum geometry in higher codimension}
\label{PrequantumInHigherCodimension}

We had indicated in section \ref{PrequantumLineBundlesOnModuliStacks} how a single
extended Lagrangian, given by a map of universal higher moduli stacks
$\mathbf{L}: {\mathbf{B}G_{\mathrm{conn}} \to \mathbf{B}^n U(1)_{\mathrm{conn}}}$,
induces, by transgression, circle $(n-k)$-bundles with connection
$$
  \mathrm{hol}_{\Sigma_k} \mathbf{Maps}(\Sigma_k, \mathbf{L})
  \;:\;
  \mathbf{Maps}(\Sigma_k, \mathbf{B}G_{\mathrm{conn}}) \longrightarrow \mathbf{B}^{n-k}U(1)_{\mathrm{conn}}
$$
on moduli stacks of field configurations over each closed $k$-manifold $\Sigma_k$.
In codimension 1, hence for $k = n-1$, this reproduces the ordinary \emph{prequantum circle bundle}
of the $n$-dimensional Chern-Simons type theory, as discussed in section 
\ref{TheSymplecticStructureOnModuliSpaceOfConnections}. 
The space of sections of the associated line bundle (restricted to the subspace of flat connections) is the space of 
\emph{prequantum states} of the theory. This becomes the space of genuine quantum states
after choosing a \emph{polarization} 
(i.e., a decomposition of the moduli space of fields into \emph{canonical coordinates}
and \emph{canonical momenta}) and restricting to polarized sections (i.e., those depending only on the 
canonical coordinates).
But moreover, for each $\Sigma_k$ we may regard $\mathrm{hol}_{\Sigma_k} \mathbf{Maps}(\Sigma_k, \mathbf{L})$ 
as a \emph{higher prequantum bundle} of the theory in higher codimension 
hence consider its prequantum geometry
in higher codimension. 

We discuss now some generalities of such a higher geometric prequantum theory and then 
show how this perspective sheds a useful light on the gauge coupling of the open string,
as part of the transgression of prequantum 2-states of Chern-Simons theory in codimension 2
to prequantum states in codimension 1.

\subsubsection{Higher prequantum states and prequantum operators}

We indicate here the basic concepts of higher extended prequantum theory and how they
reproduce traditional prequantum theory.\footnote{A discussion of this and the following
can be found in sections 3.9.13 and 4.4.19 of \cite{survey}; see also \cite{Prequantum1,Prequantum2}.}

Consider a (pre)-$n$-plectic form, given by 
a map
$$
  \omega : X \longrightarrow \Omega^{n+1}(-;\mathbb{R})_{\mathrm{cl}}
$$
in $\mathbf{H}$. A \emph{$n$-plectomorphism} of $(X,\omega)$ is an 
auto-equivalence of $\omega$ regarded as an object in the slice $\mathbf{H}_{/\Omega^{n+1}_{\mathrm{cl}}}$,
hence a diagram of the form
$$
  \raisebox{20pt}{
  \xymatrix{
    X \ar[dr]_{\omega} \ar[rr]^{\simeq} && X \ar[dl]^{\omega}
	\\
	& \Omega^{n+1}(-;\mathbb{R})_{\mathrm{cl}}\;.
  }
  }
$$ 
A \emph{prequantization} of $(X, \omega)$ is a choice of prequantum line bundle, hence a choice of 
lift $\nabla$ in 
$$
  \raisebox{42pt}{
  \xymatrix{
    & \mathbf{B}^n U(1)_{\mathrm{conn}}
	  \ar[d]^{F_{(- )}}
    \\
    X \ar[r]_{\omega} \ar[ur]^{\nabla} & \Omega^{n+1}_{\mathrm{cl}}
  }
  }\,,
$$
modulating a circle $n$-bundle with connection on $X$. We write
$\mathbf{c}(\nabla) : X \xrightarrow{\nabla}\mathbf{B}^n U(1)_{\mathrm{conn}} \to\mathbf{B}^n U(1)$
for the underlying principal $U(1)$-$n$-bundle.
An autoequivalence  
$$
  \hat O : \nabla \stackrel{\simeq}{\longrightarrow} \nabla
$$
of the prequantum $n$-bundle
regarded as an object in the slice $\mathbf{H}_{/\mathbf{B}^n U(1)_{\mathrm{conn}}}$, 
i.e., a diagram in $\mathbf{H}$ of the form
$$
  \xymatrix{
    X \ar[rr]^\simeq_{\ }="s" \ar[dr]_{\nabla}^{\ }="t" && X \ar[dl]^{\nabla}
	\\
	& \mathbf{B}^n U(1)_{\mathrm{conn}}
	\ar@{=>}^O "s"; "t"
  }
$$
is an (exponentiated) \emph{prequantum operator} or \emph{quantomorphism}  or \emph{regular contact transformation}
of the prequantum geometry $(X, \nabla)$. These form an $\infty$-group in $\mathbf{H}$. 
The $L_\infty$-algebra of this 
\emph{quantomorphism $\infty$-group} 
is the higher \emph{Poisson bracket} Lie algebra of the system. If $X$ is equipped with
group structure then the quantomorphisms covering the action of $X$ on itself form the 
\emph{Heisenberg $\infty$-group}. The homotopy labeled $O$ above diagram is the \emph{Hamiltonian}
of the prequantum operator. The image of the quantomorphisms in the symplectomorphisms
(given by composition the above diagram with the curvature morphism 
$F_{(-)} : \mathbf{B}^n U(1)_{\mathrm{conn}} \to \Omega^{n+1}_{\mathrm{cl}}$) is the 
group of \emph{Hamiltonian $n$-plectomorphisms}. A lift of an $\infty$-group action
$G \to \mathbf{Aut}(X)$ on $X$ from automorphisms of $X$ (i.e., diffeomorphisms) to quantomorphisms
is a \emph{Hamiltonian action}, infinitesimally (and dually) a \emph{momentum map}.

To define higher prequantum states, we fix a linear representation $(V, \rho)$ of the circle $n$-group $\mathbf{B}^{n-1}U(1)$ on some higher vector space $V$, i.e., a morphism $\rho:\mathbf{B}^nU(1)\to \mathbf{B}\mathrm{Aut}(V)$.
By the general results in \cite{NSSa} this is equivalent to fixing a homotopy fiber sequence of the form
$$
  \raisebox{20pt}{
  \xymatrix{
    \underline{V} \ar[r] & \underline{V}/\!/\mathbf{B}^{n-1}U(1)
	\ar[d]^{\mathbf{\rho}}
	\\
	& \mathbf{B}^n U(1)
  }
  }
$$
in $\mathbf{H}$. The vertical morphism here is the \emph{universal $\rho$-associated $V$-fiber $\infty$-bundle}
and characterizes $\rho$ itself. Given such, a section of the $V$-fiber bundle which is 
$\rho$-associated to $\mathbf{c}(\nabla)$ is equivalently a map
$$
  \Psi : \mathbf{c}(\nabla) \longrightarrow \mathbf{\rho}
$$
in the slice $\mathbf{H}_{/\mathbf{B}^n U(1)}$. This is a higher \emph{prequantum state} of the
prequantum geometry $(X, \nabla)$. Since every prequantum operator $\hat O$ as above
in particular is an auto-equivalence of the underlying prequantum bundle 
$\hat O : \mathbf{c}(\nabla) \stackrel{\simeq}{\longrightarrow} \mathbf{c}(\nabla)$ it
canonically acts on prequantum states given by maps as above simply by precomposition
$$
  \Psi \mapsto \hat O \circ \Psi
  \,.
$$
Notice also that from the perspective of section \ref{FieldsInSlices} all this has
an equivalent interpretation in terms of twisted cohomology: a preqantum state is a cocycle
in twisted $V$-cohomology, with the twist being the prequantum bundle. And a 
prequantum operator/quantomorphism is equivalently a twist automorphism 
(or ``generalized local diffeomorphism'').

For instance if $n = 1$ then $\omega$ is an ordinary (pre)symplectic form 
and $\nabla$ is the connection on a circle bundle. In this case the above notions
of prequantum operators, quantomorphism group, Heisenberg group and Poisson bracket Lie algebra
reproduce exactly all the traditional notions if $X$ is a smooth manifold,
and generalize them to the case that $X$ is for instance an orbifold or even itself a higher moduli stack, 
as we have seen. The canonical representation of the circle group $U(1)$ on the complex numbers
yields a homotopy fiber sequence
$$
  \raisebox{42pt}{
  \xymatrix{
    \underline{\mathbb{C}}
	\ar[r]
	&
	\underline{\mathbb{C}}/\!/\underline{U}(1) \ar[d]^{\mathbf{\rho}}
	\\
	&
	\mathbf{B}U(1)
  }
  }
  \,,
$$
where $\underline{\mathbb{C}}/\!/\underline{U}(1)$ is the stack corresponding to the ordinary action groupoid of the action of $U(1)$ on $\mathbb{C}$, 
and where the vertical map is the canonical functor forgetting the data of the local $\mathbb{C}$-valued functions. 
This is the \emph{universal complex line bundle} associated to the universal $U(1)$-principal bundle.
One readily checks that a prequantum state $\Psi : \mathbf{c}(\nabla) \to \mathbf{\rho}$,
hence a diagram of the form
$$
  \raisebox{20pt}{
  \xymatrix{
    X \ar[rr]^\sigma \ar[dr]_{\mathbf{c}(\nabla)} && \underline{\mathbb{C}}/\!/\underline{U}(1) \ar[dl]^{\mathbf{\rho}}
	\\
	& \mathbf{B}U(1)
  }
  }
$$
in $\mathbf{H}$ is indeed equivalently a section of the complex line bundle canonically associated to 
$\mathbf{c}(\nabla)$ and that under this equivalence the pasting composite
$$
  \xymatrix{
    X \ar[r]^\simeq_>{\ }="s" \ar[dr]_{\mathbf{c}(\nabla)}^{\ }="t" & X \ar[r] \ar[d]|{\mathbf{c}(\nabla)} & \underline{\mathbb{C}}/\!/\underline{U}(1)
	\ar[dl]^{\mathbf{\rho}}
	\\
	& \mathbf{B}U(1)
	\ar@{=>}_O "s"; "t"
  }
$$
is the result of the traditional formula for the action of the prequantum operator $\hat O$ on $\Psi$.

Instead of forgetting the connection on the prequantum bundle in the above composite, 
one can equivalently equip the prequantum state with a differential refinement, namely with its
\emph{covariant derivative} and then exhibit the prequantum operator action directly. 
Explicitly, let 
$\mathbb{C}/\!/U(1)_{\mathrm{conn}}$ denote the quotient stack
  $(\underline{\mathbb{C}}\times \Omega^1(-,\mathbb{R}) )/\!/\underline{U}(1)
  \,,$
with $U(1)$ acting diagonally. This sits in a 
homotopy fiber sequence
$$
  \xymatrix{
   \underline{\mathbb{C}} \ar[r] & 
      \underline{\mathbb{C}}/\!/\underline{U}(1)_{\mathrm{conn}} \ar[d]^{\rho_{\mathrm{conn}}}
	\\
	& \mathbf{B}U(1)_{\mathrm{conn}}
  }
$$
which may be thought of as the differential refinement of the above fiber sequence
$\underline{\mathbb{C}} \to \underline{\mathbb{C}}/\!/\underline{U}(1) \to \mathbf{B}U(1)$. 
(Compare this to section \ref{WithWilsonLoops}, where we had similarly 
seen the differential refinement of the 
fiber sequence $\underline{G}/\underline{T}_\lambda \to \mathbf{B}T_\lambda \to \mathbf{B}G$, 
which analogously characterizes the canonical action of $G$ on the coset space $G/T_{\lambda}$.)
Prequantum states are now equivalently maps
$$
  \widehat{\mathbf{\Psi}}
  :
  \nabla \longrightarrow \mathbf{\rho}_{\mathrm{conn}}
$$
in $\mathbf{H}_{/\mathbf{B}U(1)_{\mathrm{conn}}}$.
This formulation realizes a section of an associated line bundle equivalently as a connection on 
what is sometimes called a groupoid bundle. As such, $\widehat{\mathbf{\Psi}}$
has not just a 2-form curvature (which is that of the prequantum bundle) but also a 1-form
curvature: this is the covariant derivative $\nabla \sigma$ of the section.

Such a relation between sections of higher associated bundles and higher covariant derivatives
holds more generally. 
In the next degree for $n = 2$ one finds that the quantomorphism 2-group is the Lie 2-group
which integrates the \emph{Poisson bracket Lie 2-algebra} of the underlying 2-plectic geometry
as introduced in \cite{Rogers}.
In the next section we look at an example for $n = 2$ in more detail and show how it interplays
with the above example under transgression.

The above higher prequantum theory becomes a genuine quantum theory
after a suitable higher analog of 
a choice of \emph{polarization}.
In particular, for $\mathbf{L} : X \to \mathbf{B}^n U(1)_{\mathrm{conn}}$ an extended Lagrangian
of an $n$-dimensional quantum field theory as discussed in all our examples here, and for $\Sigma_k$ any closed
manifold, the polarized prequantum states of the transgressed prequantum bundle
$\mathrm{hol}_{\Sigma_k}\mathbf{Maps}(\Sigma_k, \mathbf{L})$ 
should form the 
$(n-k)$-vector spaces of higher quantum states in codimension $k$.
These states 
 would be assigned to $\Sigma_k$ by the 
\emph{extended quantum field theory}, in the sense of \cite{LurieTQFT}, 
obtained from the extended Lagrangian $\mathbf{L}$ by extended geometric quantization.
There is an equivalent reformulation of this last step for $n = 1$ given simply by the 
push-forward of the prequantum line bundle in K-theory 
(see section 6.8 of \cite{GinzburgGuilleminKarshon}) 
and so one would expect that accordingly
the last step of higher geometric quantization involves similarly a push-forward
of the associated $V$-fiber $\infty$-bundles above
in some higher generalized cohomology theory. But this remains to be investigated.

\subsubsection{Example: The anomaly-free gauge coupling of the open string}
\label{OpenStringGaugeCoupling}

As an example of these general phenomena, we close by briefly indicating how the
higher prequantum states of 3d Chern-Simons theory in codimension 2 reproduce the 
\emph{twisted Chan-Paton gauge bundles} of open string backgrounds, and how their
transgression to codimension 1 reproduces the cancellation of the 
Freed-Witten-Kapustin anomaly of the open string.

By the above, the Wess-Zumino-Witten gerbe 
$\mathbf{wzw} : {G \to \mathbf{B}^2 U(1)_{\mathrm{conn}}}$ as discussed in section
\ref{TheWZWGerbe} may be regarded as the \emph{prequantum 2-bundle} of Chern-Simons theory
in codimension 2 over the circle. Equivalently,  if we consider the WZW $\sigma$-model for the 
string on $G$ and take the limiting TQFT case obtained by 
sending the kinetic term to 0 while keeping
only the gauge coupling term in the action, then it is the extended Lagrangian of the string
$\sigma$-model: its transgression to the mapping space
out of a \emph{closed}  worldvolume $\Sigma_2$ of the string is the topological piece of the exponentiated
WZW $\sigma$-model action. For $\Sigma_2$ with boundary the situation is more interesting,
and this we discuss now.

The \emph{Heisenberg 2-group} of the prequantum geometry $(G,\mathbf{wzw})$ 
is\footnote{This follows for instance as the Lie integration by \cite{Prequantum1}
of the result in \cite{BaezRogers,Prequantum2} that the Heisenberg Lie 2-algebra here is the 
$\mathfrak{string}(\mathfrak{g})$ Lie 2-algebra.} the \emph{String 2-group} (see the appendix of \cite{FiorenzaSatiSchreiberI} for a review),
the smooth 2-group $\mathrm{String}(G)$ which is, up to equivalence, the loop space object of the
homotopy fiber of the smooth universal class $\mathbf{\mathbf{c}}$
$$
  \xymatrix{
    \mathbf{B}\mathrm{String}(G)
	\ar[r]
	&
	\mathbf{B}G
	\ar[r]^-{\mathbf{c}}
	&
	\mathbf{B}^3 U(1)
  }
  \,.
$$
The canonical representation of the 2-group $B U(1)$ is on the complex K-theory spectrum,
whose smooth (stacky) refinement is given by 
$\mathbf{B}U := \underset{\longrightarrow}{\lim}_n \mathbf{B}U(n)$ in $\mathbf{H}$
(see section 5.4.3 of \cite{survey} for more details). On any
component for fixed $n$ the action of the smooth 2-group $\mathbf{B}U(1)$ 
is exhibited by the long homotopy fiber sequence
$$
  \xymatrix{
    U(1) \longrightarrow U(n) \to \mathrm{PU}(n) \longrightarrow \mathbf{B}U(1)
	\longrightarrow \mathbf{B}U(n) \longrightarrow \mathbf{B}\mathrm{PU}(n)
    \ar[r]^-{\mathbf{dd}_n} & \mathbf{B}^2 U(1)	
  }
$$
in $\mathbf{H}$, in that $\mathbf{dd}_n$ is the universal $(\mathbf{B}U(n))$-fiber 2-bundle which is
associated by this action to the universal $(\mathbf{B}U(1))$-2-bundle.\footnote{
 The notion of $(\mathbf{B}U(n))$-fiber 2-bundle is equivalently that of 
 nonabelian $U(n)$-\emph{gerbes} in the original sense of Giraud, see \cite{NSSa}.
 Notice that for $n = 1$ this is more general than then notion of $U(1)$-bundle gerbe:
 a $G$-gerbe has structure 2-group $\mathbf{Aut}(\mathbf{B}G)$, but a $U(1)$-bundle gerbe
 has structure 2-group only in the left inclusion of the fiber sequence
 $\mathbf{B}U(1) \hookrightarrow \mathbf{Aut}(\mathbf{B}U(1)) \to \mathbb{Z}_2$.
}
The two d's in $\mathbf{dd}_n$ stand for Dixmier-Douady; namely, in the limit for $n\to \infty$ and under topological realization, the morphisms $\mathbf{dd}_n$ induce the Dixmier-Douady isomorphism $BPU\simeq K(\mathbb{Z},3)$.
Using the general higher representation theory 
in $\mathbf{H}$ as developed in \cite{NSSa},  a local section
of the $(\mathbf{B}U(n))$-fiber prequantum 2-bundle which is $\mathbf{dd}_n$-associated to 
the prequantum 2-bundle $\mathbf{wzw}$, hence a local prequantum 2-state, is, equivalently, a map
$$
  \mathbf{\Psi} : 
    \mathbf{wzw}|_Q
	\longrightarrow
	\mathbf{dd}_n
$$
in the slice $\mathbf{H}_{/\mathbf{B}^2 U(1)}$, where $\iota_Q : Q \hookrightarrow G$ is some subspace.
Equivalently (compare with the general discussion in section \ref{FieldsInSlices}), this is a map 
$$
  (\mathbf{\Psi}, \mathbf{wzw}) 
  :
  \iota_Q 
  \longrightarrow
  \mathbf{dd}_n
$$
in $\mathbf{H}^{(\Delta^1)}$, hence a diagram in $\mathbf{H}$ of the form
$$
  \raisebox{20pt}{
  \xymatrix{
    Q \ar[rr]^-{\mathbf{\Psi}}_>{\ }="s" \ar@{^{(}->}[d]_{\iota_Q}^>{\ }="t" 
    && \mathbf{B}\mathrm{PU}(n) \ar[d]^{\mathbf{dd}_n}
	\\
	G \ar[rr]_-{\mathbf{wzw}} 
	&& \mathbf{B}^2 U(1)\;.
	\ar@{=>} "s"; "t"
  }
  }
$$
One finds (section 5.4.3 of \cite{survey}) 
that this equivalently modulates a unitary bundle on $Q$ which is \emph{twisted}
by the restriction of $\mathbf{wzw}$ to $Q$ as in twisted K-theory
(such a twisted bundle is also called a \emph{gerbe module} if $\mathbf{wzw}$ is thought of in terms
of bundle gerbes \cite{BCMMS}). So  
$$
  \mathbf{dd}_n \;\in \mathbf{H}_{/\mathbf{B}^2 U(1)}
$$ 
is the moduli stack for twisted rank-$n$ unitary bundles.
As with the other moduli stacks before,  one finds a differential refinement of this moduli stack,
which we write
$$
  (\mathbf{dd}_n)_{\mathrm{conn}}
  \;:\;
  (\mathbf{B}U(n)/\!/\mathbf{B}U(1))_{\mathrm{conn}}
  \to
  \mathbf{B}^2 U(1)_{\mathrm{conn}}
  \,,
$$ 
and which modulates twisted unitary bundles with twisted connections
(bundle gerbe modules with connection). Hence 
a differentially refined state is a map
$\widehat {\mathbf{\Psi}}
  \;:\;
    \mathbf{wzw}|_Q
	\to
	(\mathbf{dd}_n)_{\mathrm{conn}}
$
in $\mathbf{H}_{/\mathbf{B}^2 U(1)_{\mathrm{conn}}}$; and this is precisely a 
twisted gauge field on a D-brane $Q$ on which open strings in $G$ may end. Hence these are the
\emph{prequantum 2-states} of Chern-Simons theory in codimension 2. 
Precursors of this perspective of Chan-Paton bundles over D-branes 
as extended prequantum 2-states can be found in \cite{Schreiber07, RogersQuantization}.

Notice that by the above discussion, together the discussion in 
section \ref{FieldsInSlices}, an equivalence
$$
  \hat O 
  : \xymatrix{\mathbf{wzw} \ar[r]^\simeq & \mathbf{wzw}} 
$$
in $\mathbf{H}_{/\mathbf{B}^2 U(1)_{\mathrm{conn}}}$
has two different, but equivalent, important interpretations:
\begin{enumerate}
 \item it is an element of the \emph{quantomorphism 2-group} 
 (i.e. the possibly non-linear generalization of the Heisenberg 2-group) of 2-prequantum operators;
 \item it is a twist automorphism analogous to the generalized diffeomorphisms for the fields in 
 gravity.
\end{enumerate}
Moreover, such a transformation is locally a structure well familiar from the literature on D-branes:
it is locally (on some cover) given by a transformation of the B-field 
of the form $B \mapsto B + d_{\mathrm{dR} } a$ for a local 1-form $a$ 
(this is the \emph{Hamiltonian 1-form} in the interpretation of this transformation in
higher prequantum geometry)
and its prequantum operator action on prequantum 2-states, hence on Chan-Paton gauge fields
$\widehat{\mathbf{\Psi}} : \xymatrix{\mathbf{wzw} \ar[r] & (\mathbf{dd}_n)_{\mathrm{conn}}}$ (by precomposition) 
is given by shifting the connection on a twisted Chan-Paton bundle (locally) by this
local 1-form $a$. This local gauge transformation data
$$
    B  \mapsto B + d a\;, \qquad \qquad
	A  \mapsto A + a\;,
$$
is familiar from string theory and D-brane gauge theory 
(see e.g. \cite{Pol}). The 2-prequantum operator action
$\Psi \mapsto \hat O \Psi$ which we see here is the fully globalized refinement of this
transformation.

\paragraph{Surface transport and the twisted bundle part of Freed-Witten-Kapustin anomalies.}
The map $\widehat{\mathbf{\Psi}} : (\iota_Q, \mathbf{wzw}) \to (\mathbf{dd}_n)_{\mathrm{conn}}$ above is 
the gauge-coupling part of the extended Lagrangian of the 
\emph{open} string on $G$ in the presence of a D-brane $Q \hookrightarrow G$. We indicate what this
means and how it works. Note that for all of the following the target space $G$ and background gauge field
$\mathbf{wzw}$ could be replaced by any
target space with any circle 2-bundle with connection on it.


The object $\iota_Q$ in $\mathbf{H}^{(\Delta^1)}$
is the target space for the open string. The worldvolume of that string is a smooth compact manifold
$\Sigma$ with boundary inclusion $\iota_{\partial \Sigma} : \partial \Sigma \to \Sigma$, also
regarded as an object in $\mathbf{H}^{(\Delta^1)}$. A field configuration of the 
string $\sigma$-model is then a map
$$
  \phi : \iota_\Sigma \to \iota_Q
$$
in $\mathbf{H}^{(\Delta^1)}$, hence a diagram
$$
  \xymatrix{
    \partial \Sigma \ar[rr] \ar@{^{(}->}[d]_{\iota_{\partial \Sigma}} 
    && Q \ar@{^{(}->}[d]^{\iota_Q}
	\\
	\Sigma \ar[rr]^\phi 
	&& G
  }
$$
in $\mathbf{H}$, hence a smooth function $\phi : \Sigma \to G$ subject to the constraint that the boundary
of $\Sigma$ lands on the D-brane $Q$. Postcomposition with the background gauge field 
$\widehat {\mathbf{\Psi}}$ 
yields the diagram
$$
  \xymatrix{
    \partial \Sigma \ar[rr] \ar@{^{(}->}[d]_{\iota_{\partial \Sigma}} 
	&& Q \ar@{^{(}->}[d]^{\iota_Q} \ar[r]^-{\widehat {\mathbf{\Psi}}} 
	& (\mathbf{B}U(n)/\!/U(1))_{\mathrm{conn}}
	\\
	\Sigma \ar[rr]^\phi && G \ar[r]_-{\mathbf{wzw}} & \mathbf{B}^2 U(1)_{\mathrm{conn}}\;.	
  }
$$
Comparison with the situation of Chern-Simons theory with Wilson lines in section \ref{WithWilsonLoops}
shows that the total action functional for the open string should be the product of the
fiber integration of the top composite morphism with that of the bottom composite morphisms.
 Hence that functional is the product
of the surface parallel transport of the $\mathbf{wzw}$ $B$-field over $\Sigma$ with the 
line holonomy of the twisted Chan-Paton bundle over $\partial \Sigma$. 

This is indeed again true, but for more subtle reasons this time,
since  the fiber integrations
here are \emph{twisted}. For the surface parallel transport we mentioned this already at the end of 
section \ref{Sigma-Models}: since $\Sigma$ has a boundary, 
parallel transport over $\Sigma$ does not yield a function on the mapping space out of $\Sigma$, but
rather a section of the line bundle on the mapping space out of $\partial \Sigma$, pulled back to this
larger mapping space.

Furthermore, the connection on a twisted unitary bundle  does not quite 
have a well-defined traced holonomy in $\mathbb{C}$, but rather a well defined traced
holonomy up to a coherent twist. More precisely, the transgression of the 
WZW 2-connection to maps
out of the circle as in section \ref{PrequantumLineBundlesOnModuliStacks} 
fits into a diagram of moduli stacks in $\mathbf{H}$ of the form
$$
  \raisebox{20pt}{
  \xymatrix{
    \mathbf{Maps}(S^1, (\mathbf{B}U(n)/\!/\mathbf{B}U(1))_{\mathrm{conn}})
	\ar[dd]|{\mathbf{Maps}(S^1, (\mathbf{dd}_n)_{\mathrm{conn}})}
	\ar[rr]^-{\mathrm{tr}\,\mathrm{hol}_{S^1}}
	&&
	\underline{\mathbb{C}}/\!/\underline{U}(1)_{\mathrm{conn}}
	\ar[dd]
	\\
	\\
	\mathbf{Maps}(S^1 , \mathbf{B}^2 U(1)_{\mathrm{conn}})
	\ar[rr]^-{\mathrm{hol}_{S^1}}
	&&
	\mathbf{B}U(1)_{\mathrm{conn}}\;.
  }
  }
$$
This is a transgression-compatibility of the form that we have already seen in 
section \ref{TheWZWGerbe}. 

In summary, we obtain the transgression of the extended Lagrangian of the 
open string in the background of B-field and Chan-Paton bundles 
as the following pasting diagram of moduli stacks in $\mathbf{H}$
(all squares are filled with homotopy 2-cells, which are notationally suppressed for readability)

$$
\hspace{-1cm}
  \xymatrix{
    \mathbf{Fields}_{\mathrm{OpenString}}(\iota_{\partial \Sigma})
	\ar[rr]
	\ar[dd]
	&&
	\mathbf{Maps}(\Sigma,G)
	\ar[rr]^{\exp(2 \pi i \int_\Sigma [\Sigma, \mathbf{wzw}] )}
	\ar[dd]|{~~~\mathbf{Maps}(\iota_{\partial \Sigma}, G)}
	&&
	\underline{\mathbb{C}}/\!/\underline{U}(1)_{\mathrm{conn}}
	\ar[dddd]
	\\
	\\
	\mathbf{Maps}(S^1, Q)
	\ar[rr]^-{\mathbf{Maps}(S^1, \iota_Q)} 
	\ar[d]|-{~~\mathbf{Maps}(S^1, \widehat{\mathbf{\Psi}})}
	&&
	\mathbf{Maps}(S^1, G)
	\ar[dr]|{\mathbf{Maps}(S^1, \mathbf{wzw})}
	\\
	\mathbf{Maps}(S^1, (\mathbf{B}U(n)/\!/\mathbf{B}U(1))_{\mathrm{conn}})
	\ar[rrr]^{\mathbf{Maps}(S^1, (\mathbf{dd}_n)_{\mathrm{conn}})}
	\ar[d]|{\mathrm{tr}\,\mathrm{hol}_{S^1}}
	&&&
	\mathbf{Maps}(S^1, \mathbf{B}^2 U(1)_{\mathrm{conn}})
	\ar[dr]|{\mathrm{hol}_{S^1}}
	\\
	\underline{\mathbb{C}}/\!/\underline{U}(1)_{\mathrm{conn}}
	\ar[rrrr]
	&&
	&&
	\mathbf{B}U(1)_{\mathrm{conn}}
  }
$$
Here
\begin{itemize}
  \item 
    the top left square is the homotopy pullback square that computes the mapping stack
	$\mathbf{Maps}(\iota_{\partial \Sigma}, \iota_Q)$ in $\mathbf{H}^{(\Delta^1)}$, which here
	is simply the smooth space of string configurations $\Sigma \to G$ which are such that the string boundary
	lands on the D-brane $Q$;
  \item 
    the top right square is the twisted fiber integration of the $\mathbf{wzw}$ background 2-bundle
	with connection: this exhibits the parallel transport of the 2-form connection over the 
	worldvolume $\Sigma$ with boundary $S^1$ as a section of the pullback of the 
	transgression line bundle
	on loop space to the space of maps out of $\Sigma$;
  \item
    the bottom square is the above compatibility between the twisted traced holonomy of twisted unitary bundles
	and the trangression of their twisting 2-bundles.
\end{itemize}
The total diagram obtained this way exhibits a difference between two section of a single complex
line bundle on $\mathbf{Fields}_{\mathrm{OpenString}}(\iota_{\partial \Sigma})$
(at least one of them non-vanishing), hence a
map
$$
  \exp\left(2 \pi i \int_{\Sigma} [\Sigma, \mathbf{wzw}]  \right)
  \cdot
  \mathrm{tr}\,\mathrm{hol}_{S^1}([S^1, \widehat {\mathbf{\Psi}}])
  \;:\;
  \mathbf{Fields}_{\mathrm{OpenString}}(\iota_{\partial \Sigma})
  \longrightarrow
 \underline{ \mathbb{C}}
  \,.
$$
This is the well-defined action functional of the open string with endpoints on the D-brane $Q \hookrightarrow G$,
charged under the background $\mathbf{wzw}$ B-field and under the twisted Chan-Paton gauge bundle
$\widehat {\Psi}$. 

Unwinding the definitions, one finds that this phenomenon is precisely the 
twisted-bundle-part, due to Kapustin \cite{Kapustin}, 
of the Freed-Witten anomaly cancellation for open strings on D-branes, 
hence is the Freed-Witten-Kapustin anomaly cancellation
mechanism either for the open bosonic string or else for the open type II superstring on 
$\mathrm{Spin}^c$-branes. Notice how in the traditional discussion the existence of
twisted bundles on the D-brane is identified just as \emph{some} construction
that happens to cancel the B-field anomaly. Here, in the perspective of extended quantization, we 
see that this choice follows uniquely from the general theory of extended prequantization, once
we recognize that $\mathbf{dd}_n$ above is (the universal associated 2-bundle induced by) 
the canonical representation of the circle 2-group $\mathbf{B}U(1)$, just as in one codimension up
$\mathbb{C}$ is the canonical representation of the circle 1-group $U(1)$.

\newpage
\noindent
{\bf \large Acknowledgements}\\

\noindent D.F. thanks ETH Z\"urich for hospitality. The research of H.S.  is supported by NSF Grant PHY-1102218. 
U.S. thanks the University of Pittsburgh for an invitation in December 2012, during which part of this
work was completed.


\end{document}